\documentclass[rmp,twocolumn,aps]{revtex4}
\usepackage{graphicx}
\begin{document}
\newcommand{\ket}[1]{\mbox{$|\!#1\;\!\rangle$}}
\newcommand{\bra}[1]{\mbox{$\langle#1|$}}
\newcommand{\aver}[1]{\mbox{$<\!#1\!\!>$}}
\def\ua{\uparrow}
\def\da{\downarrow}
\def\be{\begin{equation}}
\def\ee{\end{equation}}

\def\bea{\begin{eqnarray}}
\def\eea{\end{eqnarray}}

\def\bfig{\begin{figure}[htbp]}
\def\efig{\end{figure}}

\def\<{\langle}
\def\>{\rangle}

\def\up{\uparrow}
\def\down{\downarrow}

\title{Spins in few-electron quantum dots}

\author{R. Hanson}
\email{hanson@physics.ucsb.edu}
\affiliation{Center for Spintronics and Quantum Computation,
University of California, Santa Barbara, California 93106, USA}
\affiliation{Kavli Institute of Nanoscience, Delft University of
Technology, PO Box 5046, 2600 GA Delft, The Netherlands}
\author{L. P. Kouwenhoven}
\affiliation{Kavli Institute of Nanoscience, Delft University of
Technology, PO Box 5046, 2600 GA Delft, The Netherlands}
\author{J. R. Petta}
\affiliation{Department of Physics, Princeton University,
Princeton, New Jersey 08544, USA}
\affiliation{Department of Physics, Harvard University, Cambridge, Massachusetts 02138, USA}
\author{S. Tarucha}
\affiliation{Department of Applied Physics and ICORP-JST,
The University of Tokyo, Hongo, Bunkyo-ku, Tokyo, 113-8656, Japan}
\author{L. M. K. Vandersypen}
\affiliation{Kavli Institute of NanoScience, Delft University of
Technology, PO Box 5046, 2600 GA Delft, The Netherlands}

\date{\today}

\begin{abstract}
The canonical example of a quantum mechanical two-level system is spin. The simplest picture of spin is a magnetic moment pointing up or down. The full quantum properties of spin become apparent in phenomena such as superpositions of spin states, entanglement among spins and quantum measurements. Many of these phenomena have been observed in experiments performed on ensembles of particles with spin. 
Only in recent years systems have been realized in which individual electrons can be trapped and their quantum properties can be studied, thus avoiding unnecessary ensemble averaging. 
This review describes experiments performed with quantum dots, which are nanometer-scale boxes defined in a semiconductor host material. Quantum dots can hold a precise, but tunable number of electron spins starting with 0, 1, 2, etc. Electrical contacts can be made for charge transport measurements and electrostatic gates can be used for controlling the dot potential. This system provides virtually full control over individual electrons. This new, enabling technology is stimulating research on individual spins.
This review describes the physics of spins in quantum dots containing one or two electrons, from an experimentalist's viewpoint. Various methods for extracting spin properties from experiment are presented, restricted exclusively to electrical measurements. Furthermore, experimental techniques are discussed that allow for: (1) the rotation of an electron spin into a superposition of up and down, (2) the measurement of the quantum state of an individual spin and (3) the control of the interaction between two neighbouring spins by the Heisenberg exchange interaction. Finally, the physics of the relevant relaxation and dephasing mechanisms is reviewed and experimental results are compared with theories for spin-orbit and hyperfine interactions. All these subjects are directly relevant for the fields of quantum information processing and spintronics with single spins (i.e. single-spintronics). 
\end{abstract}

\maketitle

\tableofcontents

\section{Introduction}
\label{Introduction} The spin of an electron remains a somewhat
mysterious property. The first derivations in 1925 of the spin
magnetic moment, based on a rotating charge distribution of finite
size, are in conflict with special relativity theory. Pauli
advised the young Ralph Kronig not to publish his theory since
``it has nothing to do with reality''. More fortunate were Samuel
Goudsmit and George Uhlenbeck, who were supervised by Ehrenfest:
``Publish, you are both young enough to be able to afford a
stupidity!''~\footnote{See http://www.lorentz.leidenuniv.nl/history/spin/goudsmit.html.}. It requires Dirac's equation to
find that the spin eigenvalues correspond to one-half times
Planck's constant, $\hbar$, while considering the electron as a
point particle. The magnetic moment corresponding to spin is
really very small and in most practical cases it can be ignored.
For instance, the most sensitive force sensor to date has only
recently been able to detect some effect from the magnetic moment
of a single electron spin~\cite{rugar04}. In solids, spin can
apparently lead to strong effects, given the existence of
permanent magnets. Curiously, this has little to do with the
strength of the magnetic moment. Instead, the fact that spin is
associated with its own quantum number, combined with Pauli's
exclusion principle that quantum states can at most be occupied
with one fermion, leads to the phenomenon of exchange interaction.
Because the exchange interaction is a correction term to the
strong Coulomb interaction, it can be of much larger strength in
solids than the dipolar interaction between two spin magnetic
moments at an atomic distance of a few Angstroms. It is the
exchange interaction that forces the electron spins in a
collective alignment, together yielding a macroscopic
magnetization ~\cite{ashcroft}. It remains striking, that an abstract concept as
(anti-)symmetrization in the end gives rise to magnets.

The magnetic state of solids has found important applications in
electronics, in particular for memory devices. An important field
has emerged in the last two decades known as spintronics.
Phenomena like Giant Magneto Resistance or Tunneling Magneto
Resistance form the basis for magnetic heads for reading out the
magnetic state of a memory cell. Logic gates have been realized
based on magnetoresistance effects as well~\cite{wolf01,zutic04}.
In addition to applications, important scientific discoveries have been
made in the field of spintronics~\cite{awsch07}, including magnetic semiconductors~\cite{ohno98} and the spin Hall
effect~\cite{awschalom05}. It is important to note that all the
spintronics phenomena consider macroscopic numbers of spins.
Together these spins form things like spin densities or a
collective magnetization. Although the origin of spin densities
and magnetization is quantum mechanical, these collective,
macroscopic variables behave entirely classically. For instance, the
magnetization of a micron-cubed piece of Cobalt is a classical
vector. The quantum state of this vector dephases so rapidly that
quantum superpositions or entanglement between vectors is never
observed. One has to go to systems with a small number of spins,
for instance in magnetic molecules, in order to find quantum
effects in the behaviour of the collective
magnetization (for an overview, see e.g.,~\textcite{gunther91}).

The technological drive to make electronic devices continuously
smaller has some interesting scientific consequences. For
instance, it is now routinely possible to make small electron
``boxes'' in solid state devices that contain an integer number of
conduction electrons. Such devices are usually operated as
transistors (via field-effect gates) and are therefore named
single electron transistors. In semiconductor boxes the number of
trapped electrons can be reduced all the way to zero, or one, two,
etc. Such semiconductor single electron transistors are called
quantum dots ~\cite{kouwenhoven01}. Electrons are trapped in a quantum
dot by repelling electric fields imposed from all sides. The final
region in which a small number of electrons can still exist is
typically at the scale of tens of nanometers. The eigenenergies in
such boxes are discrete. Filling these states with electrons
follows the rules from atomic physics, including Hund's rule,
shell filling, etc. 

Studies with quantum dots have been performed successfully
during the nineties. By now it has become standard technology to confine single electron charges. Electrons
can be trapped as long as one desires. Changes in charge when one
electron tunnels out of the quantum dot can be measured on a
microsecond timescale. Compared to this control of charge, it is
very difficult to control individual spins and measure the spin of
an individual electron. Such techniques have been developed only
over the past few years.

In this review we describe experiments in which individual spins
are controlled and measured. This is mostly an experimental review
with explanations of the underlying physics. This review is
strictly limited to experiments that involve one or two electrons
strongly confined to single or double quantum dot devices. The
experiments show that one or two electrons can be trapped in a
quantum dot; that the spin of an individual electron can be put in
a superposition of up and down states; that two spins can be made
to interact and form an entangled state such as a spin singlet or triplet state; and that the result of such manipulation can be measured on individual spins.

These abilities of almost full control over the spin of individual
electrons enable the investigation of a new regime: single spin
dynamics in a solid state environment. The dynamics are fully
quantum mechanical and thus quantum coherence can be studied on an
individual electron spin. The exchange interaction is now also
controlled on the level of two particular spins that are brought
into contact simply by varying some voltage knob.

In a solid the electron spins are not completely decoupled from
other degrees of freedom. First of all, spins and orbits are
coupled by the spin-orbit interaction. Second, the electron spins
have an interaction with the spins of the atomic nuclei, i.e. the
hyperfine interaction. Both interactions cause the life time of a
quantum superposition of spin states to be finite. We therefore
also describe experiments that probe spin-orbit and hyperfine
interactions by measuring the dynamics of individual spins.

The study of individual spins is motivated by an interest in
fundamental physics, but also by possible applications. First of
all, miniaturized spintronics is developing towards single spins.
In this context, this field can be denoted as
single-spintronics~\footnote{Name coined by Stu Wolf, private communication.} in analogy to single-electronics.
A second area of applications is quantum information science. Here
the spin states form the qubits. The
original proposal by Loss and DiVincenzo~\cite{LossDiVincenzo} has
been the guide in this field. In the context of quantum
information, the experiments described in this review demonstrate
that the five DiVincenzo criteria for universal quantum
computation using single electron spins have been fulfilled to a large extent~\cite{DiVincenzo_criteria}:
initialization, one- and two-qubit operations, long coherence times
and readout. Currently, the state of the art is at the level of
single and double quantum dots and much work is required to build
larger systems.

In this review the system of choice are quantum dots in GaAs
semiconductors, simply because these have been most successful.
Nevertheless, the physics is entirely general and can be fully
applied to new material systems such as silicon based transistors,
carbon nanotubes, semiconductor nanowires, graphene devices, etc.
These other host materials may have advantageous spin properties.
For instance, carbon-based devices can be purified with the
isotope $^{12}$C in which the nuclear spin is zero, thus entirely
suppressing spin dephasing by hyperfine interaction. This kind of
hardware solution to engineer a long-lived quantum system will be
discussed at the end of this review.
Also, we here restrict ourselves exclusively to electron transport
measurements of quantum dots, leaving out optical spectroscopy of
quantum dots, which is a very active field in its own~\footnote{see e.g. ~\textcite{greilich06a,AtatureScience2006,KrennerPRL2006,BerezovskyScience2006} and references therein}. Again, much of the physics discussed in this review also applies to optically 
measured quantum dots.

Section~\ref{Section:Basics} starts with an introduction on
quantum dots including the basic model of Coulomb blockade to
describe the relevant energies. These energies can be visualized
in transport experiments and the relation between experimental
spectroscopic lines and underlying energies are explained in
section~\ref{Section:SpinSpectroscopy}. This spectroscopy is
specifically applied to spin states in single quantum dots in
section ~\ref{Section:SingleDotSpin}. Section
~\ref{Section:ChargeSensing} introduces a charge-sensing technique
that is used in section~\ref{Section:Readout} to read out the spin
state of individual electrons. Section~\ref{Spinenvironment}
provides an extensive description of spin-orbit and hyperfine
interactions. In section ~\ref{Section:DoubleDotSpin}, spin states
in double quantum dots are introduced and the important concept of
Pauli spin blockade is discussed. Quantum coherent manipulations
of spins in double dots are discussed in
section~\ref{Section:Coherent}. Finally, a perspective is outlined
in section X.

\section{Basics of quantum dots}
\label{Section:Basics}
\subsection{Introduction to quantum dots}
\begin{figure}[htb]
\includegraphics[width=3.4in, clip=true]{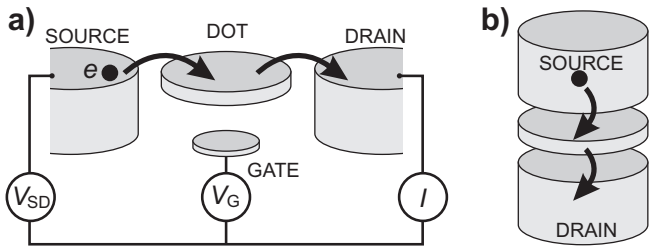}
\caption{ Schematic picture of a quantum dot in (a) a lateral
geometry and (b) in a vertical geometry. The quantum dot
(represented by a disk) is connected to source and drain
reservoirs via tunnel barriers, allowing the current through the
device, $I$, to be measured in response to a bias voltage,
$V_{SD}$ and a gate voltage, $V_G$. } \label{fig1:lateraldot}
\end{figure}

A quantum dot is an artificially structured system that can be
filled with electrons (or holes). The dot can be coupled via
tunnel barriers to reservoirs, with which electrons can be
exchanged (see Fig.~\ref{fig1:lateraldot}). By attaching current
and voltage probes to these reservoirs, we can measure the
electronic properties. The dot is also coupled capacitively to one
or more `gate' electrodes, which can be used to tune the
electrostatic potential of the dot with respect to the reservoirs.

Because a quantum dot is such a general kind of system, there
exist quantum dots of many different sizes and materials: for
instance single molecules trapped between electrodes
\cite{Pasupathy}, normal metal\cite{Petta_spin_orbit},
superconducting \cite{Ralph_PRL,RalphReview01} or ferromagnetic
nanoparticles~\cite{Gueron}, self-assembled quantum dots
\cite{McEuen}, semiconductor lateral~\cite{Kouwenhoven97} or
vertical dots~\cite{kouwenhoven01}, and also semiconducting nanowires
or carbon nanotubes~\cite{Bjork_nanowire_dot,dekker99,mceuen2000}.

The electronic properties of quantum dots are dominated by two
effects. First, the Coulomb repulsion between the electrons on the
dot leads to an energy cost for adding an extra electron to the
dot. Due to this \textit{charging energy}, tunneling of electrons
to or from the reservoirs can be dramatically suppressed at low
temperatures; this phenomena is called \textit{Coulomb blockade}
\cite{Beenakker}. Second, the confinement in all three directions
leads to quantum effects that strongly influence the electron
dynamics. Due to the resulting discrete energy spectrum, quantum
dots behave in many ways as \emph{artificial
atoms}~\cite{kouwenhoven01}.

The physics of dots containing more than two electrons has been
reviewed before~\cite{Kouwenhoven97,ReimannRMP}. Therefore, we
focus on single and coupled quantum dots containing only one or
two electrons. These systems are particularly important as they
constitute the building blocks of proposed electron spin-based
quantum information
processors~\cite{LossDiVincenzo,DiVincenzoNature2000,Levy,wulidar02a, wulidar02b,ByrdLidar,Meier,Kyriakidis2005,TaylerNaturePhysics2005,hanson06}.

\subsection{Fabrication of gated quantum dots}
\label{Sec:Fabrication}
The bulk of the experiments discussed in this review was performed
on electrostatically defined quantum dots in GaAs. These devices
are sometimes referred to as ``lateral dots'' because of the
lateral gate geometry.

\begin{figure}[htb]
\includegraphics[width=3.4in, clip=true]{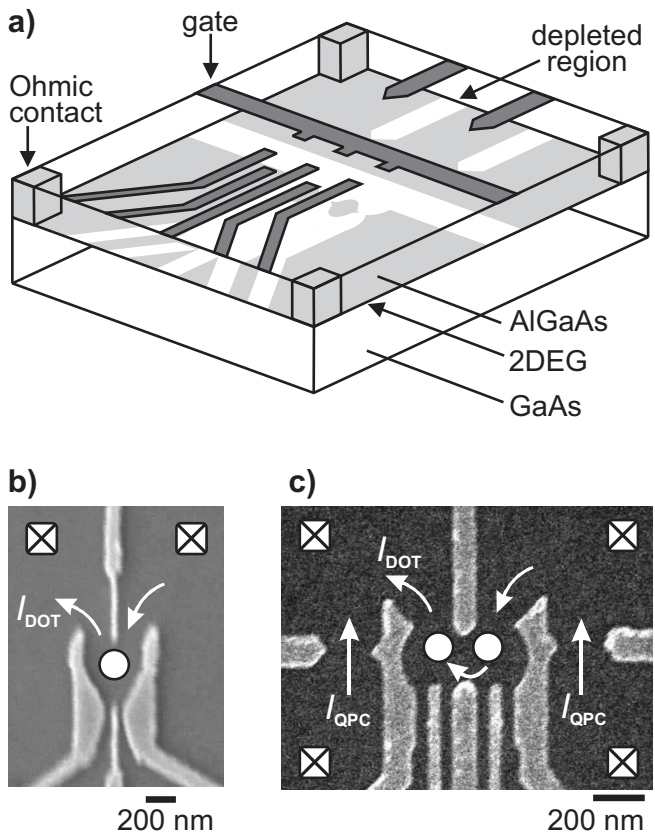}
\caption{Lateral quantum dot device defined by metal surface
electrodes. \textbf{(a)} Schematic view. Negative voltages applied
to metal gate electrodes (dark gray) lead to depleted regions
(white) in the 2DEG (light gray). Ohmic contacts (light gray
columns) enable bonding wires (not shown) to make electrical
contact to the 2DEG reservoirs. \textbf{(b)-(c)} Scanning electron
micrographs of a few-electron single-dot device (b) and a
double-dot device (c), showing the gate electrodes (light gray) on
top of the surface (dark gray). The white dots indicate the
location of the quantum dots. Ohmic contacts are shown in the
corners. White arrows outline the path of current $I_{DOT}$ from
one reservoir through the dot(s) to the other reservoir. For the
device in (c), the two gates on the side can be used to create two
quantum point contacts, which can serve as electrometers by
passing a current $I_{QPC}$. Note that this device can also be
used to define a single dot. Image in (b) courtesy of A.
Sachrajda.} \label{fig:structure}
\end{figure}

Lateral GaAs quantum dots are fabricated from heterostructures of
GaAs and AlGaAs grown by molecular beam epitaxy, (see
Fig.~\ref{fig:structure}). By doping the AlGaAs layer with Si,
free electrons are introduced. These accumulate at the GaAs/AlGaAs
interface, typically 50-100 nm below the surface, forming a
two-dimensional electron gas (2DEG) -- a thin ($\sim$10 nm) sheet
of electrons that can only move along the interface. The 2DEG can
have a high mobility and relatively low electron density
(typically $10^5-10^7$ cm$^2$/Vs and $\sim 1-5\times 10^{15}$
m$^{-2}$, respectively). The low electron density results in a
large Fermi wavelength ($\sim 40$ nm) and a large screening
length, which allows us to locally deplete the 2DEG with an
electric field. This electric field is created by applying
negative voltages to metal gate electrodes on top of the
heterostructure (see Fig.~\ref{fig:structure}a).

Electron-beam lithography enables fabrication of gate structures
with dimensions down to a few tens of nanometers
(Fig.~\ref{fig:structure}), yielding local control over the
depletion of the 2DEG with roughly the same spatial resolution.
Small islands of electrons can be isolated from the rest of the
2DEG by choosing a suitable design of the gate structure, thus
creating quantum dots. Finally, low-resistance (Ohmic) contacts
are made to the 2DEG reservoirs. To access the quantum phenomena
in GaAs gated quantum dots, they have to be cooled down to well
below 1~K. All experiments that are discussed in this review are
performed in dilution refrigerators with typical base temperatures
of 20~mK.

In so-called vertical quantum dots, control over the number of electrons
down to zero was already achieved in the 1990s~\cite{kouwenhoven01}. In
lateral gated dots this proved to be more difficult, since
reducing the electron number by driving the gate voltage to more
negative values tends to decrease the tunnel coupling to the
leads. The resulting current through the dot can then become
unmeasurably small before the few-electron regime is reached.
However, by proper design of the surface gate geometry the
decrease of the tunnel coupling can be compensated for.

In 2000, Ciorga~\textit{et al.} reported measurements on the first
lateral few-electron quantum dot~\cite{CiorgaPRB2000}. Their
device, shown in Fig.~\ref{fig:structure}b, makes use of two types
of gates specifically designed to have different functionalities.
The gates of one type are big and largely enclose the quantum dot.
The voltages on these gates mainly determine the dot potential.
The other type of gate is thin and just reaches up to the barrier
region. The voltage on this gate has a very small effect on the
dot potential but it can be used to set the tunnel barrier. The
combination of the two gate types allows the dot potential (and
thereby electron number) to be changed over a wide range while
keeping the tunnel rates high enough for measuring electron
transport through the dot.

Applying the same gate design principle to a double quantum dot,
Elzerman~\textit{et al.} demonstrated in 2003 control over the
electron number in both dots while maintaining tunable tunnel
coupling to the reservoir~\cite{ElzermanPRB2003}. Their design is
shown in Fig.~\ref{fig:structure}c (for more details on design
considerations and related versions of this gate design, see
~\textcite{HansonThesis2005}). In addition to the coupled
dots, two quantum point contacts (QPCs) are incorporated in this
device to serve as charge sensors. The QPCs are placed close to
the dots, thus ensuring a good charge sensitivity. This design has
become the standard for lateral coupled quantum dots and is used
with minor adaptions by several research
groups~\cite{PettaPRL2004,PioroPRB2005}; one noticable improvement
has been the electrical isolation of the charge sensing part of
the circuit from the reservoirs that connect to the
dot~\cite{HansonPRL2005}.

\subsection{Measurement techniques}
In this review, two all-electrical measurement techniques are
discussed: i) measurement of the current due to transport of
electrons through the dot, and ii) detection of changes in the
number of electrons on the dot with a nearby electrometer,
so-called \textit{charge sensing}. With the latter technique, the
dot can be probed \textit{non-invasively} in the sense that no
current needs to be sent through the dot.

The potential of charge sensing was first demonstrated in the
early 1990s~\cite{AshooriPRL92,FieldPRL93}. But whereas current
measurements were already used extensively in the first
experiments on quantum dots ~\cite{Kouwenhoven97}, charge sensing
has only recently been fully developed as a spectroscopic
tool~\cite{ElzermanAPL2004,JohnsonPRB2005}. Several
implementations of electrometers coupled to a quantum dot have
been demonstrated: a single-electron transistor fabricated on top
of the heterostructure~\cite{AshooriPRL92,Lu03a}, a second
electrostatically defined quantum
dot~\cite{HoffmanPRB1995,Fujisawa04a} and a quantum point contact
(QPC)~\cite{FieldPRL93,Sprinzak2002}. The QPC is the most widely
used because of its ease of fabrication and experimental
operation. We discuss the QPC operation and charge sensing
techniques in more detail in section~\ref{QPC} .

We briefly compare charge sensing to electron transport
measurements. The smallest currents that can be resolved in optimized setups and devices are roughly
10~fA, which sets a lower bound of order 10 fA/$e \approx$ 100 kHz on the
tunnel rate to the reservoir, $\Gamma$, for which transport
experiments are possible (see e.g. ~\textcite{VandersypenAPL2004} for a discussion on noise sources). For $\Gamma<$~100~kHz the charge detection technique can be used to resolve electron tunneling in
real time. Because the coupling to the leads is a source of
decoherence and relaxation (most notably via cotunneling), charge
detection is preferred for quantum information purposes since it
still functions for very small couplings to a (single) reservoir.

Measurements using either technique are conveniently understood
with the Constant Interaction model. In the next section we
use this model to describe the physics of single dots and show
how relevant spin parameters can be extracted from measurements.

\subsection{The Constant Interaction model}
\label{CImodel}
We briefly outline the main ingredients of the Constant Interaction model; for more
extensive discussions see~\textcite{kouwenhoven01,Beenakker,Kouwenhoven97}. The model is
based on two assumptions. First, the Coulomb interactions among
electrons in the dot, and between electrons in the dot and those
in the environment, are parameterized by a single, constant
capacitance, $C$. This capacitance is the sum of the capacitances
between the dot and the source, $C_S$, the drain, $C_D$, and the
gate, $C_G$: $C = C_S + C_D + C_G$. (In general, capacitances to
multiple gates and other parts of the 2DEG will also play a role;
they can simply be added to $C$). The second assumption is that
the single-particle energy level spectrum is independent of these
interactions and therefore of the number of electrons. Under these
assumptions, the total energy $U(N)$ of a dot with $N$ electrons
in the ground state, with voltages $V_{S}$, $V_D$ and $V_G$
applied to  the source, drain and gate respectively, is given by

\begin{eqnarray}
U(N) &=& \frac{[-|e|(N\!-\!N_0) + C_S V_{S}+C_D V_{D}+ C_G V_G]^2}{2C} \nonumber \\
&&+ \sum_{n=1}^N E_n(B) \label{introenergy}
\end{eqnarray}

\noindent where $-|e|$ is the electron charge, $N_0 |e|$ is the charge in the dot compensating the positive background charge originating from the donors in the heterostructure, and $B$ is the applied magnetic field. The terms $C_S V_{S}$, $C_D V_{D}$ and $C_G V_G$ can be changed continuously and represent an effective induced charge that changes the electrostatic potential on the dot. The last term of Eq.~\ref{introenergy} is a sum over the occupied single-particle energy levels, $E_n(B)$, which depend on the characteristics of the confinement potential.

The electrochemical potential $\mu(N)$ of the dot is defined as:
\begin{eqnarray}
&&\mu(N) \equiv \hspace{0.1cm} U(N) - U(N\!-\!1)= \nonumber \\
 &&(N\!-\! N_0 -\!\frac{1}{2})E_C\! -\! \frac{E_C}{|e|}
(C_S V_{S}\!+\! C_D V_{D}\! +\! C_G V_G)\! +\! E_N \ \ \ \ 
\label{mu}
\end{eqnarray}

\noindent where $E_C=e^2/C$ is the charging energy. The
electrochemical potential contains an electrostatic part (first
two terms) and a chemical part (last term). Here, $\mu(N)$ denotes
the transition between the $N$-electron ground state, $GS(N)$, and
the ($N\!-\!1$)-electron ground state, $GS(N\!-\!1)$. When also
excited states play a role, we have to use a more explicit
notation to avoid confusion: the electrochemical potential for the
transition between the ($N-1$)-electron state \ket{\:a} and the
$N$-electron state \ket{\:b} is then denoted as
$\mu_{a\leftrightarrow b}$, and is defined as the difference in
total energy between state \ket{\:b}, $U_b(N)$, and state
\ket{\:a}, $U_a(N\!-\!1)$:
\begin{equation}
\mu_{a\leftrightarrow b} \equiv U_b(N) -U_a(N\!-\!1)\ \
\label{mu2}
\end{equation}

Note that the electrochemical potential depends \textit{linearly}
on the gate voltage, whereas the energy has a quadratic
dependence. In fact, the dependence is the same for all $N$ and
the whole `ladder' of electrochemical potentials can be moved up
or down while the distance between levels remains
constant~\footnote{Deviations from this model are sometimes
observed in systems where the source-drain voltage and gate
voltage are varied over a very wide range; one notable example
being single molecules trapped between closely-spaced electrodes,
where the capacitances can depend on the electron state.}. It is
this property that makes the electrochemical potential the most
convenient quantity for describing electron tunneling.

The electrochemical potentials of the transitions between
successive ground states are spaced by the so-called addition
energy:
\begin{equation}
E_{add}(N) = \hspace{0.1cm} \mu(N\!+\!1) - \mu(N) = \hspace{0.1cm}
E_C + \Delta E \ \ \label{addition}
\end{equation}
\noindent The addition energy consists of a purely electrostatic
part, the charging energy $E_C$, plus the energy spacing between
two discrete quantum levels, $\Delta E$. Note that $\Delta E$ can
be zero, when two consecutive electrons are added to the same
spin-degenerate level.

Electron tunneling through the dot critically depends on the
alignment of electrochemical potentials in the dot with respect to
those of the source, $\mu_S$, and the drain, $\mu_D$. The
application of a bias voltage $V_{SD}=V_{S}-V_{D}$ between the
source and drain reservoir opens up an energy window between
$\mu_S$ and $\mu_D$ of  $\mu_S -\mu_D=-|e|V_{SD}$. This energy
window is called the \textit{bias window}. For energies within the
bias window, the electron states in one reservoir are filled
whereas states in the other reservoir are empty. Therefore, if
there is an `appropiate' electrochemical potential level within the
bias window, electrons can tunnel from one reservoir onto the dot
and off to the empty states in the other reservoir. Here, `appropriate' means that the electrochemical potential corresponds to a transition that involves the current state of the quantum dot.

In the following, we assume the temperature to be negligible
compared to the energy level spacing $\Delta E$ (for GaAs dots
this roughly means $T<$0.5~K). The size of the bias window then
separates two regimes: the low-bias regime where at most one dot
level is within the bias window ($-|e|V_{SD} < \Delta E,
E_{add}$), and the high-bias regime where multiple dot levels can
be in the bias window ($-|e|V_{SD}\ge\Delta E$ and/or
$-|e|V_{SD}\ge E_{add}$).

\subsection{Low-bias regime}
\begin{figure}[htbp]
\includegraphics[width=3in, clip=true]{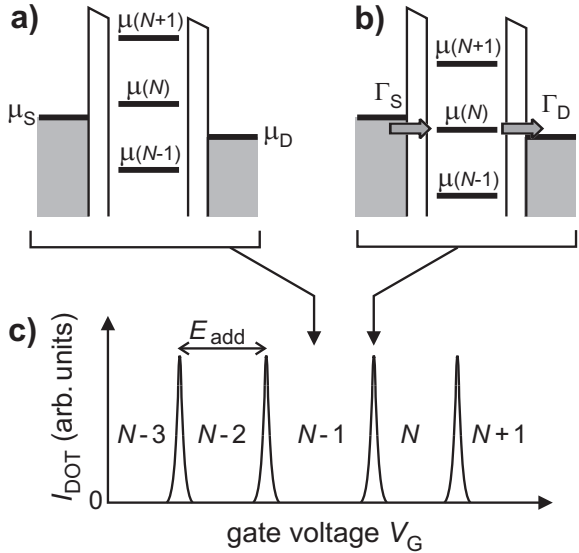}
\caption{(a)-(b)Schematic diagrams of the electrochemical
potential levels of a quantum dot in the low-bias regime.
\textbf{(a)} If no level in the dot falls within the bias window
set by $\mu_S$ and $\mu_D$, the electron number is fixed at
$N\!-\!1$ due to Coulomb blockade. \textbf{(b)} The $\mu(N)$ level
is in the bias window, so the number of electrons can alternate
between $N\!-\!1$ and $N$, resulting in a single-electron
tunneling current. The magnitude of the current depends on the
tunnel rate between the dot and the reservoir on the left,
$\Gamma_{S}$, and on the right, $\Gamma_{D}$ (see
~\textcite{Kouwenhoven97} for details). \textbf{(c)} Schematic
plot of the current $I_{DOT}$ through the dot as a function of
gate voltage $V_G$. The gate voltages where the level alignments
of (a) and (b) occur are indicated. } \label{fig:lowbias}
\end{figure}

For a quantum dot system in equilibrium, electron transport is
only possible when a level corresponding to transport between
successive ground states is in the bias window, i.e. $\mu_S \geq
\mu(N) \geq \mu_D$ for at least one value of $N$. If this
condition is not met, the number of electrons on the dot remains
fixed and no current flows through the dot. This is known as
\textit{Coulomb blockade}. An example of such a level alignment is
shown in Fig.~\ref{fig:lowbias}a.

Coulomb blockade can be lifted by changing the voltage applied to
the gate electrode, as can be seen from Eq.~\ref{mu}. When
$\mu(N)$ is in the bias window one extra electron can tunnel onto
the dot from the source (see Fig.~\ref{fig:lowbias}b), so that the
number of electrons increases from $N\!-\!1$ to $N$. After it has
tunneled to the drain, another electron can tunnel onto the dot
from the source. This cycle is known as \textit{single-electron
tunneling}.

By sweeping the gate voltage and measuring the current through the dot, $I_{DOT}$, a trace is
obtained as shown in Fig.~\ref{fig:lowbias}c. At the positions of
the peaks in $I_{DOT}$, an electrochemical potential level corresponding to
transport between successive ground states is aligned between the
source and drain electrochemical potentials and a single-electron
tunneling current flows. In the valleys between the peaks, the
number of electrons on the dot is fixed due to Coulomb blockade.
By tuning the gate voltage from one valley to the next one, the
number of electrons on the dot can be precisely controlled. The
distance between the peaks corresponds to $E_{add}$ (see
Eq.~\ref{addition}), and therefore provides insight into the
energy spectrum of the dot.

\begin{figure}[htb]
\includegraphics[width=3in, clip=true]{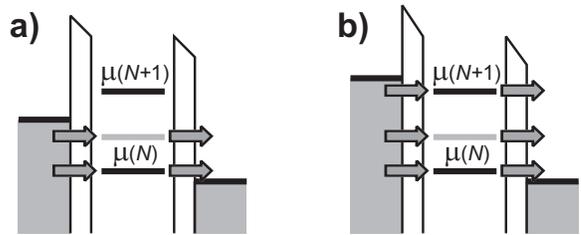}
\caption{Schematic diagrams of the electrochemical potential
levels of a quantum dot in the high-bias regime. The level in grey
corresponds to a transition involving an excited state.
\textbf{(a)} Here, $V_{SD}$ exceeds $\Delta E$ and electrons can
now tunnel via two levels. \textbf{(b)} $V_{SD}$ exceeds the
addition energy for $N$ electrons, leading to double-electron
tunneling. } \label{fig:highbias1}
\end{figure}

\begin{figure*}[htb]
\includegraphics[width=7in, clip=true]{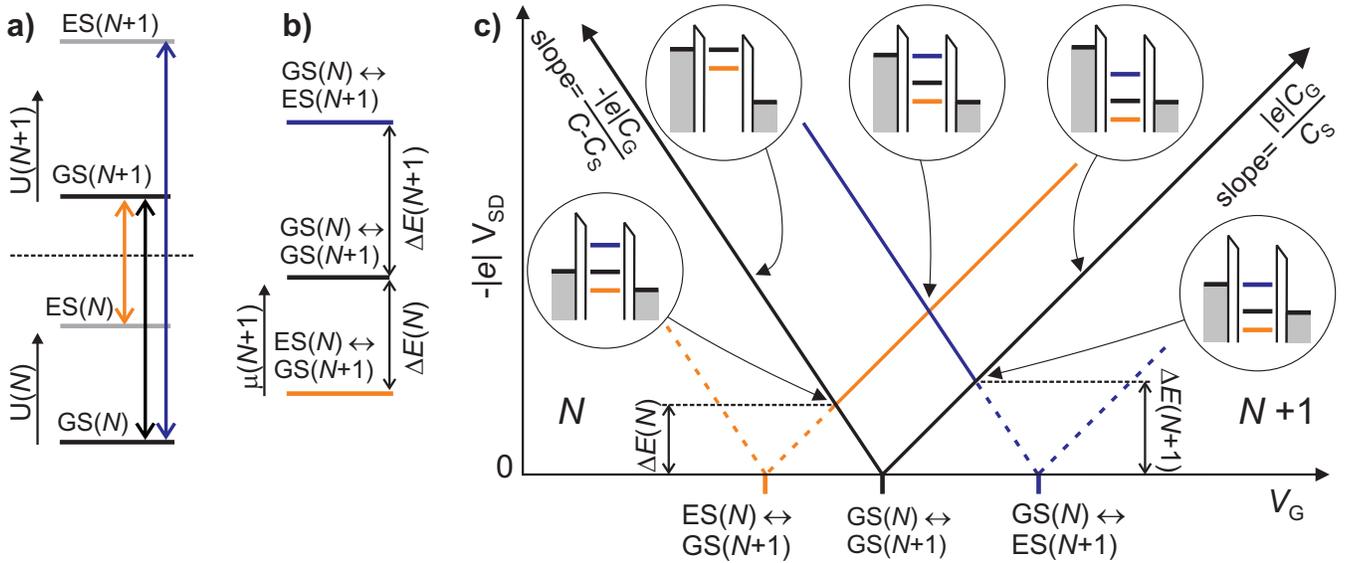}
\caption{(Color in online edition) \textbf{(a)} Energies for $N$ electrons, $U(N)$, and for
$N$+1 electrons, $U(N+1)$. Possible transitions are indicated by
arrows. \textbf{(b)} The electrochemical potential ladder for the
transitions depicted in (a). \textbf{(c)} Schematic plot of the
differential conductance $dI_{DOT}/dV_{SD}$ as a function of
$-\left|e\right|V_{SD}$ and ${V_G}$. At several positions the
level alignment is indicated with schematic diagrams. }
\label{fig:highbias2}
\end{figure*}

\subsection{High-bias regime}
We now look at the regime where the source-drain bias is so high
that multiple dot levels can participate in electron tunneling.
Typically the electrochemical potential of only one of the
reservoirs is changed in experiments, and the other one is kept
fixed. Here, we take the drain reservoir to be at ground, i.e.
$\mu_D=0$. When a negative voltage is applied between the source
and the drain, $\mu_S$ increases (since $\mu_S= -\left|e\right|
V_{SD}$). The levels of the dot also increase, due to the
capacitive coupling between the source and the dot (see
Eq.~\ref{mu}). Again, a current can flow only when a level
corresponding to a transition between ground states falls within
the bias window. When $V_{SD}$ is increased further such that also
a transition involving an \textit{excited state} falls within the
bias window, there are two paths available for electrons tunneling
through the dot (see Fig.~\ref{fig:highbias1}a). In general, this
will lead to a change in current, enabling us to perform energy
spectroscopy of the excited states. How exactly the current
changes depends on the tunnel coupling of the two levels involved.
Increasing $V_{SD}$ even more eventually leads to a situation
where the bias window is larger than the addition energy (see
Fig.~\ref{fig:highbias1}b). Here, the electron number can
alternate between $N-1$, $N$ and $N\!+\!1$, leading to a
double-electron tunneling current.

We now show how the current spectrum as a function of bias and
gate voltage can be mapped out. First, the electrochemical
potentials of all relevant transitions are calculated by applying
Eq.~\ref{mu2}. For example, consider two successive ground
states, GS($N$) and GS($N\!+\!1$), and the excited states ES($N$)
and ES($N\!+\!1$), which are separated from the GSs by $\Delta
E(N)$ and $\Delta E(N\!+\!1)$ respectively (see
Fig.~\ref{fig:highbias2}a). The resulting electrochemical
potential ladder is shown in Fig.~\ref{fig:highbias2}b (we omit
the transition between the two ESs). Note that the electrochemical
potential of the transition ES($N$)$\leftrightarrow$GS($N\!+\!1$)
is \textit{lower} than that of the transition between the two
ground states.

The electrochemical potential ladder is used to define the gate
voltage axis of the ($-\left|e\right|V_{SD},{V_G}$) plot, as in
Fig.~\ref{fig:highbias2}c. Here, each transition indicates the
gate voltage at which its electrochemical potential is aligned
with $\mu_S$ and $\mu_D$ at $V_{SD}=0$. Analogous to
Fig.~\ref{fig:lowbias}c-d, sweeping the gate voltage at low bias
will show electron tunneling only at the gate voltage indicated by
$GS(N)\leftrightarrow GS(N\!+\!1)$. For all other gate voltages
the dot is in Coulomb blockade.

Next, for each transition a V-shaped region is outlined in the
($-\left|e\right|V_{SD}$,${V_G}$)-plane, where its electrochemical
potential is within the bias window. This yields a plot like
Fig.~\ref{fig:highbias2}c. The slopes of the two edges of the
V-shape depend on the capacitances; for $V_D=0$, the two slopes
$d(-\left|e\right|V_{SD})/d{V_G}$ are $-C_G/(C\!-\!C_S)$ and
$+C_G/C_S$. The transition between the $N$-electron GS and the
($N\!+\!1$)-electron GS (black solid line) defines the regions of
Coulomb blockade (outside the V-shape) and tunneling (within the
V-shape). The other solid lines indicate where the current changes
due to the onset of transitions involving excited states. 

The set of solid lines indicate all the values in the parameter space spanned by $V_{SD}$ and ${V_G}$ where the current $I_{DOT}$ changes. Typically, the differential conductance $dI_{DOT}/dV_{SD}$ is plotted, which has a nonzero value \textit{only} at the solid lines \footnote{In practice, a dependence of the tunnel couplings on $V_G$ and $V_{SD}$ may result in a nonzero value of $dI_{DOT}/dV_{SD}$ throughout the region where current flows. Since this ``background'' of nonzero $dI_{DOT}/dV_{SD}$ is typically more uniform and much smaller than the peaks in $dI_{DOT}/dV_{SD}$ at the solid lines, the two are easily distinguished in experiments.}.

A general `rule of thumb' for the positions of the lines indicating finite differential conductance is this: \textit{if a line terminates at
the $N$-electron Coulomb blockade region, the transition
necessarily involves an $N$-electron excited state}. This is true
for any $N$. As a consequence, no lines terminate at the Coulomb
blockade region where $N$=0, as there exist no excited state for
$N$=0~\footnote{Note that energy absorption from the environment
can lead to exceptions: photon- or phonon-assisted tunneling can
give rise to lines ending in the $N$=0 Coulomb blockade region.
However, many experiments are performed at very low temperatures
where the number of photons and phonons in thermal equilibrium is
extremely small. Therefore, these processes are usually
negligible.}. For a transition between two excited states, say
ES($N$) and ES($N\!+\!1$), the position of the line depends on the
energy level spacing: for $\Delta E(N\!+\!1)>\Delta E(N)$, the
line terminates at the ($N\!+\!1$)-electron Coulomb blockade
region, and vice versa.

A measurement as shown in Fig.~\ref{fig:highbias2}c is very useful
for finding the energies of the excited states. Where a line of a
transition involving one excited state touches the Coulomb
blockade region, the bias window exactly equals the energy level
spacing. Figure~\ref{fig:highbias2}c shows the level diagrams at
these special positions for both
ES($N$)$\leftrightarrow$GS($N\!+\!1$) and
GS($N$)$\leftrightarrow$ES($N\!+\!1$). Here, the level spacings
can be read off directly on the $-\left|e\right|V_{SD}$-axis.

We briefly discuss the transition
ES($N$)$\leftrightarrow$ES($N\!+\!1$), that was neglected in the
discussion thus far. The visibility of such a transition depends
on the relative magnitudes of the tunnel rates and the relaxation
rates. When the relaxation is much \textit{faster} than the tunnel
rates, the dot will effectively be in its ground state all the
time and the transition ES($N$)$\leftrightarrow$ES($N\!+\!1$) can
therefore never occur. In the opposite limit where the relaxation
is much \textit{slower} than the tunneling, the transition
ES($N$)$\leftrightarrow$ES($N\!+\!1$) participates in the electron
transport and will be visible in a plot like in
Fig.~\ref{fig:highbias2}c. Thus, the visibility of transitions can
give information on the relaxation rates between different
levels~\cite{FujisawaPRL02}.

If the voltage is swept across multiple electron transitions and
for both signs of the bias voltage, the Coulomb blockade regions
appear as diamond shapes in the
($-\left|e\right|V_{SD}$,${V_G}$)-plane. These are the well-known
\textit{Coulomb diamonds}.

\section{Spin spectroscopy methods}
\label{Section:SpinSpectroscopy} In this section, we discuss
various methods for getting information on the spin state of the
electrons on a quantum dot. These methods make use of various
spin-dependent energy terms. First, each electron spin is
influenced directly by an external magnetic field via the Zeeman
energy $E_Z=S_z g \mu_B B$ where $S_z$ is the spin $z$-component.
Moreover, the Pauli exclusion principle forbids two electrons with
equal spin orientation to occupy the same orbital, thus forcing
one of the electrons into a different orbital. This generally
leads to a state with a different energy. Finally, the Coulomb
interaction leads to an energy difference (the exchange energy)
between states with symmetric and anti-symmetric orbital
wavefunctions. Since the total wavefunction of the electrons is
anti-symmetric, the symmetry of the orbital part is linked to that
of the spin.

\subsection{Spin filling derived from magnetospectroscopy}
The spin filling of a quantum dot can be derived from the Zeeman
energy shift of the Coulomb peaks in a magnetic field. (An
in-plane magnetic field orientation is favored to ensure minimum
disturbance of the orbital levels). On adding the $N$th electron,
the $z$-component $S_z$ of the spin on the dot is either increased
by 1/2 (if a spin-up electron is added) or decreased by 1/2 (if a
spin-down electron is added). This change in spin is reflected in
the magnetic field dependence of the electrochemical potential
$\mu$($N$) via the Zeeman term

\begin{equation}
     g \mu_B B \left[S_z(N)-S_z(N\!-\!1)\right] = g \mu_B B \left[ \Delta S_z(N)\right].
\end{equation}

As the $g$-factor in GaAs is negative (see
Appendix~\ref{App:nuclearfield}), addition of a spin-up electron
($\Delta S_z(N)$=+1/2) results in $\mu(N)$ decreasing with
increasing $B$. Spin-independent shifts of $\mu(N)$ with $B$ (e.g.
due to a change in confinement potential) are removed by looking
at the dependence of the addition energy $E_{add}$ on
$B$~\cite{WeisPRL1993}:

\begin{eqnarray}
    \frac{\partial E_{add}(N)}{\partial B}&=&\frac{\partial\mu(N)}{\partial B}-\frac{\partial\mu(N\!-\!1)}{\partial B}\nonumber \\
    &=& g \mu_B \left[ \Delta S_z(N)-\Delta S_z(N\!-\!1) \right].
\end{eqnarray}

Assuming $S_z$ only changes by $\pm\frac{1}{2}$, the possible
outcomes and the corresponding filling schemes are
\begin{eqnarray*}
 \frac{\partial E_{add}(N)}{\partial B}\ =&0 &:\ \ua, \ua \ or\ \da , \da \\
=& +g \mu_B &:\ \ua, \da \\
=& -g \mu_B &:\ \da , \ua ,
\end{eqnarray*}
where the first (second) arrow depicts the spin added in the
$N\!-\!2 \rightarrow N\!-\!1$ ($N\!-\!1 \rightarrow N$) electron
transition. Spin filling of both vertical \cite{Sasaki98} and
lateral GaAs quantum dots \cite{Duncan00,Lindmann02,PotokPRL2003}
has been determined using this method, showing clear deviations
from a simple ``Pauli'' filling ($S_z$ alternating between 0 and
$\frac{1}{2}$). Note that transitions where $S_z$ of the ground state changes by more
than $\frac{1}{2}$, which can occur due to many-body interactions in the dot, can lead to a spin blockade of the
current~\cite{Weinmann, KorkusinskiPRL2004}.

In circularly symmetric few-electron vertical dots, spin states
have been determined from the evolution of orbital states in a
magnetic field perpendicular to the plane of the dots. This
indirect determination of the spin state has allowed the
observation of a two-electron singlet-to-triplet ground state
transition and a four-electron spin filling following Hund's rule.
For a review on these experiments, see ~\textcite{kouwenhoven01}.
Similar techniques were also used in experiments on few-electron lateral dots in both weak and strong magnetic fields~\cite{CiorgaPRB2000,Kyriakidis02}.

\subsection{Spin filling derived from excited-state spectroscopy}
Spin filling can also be deduced from excited-state spectroscopy
without changing the magnetic field~\cite{Cobden98}, provided the
Zeeman energy splitting $\Delta E_Z=2\left|E_Z\right|=g \mu_B B$
between spin-up and spin-down electrons can be resolved. This
powerful method is based on the simple fact that any
single-particle orbital can be occupied by at most two electrons
due to Pauli's exclusion principle. Therefore, as we add one
electron to a dot containing $N$ electrons, there are only two
scenarios possible: either the electron moves into an empty
orbital, or it moves into an orbital that already holds one
electron. As we show below, these scenarios always correspond to
ground state filling with spin-up and spin-down, respectively.

\begin{figure}[htb]
\begin{center}
\includegraphics[width=2.5in, clip=true]{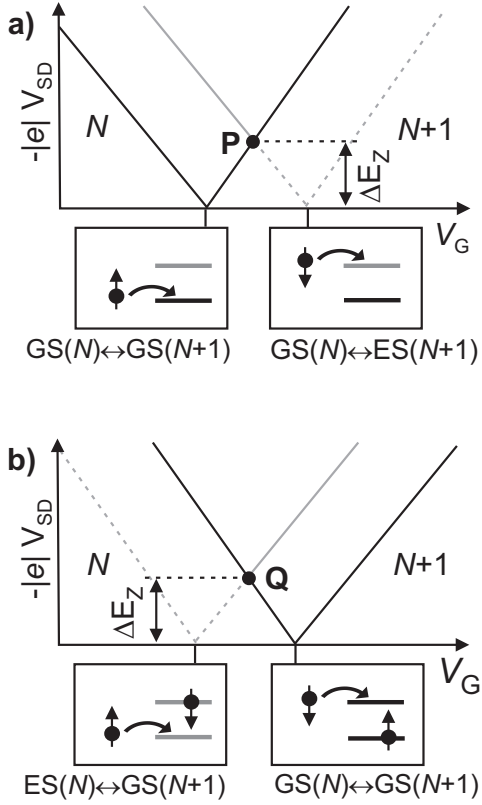}
\end{center}
\caption{\label{fig:ESspinfilling} Spin filling deduced from
high-bias excited-state spectroscopy. Shown are schematic diagrams
of $dI_{DOT}/dV_{SD}$ in the ($V_{SD},V_G$)-plane. (a)
Ground-state filling is spin-up: a line corresponding to an
($N$+1)-electron ES, separated from GS($N$+1) by $\Delta E_{Z}$,
terminates at the edge of the ($N$+1)-electron Coulomb blockade
region (point P). (b) Ground-state filling is spin-down: a line
corresponding to an $N$-electron ES, separated from GS($N$) by
$\Delta E_{Z}$, terminates at the $N$-electron Coulomb blockade
region (point Q).}
\end{figure}

First consider an electron entering an empty orbital with
well-resolved spin splitting (see Fig.~\ref{fig:ESspinfilling}a).
Here, addition of a spin-up electron corresponds to the transition
$GS(N)\leftrightarrow GS(N\!+\!1)$. In contrast, addition of a
spin-down electron takes the dot from $GS(N)$ to $ES(N\!+\!1)$,
which is $\Delta E_{Z}$ higher in energy than $GS(N\!+\!1)$. Thus
we expect a high-bias spectrum as in
Fig.~\ref{fig:ESspinfilling}a.

Now consider the case where the ($N\!+\!1$)th electron moves into
an orbital that already contains one electron (see
Fig.~\ref{fig:ESspinfilling}b). The two electrons need to have
anti-parallel spins, in order to satisfy the Pauli exclusion
principle. If the dot is in the ground state, the electron already
present in this orbital has spin-up. Therefore, the electron added
in the transition from $GS(N)$ to $GS(N\!+\!1)$ must have
spin-down. A spin-up electron can only be added if the first
electron has spin-down, i.e. when the dot starts from $ES(N)$,
$\Delta E_{Z}$ higher in energy than $GS(N)$. The high-bias
spectrum that follows is shown schematically in
Fig.~\ref{fig:ESspinfilling}b.

Comparing the two scenarios, we see that the spin filling has a
one-to-one correspondence with the excited-state spectrum: if the
spin $ES$ line terminates at the ($N\!+\!1$)-electron Coulomb
blockade region (as point $P$ in Fig.~\ref{fig:ESspinfilling}a), a
spin-up electron is added to the $GS$; if however the spin $ES$
line terminates at the $N$-electron Coulomb blockade region (as
point $Q$ in Fig.~\ref{fig:ESspinfilling}b), a spin-down electron
is added to the $GS$.

The method is valid regardless of the spin of the ground states
involved, as long as the addition of one electron changes the spin
$z$-component of the ground state by $\left|\Delta
S_z\right|=1/2$. If $|\Delta S_z|>1/2$, the $(N+1)$-electron GS
cannot be reached from the $N$-electron $GS$ by addition of a
single electron. This would cause a spin blockade of electron
transport through the dot~\cite{Weinmann}.

\begin{figure*}[hbt]
\includegraphics[width=6.9in, clip=true]{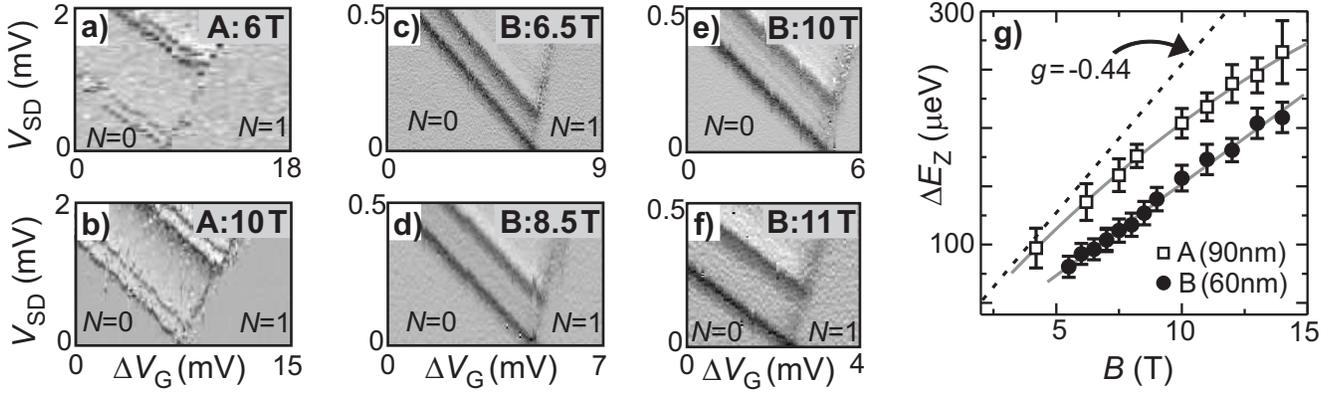}
\caption{\textbf{(a)}-\textbf{(f)} Excited-state spectroscopy on
two devices: device A is fabricated on a heterostructure with the
2DEG at 90~nm below the surface and device B with the 2DEG at
60~nm below the surface. Differential conductance
$dI_{DOT}/dV_{SD}$ is plotted as a function of $V_{SD}$ and gate
voltage near the 0$\leftrightarrow$1 electron transition, for
in-plane magnetic fields (as indicated in top right corners).
Darker corresponds to larger $dI_{DOT}/dV_{SD}$. Data on device A
shows spin splitting in both the orbital ground and first excited
state; data on device B only displays the orbital ground state.
\textbf{(g)} Zeeman splitting $\Delta E_{Z}$ as a function of $B$
extracted from (a)-(f) and similar measurements. Gray solid lines
are fits to the data. The dashed line shows $\Delta E_{Z}$
expected for the bulk GaAs $g$-factor of -0.44. Data adapted from
~\textcite{HansonPRL2003,BeverenNJP2005}}
\label{Fig:Zeeman01}
\end{figure*}

\subsection{Other methods}
If the tunnel rates for spin-up and spin-down are not equal, the
amplitude of the current can be used to determine the spin
filling. This method has been termed spin-blockade spectroscopy. This name is slightly misleading as the current is not actually blocked, but rather assumes a finite value that depends on the spin orientation of the transported electrons. This method has been
demonstrated and utilized in the quantum Hall regime, where the spatial separation of spin-split edge channels induces a large
difference in the tunnel rates of spin-up and spin-down electrons~\cite{CiorgaPRB2000,CiorgaNDR,Kupidura06}. Spin-polarized leads can also be obtained in moderate magnetic fields by changing the electron density near the dot with a gate. This concept was used to perform spin spectroscopy on a quantum dot connected to gate-tunable quasi-one-dimensional channels~\cite{HitachiPRB2006}.

Care must be taken when inferring the spin filling from the amplitude of the current as other factors, such as the orbital spread of the wavefunction, can have a large, even dominating influence on the current amplitude. A prime example is the difference in tunnel rate between the two-electron spin singlet and triplet states due to the different orbital wavefunctions of these states. In fact, this difference is large enough to allow single-shot readout of the two-electron spin state, as will be discussed in Section \ref{Subsection:TRRO}.

In zero magnetic field, a state with total spin $S$ is
($2S+1$)-fold degenerate. This degeneracy is reflected in the
current if the dot has strongly asymmetric barriers. As an example, in the transition from a one-electron $S$=1/2 state to a two-electron $S$=0 state, only a spin-up electron can tunnel onto the dot if the electron that is already on the dot is spin-down, and vice-versa. However, in the reverse transition ($S$=0 to $S$=1/2), both electrons on the dot can tunnel off. Therefore, the rate for tunneling off the dot is twice the rate for tunneling onto the dot. In general, the ratio of the currents in opposite bias directions at the $GS(N)\leftrightarrow GS(N\!+\!1)$ transition is, for spin-independent tunnel rates and for strongly asymmetric barriers, given by $[2S(N\!+\!1)+1]/[2S(N)+1]$~\cite{Akera99}. Here, $S(N)$ and $S(N\!+\!1)$ denote the total spin of $GS(N)$ and $GS(N\!+\!1)$ respectively. This
relation can be used in experiments to determine the ground state
total spin~\cite{Cobden98,Hayashi03}.

Information on the spin of the ground state can also be found from
(inelastic) cotunneling currents~\cite{KoganPRL2004} or the
current due to a Kondo resonance~\cite{Goldhaber98, Cronenwett98}. If a magnetic
field $B$ drives the onset of these currents to values of
$V_{SD}=\pm g \mu_B B/\left|e\right|$, it follows that the ground
state has nonzero spin. Since the processes in
these currents can change the spin $z$-component by at most 1, the
absolute value of the spin can not be deduced with this method,
unless the spin is zero.

We end this section with some remarks on spin filling. First, the
parity of the electron number can not be inferred from spin
filling unless the sequence of spin filling is exactly known. For
example, consider the case where the electron added in the
GS($N$)$\rightarrow$GS($N$+1) transition has spin-down. Then, if
the dot follows an alternating (Pauli) spin filling scheme, $N$ is
odd. However, if there is a deviation from this scheme such that
GS($N$) is a spin triplet state (total spin $S$=1), then $N$ is
even.

Second, spin filling measurements do not yield the absolute spin
of the ground states, but only the change in ground state spin.
However, by starting from zero electrons (and thus zero spin) and
tracking the change in spin at subsequent electron transitions,
the total spin of the ground state can be
determined~\cite{BeverenNJP2005}.

\section{Spin states in a single dot}
\label{Section:SingleDotSpin}
\subsection{One-electron spin states}
The simplest spin system is that of a single electron, which can
have one of only two orientations: spin-up or spin-down. Let
$E_{\ua,0}$ and $E_{\da,0}$ ($E_{\ua,1}$ and $E_{\da,1}$) denote
the one-electron energies for the two spin states in the lowest
(first excited) orbital. With a suitable choice of the zero of
energy we arrive at the following electrochemical potentials:
\begin{eqnarray}
\mu_{0\leftrightarrow \ua,0}&=&E_{\ua,0} \\
\mu_{0\leftrightarrow \da,0}&=&E_{\da,0}=E_{\ua,0} + \Delta E_Z \\
\mu_{0\leftrightarrow \ua,1}&=&E_{\ua,1}=E_{\ua,0} + \Delta E_{orb}\\
\mu_{0\leftrightarrow \da,1}&=&E_{\da,1}=E_{\ua,0} + \Delta
E_{orb}+ \Delta E_Z
\end{eqnarray}
where $\Delta E_{orb}$ is the orbital level spacing.

Figures~\ref{Fig:Zeeman01}a-f show excited-state spectroscopy
measurements on two devices, $A$ and $B$, via electron transport
at the $N$=$0\leftrightarrow 1$ transition, at different magnetic
fields $B_{//}$ applied in the plane of the 2DEG. A clear
splitting of both the orbital ground and first excited state is
observed, which increases with increasing magnetic
field~\cite{HansonPRL2003,PotokPRL2003,BeverenNJP2005,Haug05}. The
orbital level spacing $\Delta E_{orb}$ in device A is about
1.1~meV. Comparison with Fig.~\ref{fig:ESspinfilling} shows that a
spin-up electron is added to the empty dot to form the
one-electron ground state, as expected.

\begin{figure*}[htb]
\includegraphics[width=14cm]{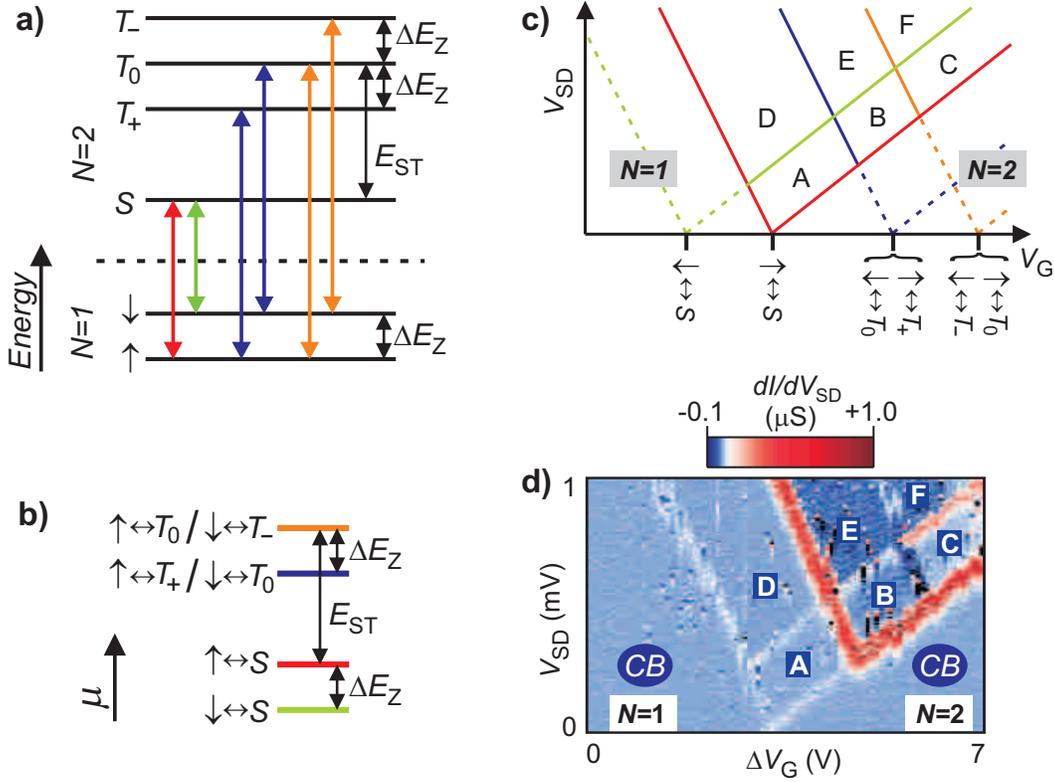}
\caption{(Color in online edition) (a) Energy diagram schematically showing the energy
levels of the one- and two-electron states. The allowed
transitions between these levels are indicated by arrows. (b)
Electrochemical potential ladder corresponding to the transitions
shown in (a), using the same color coding. Changing the gate
voltage shifts the ladder as a whole. Note that the three triplet
states appear at only two values of the electrochemical
potential.(c) Energetically allowed 1$\leftrightarrow$2 electron
transitions as a function of $V_{SD}$ and $V_{G}$. The lines
corresponding to $\ua\leftrightarrow\!S$ outline the region of
transport; outside this region, where lines are dashed, the dot is
in Coulomb blockade. (d) $dI_{DOT}/dV_{SD}$ as a function of $V_G$
and $V_{SD}$ around the 1$\leftrightarrow$2 electron transition at
$B_{/\!/}\!=\!$ 12 T in device $A$. The regions labelled with
letters A-F correspond well to those in (c). In the region labeled
A only spin-down electrons pass through the dot. Data adapted from
~\textcite{HansonPRB2004}.} \label{Fig:N1-2}
\end{figure*}

In Fig.~\ref{Fig:Zeeman01}g the Zeeman splitting $\Delta E_Z$ is
plotted as function of $B_{//}$  for the same two devices, $A$ and
$B$, which are made on different heterostructures. These
measurements allow a straightforward determination of the electron
$g$-factor. The measured $g$-factor can be affected by: (i)
extension of the electron wave function into the
Al$_{0.3}$Ga$_{0.7}$As region, where $g\!=\!+0.4$
\cite{Snelling,salis01}, (ii) thermal nuclear polarization,
which decreases the effective magnetic field through the hyperfine
interaction \cite{BookOptical}, (iii) dynamic nuclear polarization
due to electron-nuclear flip-flop processes in the dot, which
enhances the effective magnetic field \cite{BookOptical}, (iv) the
nonparabolicity of the GaAs conduction band~\cite{Snelling}, (v)
the spin-orbit coupling~\cite{falko05}, and (vi) the confinement
potential~\cite{ WeisbuchPRB77,BjorkPRB05}. The effect of the
nuclear field on the measured $g$-factor is discussed in more
detail in Appendix~\ref{App:nuclearfield}. More experiments are
needed to separate these effects, e.g. by measuring the dependence
of the $g$-factor on the orientation of the in-plane magnetic
field with respect to the crystal axis~\cite{falko05}.


\subsection{Two-electron spin states}

The ground state of a two-electron dot in zero magnetic field is
always a spin singlet (total spin quantum number $S\!=\!$ 0)
\cite{ashcroft}, formed by the two electrons occupying the lowest
orbital with their spins anti-parallel:
$\ket{\:S}\!=\!(\ket{\ua\da}\!-\!\ket{\da\ua})/\sqrt{2}$. The
first excited states are the spin triplets ($S\!=\!$ 1), where the
antisymmetry of the total two-electron wave function requires one
electron to occupy a higher orbital. Both the antisymmetry of the
orbital part of the wavefunction and the occupation of different
orbitals reduce the Coulomb energy of the triplet states with
respect to the singlet with two electrons in the same
orbital~\cite{kouwenhoven01}. We include this change in Coulomb energy
by the energy term $E_K$. The three triplet states are
degenerate at zero magnetic field, but acquire different Zeeman
energy shifts $E_{Z}$ in finite magnetic fields because their spin
$z$-components differ: $S_z\!=\!+1$ for
$\ket{\:T_{+\!}}\!=\!\ket{\ua\ua}$, $S_z\!=\!0$ for
$\ket{\:T_{0}}\!=\!(\ket{\ua\da}\!+\!\ket{\da\ua})/\sqrt{2}$ and
$S_z\!=\!-1$ for $\ket{\:T_{-\!}}\!=\!\ket{\da\da}$.

Using the Constant Interaction model, the energies of the states can be expressed in
terms of the single-particle energies of the two electrons plus a
charging energy $E_{C}$ which accounts for the Coulomb
interactions:
\begin{eqnarray*}
    &&\!\!\!\!\!\!\!\!U_S\ =\! E_{\ua,0}+ E_{\da,0} + E_{C} = 2 E_{\ua,0}+ \Delta E_{Z} + E_{C}\\
    &&\!\!\!\!\!\!\!\!U_{T_+}\!= E_{\ua,0}+ E_{\ua,1} - E_{K} + E_{C} \\
    && = 2 E_{\ua,0}\!+\! \Delta E_{orb}\! -\! E_{K}\! +\! E_{C} \\
    && = 2 E_{\ua,0} + E_{ST} + \!E_{C}\\
    &&\!\!\!\!\!\!\!\!U_{T_0} =\! E_{\ua,0}\! + \! E_{\da,0}\! + \! E_{ST}\!+\!E_{C}\\
    &&=2 E_{\ua,0}\!+\! \Delta E_{orb}\! -\! E_{K}\! +\! \Delta E_{Z}\! +\! E_{C} \\
    && =2 E_{\ua,0}\!+\! E_{ST}\! +\! \Delta E_{Z}\! + \! E_{C}\\
    &&\!\!\!\!\!\!\!\!U_{T_-}\! =\! 2E_{\da,0}\!+\!E_{ST}\!+\!E_{C}\\
    &&=2 E_{\ua,0}\!+\! \Delta E_{orb}\! -\! E_{K}\! +\! 2 \Delta E_{Z}\! +\! E_{C} \\
    &&=2 E_{\ua,0}\!+\!E_{ST}\!+\! 2 \Delta E_{Z} \!+\!E_{C},
\end{eqnarray*}
with $E_{ST}$ denoting the singlet-triplet energy difference in the absence of the Zeeman splitting $\Delta E_{Z}$: $E_{ST}=\Delta E_{orb}-E_{K}$.

We first consider the case of an in-plane magnetic field $B_{//}$.
Here, $E_{ST}$ is almost independent of $B_{//}$ and the ground
state remains a spin singlet for all fields attainable in the lab.
The case of a magnetic field perpendicular to the plane of the
2DEG will be treated below.

Fig.~\ref{Fig:N1-2}a shows the possible transitions between the
one-electron spin-split orbital ground state and the two-electron
states. The transitions $\ua\leftrightarrow\!T_-$ and
$\da\leftrightarrow\!T_+$ are omitted, since these require a
change in the spin $z$-component of more than $\frac{1}{2}$ and
are thus spin-blocked \cite{Weinmann}. From the energy diagram the
electrochemical potentials can be deduced (see
Fig.~\ref{Fig:N1-2}b):

\begin{eqnarray*}
    \mu_{\ua,0\leftrightarrow S} &=& E_{\ua,0} + \Delta E_{Z} + E_{C}\\
    \mu_{\ua,0\leftrightarrow T_+} &=& E_{\ua,0}+ E_{ST} + E_{C}\\
    \mu_{\ua,0\leftrightarrow T_0} &=& E_{\ua,0}+ E_{ST} + \Delta E_Z + E_{C}\\
    \mu_{\da,0\leftrightarrow S} &=& E_{\ua,0} + E_{C}\\
    \mu_{\da,0\leftrightarrow T_0} &=& E_{\ua,0}+ E_{ST} + E_{C}\\
    \mu_{\da,0\leftrightarrow T_-} &=& E_{\ua,0}+ E_{ST} + \Delta E_Z + E_{C}
\end{eqnarray*}

Note that $\mu_{\ua,0 \leftrightarrow T_+} = \mu_{\da,0
\leftrightarrow T_0}$ and $\mu_{\ua,0 \leftrightarrow \!T_0} =
\mu_{\da,0 \leftrightarrow \!T_-}$. Consequently, the
\textit{three} triplet states change the first-order transport
through the dot at only \textit{two} values of $V_{SD}$. The
reason is that the first-order transport probes the energy
difference between states with \textit{successive} electron
number. In contrast, the onset of second-order (cotunneling)
currents is governed by the energy difference between states with
the \textit{same} number of electrons. Therefore, the triplet
states change the second-order (cotunneling) currents at three
values of $V_{SD}$ if the ground state is a singlet~\footnote{If
the ground state is a triplet, the cotunneling current only
changes at two values of $V_{SD}$ (0 and $\Delta E_Z/|e|$), due to
the spin selection rules.}~\cite{PaaskeNatPhys2006}.

In Fig.~\ref{Fig:N1-2}c we map out the positions of the
electrochemical potentials as a function of $V_G$ and $V_{SD}$.
For each transition, the two lines originating at $V_{SD}\!=\!$ 0
span a V-shaped region where the corresponding electrochemical
potential is in the bias window. In the region labeled A, only
transitions between the one-electron ground state, \ket{\ua,0},
and the two-electron ground state, \ket{\:S}, are possible, since
only $\mu_{\ua,0 \leftrightarrow S}$ is positioned inside the bias
window. In the other regions several more transitions are possible
which leads to a more complex, but still understandable behavior
of the current. Outside the V-shaped region spanned by the ground
state transition $\mu_{\ua,0\leftrightarrow S}$, Coulomb blockade
prohibits first order electron transport.

Experimental results from device $A$, shown in
Fig.~\ref{Fig:N1-2}d, are in excellent agreement with the
predictions of Fig.~\ref{Fig:N1-2}c. Comparison of the data with
Fig.~\ref{fig:ESspinfilling} indicates that indeed a spin-down
electron is added to the one-electron (spin-up) ground state to
form the two-electron singlet ground state. From the data the
singlet-triplet energy difference $E_{ST}$ is found to be
$\approx 520~\mu$eV. The fact that $E_{ST}$ is about half the
single-particle level spacing ($\Delta E_{orb}=1$~meV) indicates
the importance of Coulomb interactions. The Zeeman energy, and
therefore the $g$-factor, is found to be the same for the
one-electron states as for the two-electron states (within the
measurement accuracy of $\approx\!5\%$) on both device $A$ and
$B$. We note that the large variation in differential conductance
observed in Fig. \ref{Fig:N1-2}d, can be explained by a sequential
tunneling model with spin- and orbital-dependent tunnel rates
\cite{RonaldMoriond}.

\begin{figure}[htb]
\includegraphics[width=3.4in]{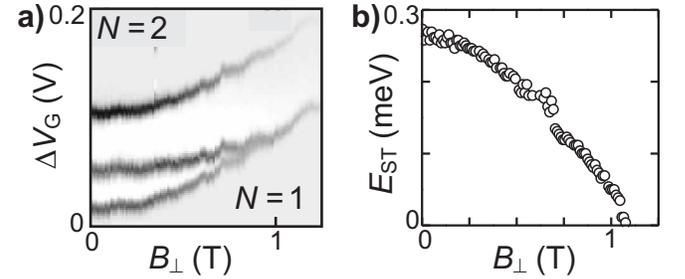}
\caption{Single-triplet ground state transition in a two-electron
quantum dot. \textbf{(a)} Differential conductance
$dI_{DOT}/dV_{SD}$ versus gate voltage, $V_G$, and perpendicular
magnetic field, $B_{\bot}$. Dark (light) corresponds to high (low)
value for $dI_{DOT}/dV_{SD}$. Within the stripe of finite
conductance, set by the source-drain bias voltage, the evolution
of the energy difference between the singlet state (ground state
at zero field) and the triplet state is visible. At around 1.1 T
the singlet and triplet states cross and the ground state becomes
a spin triplet. \textbf{(b)} Energy difference between the singlet
and the triplet states, $E_{ST}$, as a function of $B_{\bot}$,
extracted from (a). Data adapted from
~\textcite{Kyriakidis02}.} \label{Fig:STcrossing}
\end{figure}

By applying a large magnetic field \textit{perpendicular} to the
plane of the 2DEG a spin singlet-triplet ground state transition
can be induced, see Fig.~\ref{Fig:STcrossing}. This transition is
driven by two effects: (i) the magnetic field reduces the energy
spacing between the ground and first excited orbital state and
(ii) the magnetic field increases the Coulomb interactions which
are larger for two electrons in a single orbital (as in the
singlet state) than for two electrons in different orbitals (as in
a triplet state). Singlet-triplet transitions were first observed
in vertical dots~\cite{SuPRB1992,kouwenhoven01}. In lateral dots, the
gate-voltage dependence of the confinement potential has allowed
electrical tuning of the singlet-triplet transition
field~\cite{Kyriakidis02,ZumbuhlPRL2004}.

In very asymmetric lateral confining potentials with large Coulomb interaction energies, the simple single-particle picture breaks down. Instead, the two electrons in the ground state spin singlet in such dots will tend to avoid each other spatially, thus forming a quasi-double dot state. Experiments and calculations indicating this double-dot-like behaviour in asymmetric dots have been reported~\cite{ZumbuhlPRL2004,EllenbergerPRL2006}.

\subsection{Quantum dot operated as a bipolar spin filter}
\begin{figure}[htb]
\includegraphics[width=3in]{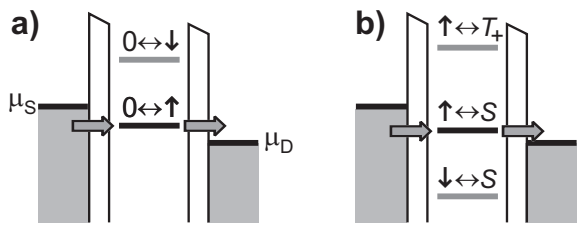}
\caption{Few-electron quantum dot operated as a bipolar spin
filter. Schematic diagrams show the level arrangement for ground
state transport at \textbf{(a)} the $0\leftrightarrow1$ electron
transition, where the dot filters for spin-up electrons, and
\textbf{(b)} at the $1\leftrightarrow2  $ electron transition,
where the dot only transmits spin-down electrons.}
\label{fig:spinfilter}
\end{figure}

If the Zeeman splitting exceeds the width of the energy levels
(which in most cases is set by the thermal energy), electron
transport through the dot is (for certain regimes) spin-polarized
and the dot can be operated as a spin
filter~\cite{RecherPRL2000,HansonPRB2004}. In particular, the
electrons are spin-up polarized at the $N=0\leftrightarrow1$
transition when only the one-electron spin-up state is
energetically accessible, as in Fig.~\ref{fig:spinfilter}a. At the
$N=1\leftrightarrow2$ transition, the current is spin-down
polarized if no excited states are accessible (region A in
Fig.~\ref{Fig:N1-2}c), see Fig.~\ref{fig:spinfilter}b. Thus, the
polarization of the spin filter can be reversed electrically, by
tuning the dot to the relevant transition.

Spectroscopy on dots containing more than two electrons has shown
important deviations from an alternating spin filling scheme.
Already for four electrons, a spin ground state with total spin
$S$=1 in zero magnetic field has been observed in both
vertical~\cite{kouwenhoven01} and lateral dots~\cite{BeverenNJP2005}.

\section{Charge sensing techniques}
\label{Section:ChargeSensing} \label{QPC} The use of local charge
sensors to determine the number of electrons in single or double
quantum dots is a recent technological improvement that has
enabled a number of experiments that would have been difficult, or
impossible to perform using standard electrical transport
measurements \cite{FieldPRL93}. In this section, we briefly
discuss relevant measurement techniques based on charge sensing.
Much of the same information as found by measuring the current can
be extracted from a measurement of the charge on the dot,
$Q_{DOT}$, using a nearby
electrometer, such as a quantum point contact (QPC).
In contrast to a measurement of the current through the dot, a charge
measurement can be also used if the dot is
connected to only one reservoir.

\begin{figure}[htb]
\includegraphics[width=3.4in]{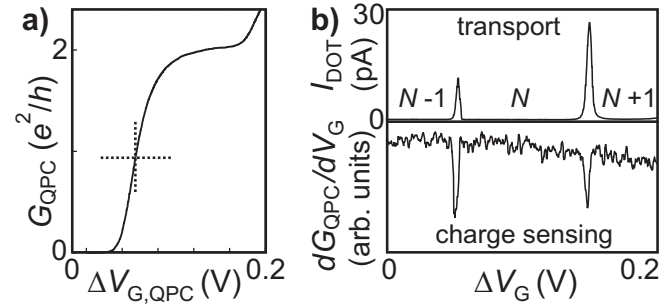}
\caption{Quantum point contact operated as an electrometer. A typical device, with the current paths through the dot and through the QPC, is shown in Fig.~\ref{fig:structure}c.
\textbf{(a)} QPC conductance $G_{QPC}$ vs. gate voltage on one of
the two gates that defines the QPC, $V_{G,QPC}$. Halfway the last
conductance step, at $G_{QPC}\approx e^2/h$ (indicated by a
cross), the QPC is very sensitive to the charge on the dot.
\textbf{(b)} Direct comparison between current measurement (top
panel) and charge sensing (bottom panel). Data adapted from
~\textcite{ElzermanPRB2003}.} \label{fig:chargesensing}
\end{figure}

The conductance $G_{QPC}$ through a QPC is
quantized~\cite{WeesQPC88,WharamQPC88}. At the transitions between
quantized conductance plateaus, $G_{QPC}$ is very sensitive to the
electrostatic environment including the number of electrons $N$ on
a nearby quantum dot (see Fig.~\ref{fig:chargesensing}a). This
property can be exploited to determine the absolute number of
electrons in single~\cite{Sprinzak2002} and coupled quantum
dots~\cite{ElzermanPRB2003}, even when the tunnel coupling is so
small that no current through the dot is detected.
Figure~\ref{fig:chargesensing}b shows measurements of the current
and of $dG_{QPC}/dV_{G}$ over the same range of $V_{G}$. Dips in
$dG_{QPC}/dV_{G}$ coincide with the current peaks, demonstrating
the validity of charge sensing. The sign of $dG_{QPC}/dV_{G}$ is
understood as follows. On increasing $V_G$, an electron is added
to the dot. The electric field created by this extra electron
reduces the conductance of the QPC, and therefore
$dG_{QPC}/dV_{G}$ dips. The sensitivity of the charge sensor to
changes in the dot charge can be optimized using an appropriate
gate design~\cite{zhang04}.

We should mention here that charge sensing fails when the tunnel time is much longer than the measurement time. In this case, no change in electron number will be observed when the gate voltage is swept and the equilibrium charge can not be probed~\cite{RushforthPRB2004}. Note that a quantum dot with very large tunnel barriers can trap electrons for minutes or even hours under non-equilibrium conditions~\cite{CooperPhysE2000}. This again emphasizes the importance of \textit{tunable} tunnel barriers (see Section~\ref{Sec:Fabrication}). Whereas the regime where the tunnel time largely exceeds the measurement time is of little interest for this review, the regime where the two are of the same order is actually quite useful, as we explain below.

We can get information on the dot energy level spectrum from QPC measurements, by monitoring the average charge on the
dot while applying short gate voltage pulses that bring the dot
out of its charge equilibrium. This is the case when the voltage
pulse pulls $\mu(N)$ from above to below the electrochemical
potential of the reservoir $\mu_{res}$. During the pulse with
amplitude $V_P>0$, the lowest energy state is $GS(N)$, whereas
when the pulse is off ($V_P=0$), the lowest energy state is
$GS(N-1)$. If the pulse length is much longer than the tunnel
time, the dot will effectively always be in the lowest-energy
charge configuration. This means that the number of electrons
fluctuates between $N-1$ and $N$ at the pulse frequency. If,
however, the pulse length is much \textit{shorter} than the tunnel
time, the equilibrium charge state is not reached during the pulse
and the number of electrons will not change. Measuring the average
value of the dot charge as a function of the pulse length thus
yields information on the tunnel time. In between the two limits,
i.e. when the pulse length is comparable to the tunnel time, the
average value of the dot charge is very sensitive to changes in
the tunnel rate.

\begin{figure}[htb]
\includegraphics[width=3.4in]{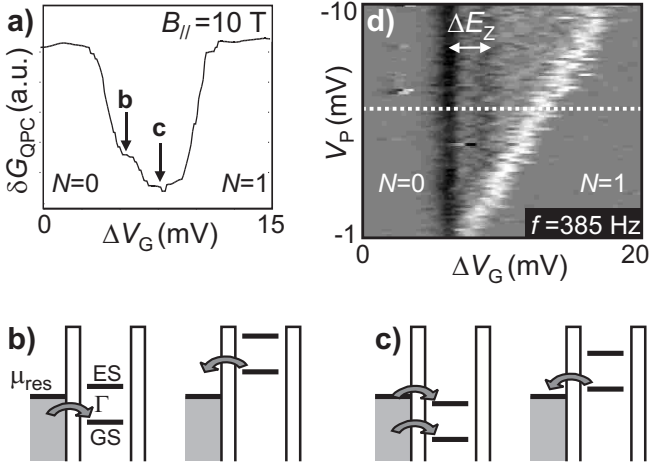}
\caption{Excited-state spectroscopy on a one-electron dot using
charge sensing. (a) $\delta G_{QPC}$ at $f = 385$ Hz versus $V_G$,
with $V_{P} = 6$~mV. Here, $\Gamma \approx 2.4$ kHz. (b) Schematic
electrochemical potential diagrams for the case that only the
\textit{GS} is pulsed across $\mu_{res}$. (c) Idem when both the
\textit{GS} and an \textit{ES} are pulsed across $\mu_{res}$. (d)
Derivative of $\delta G_{QPC}$ with respect to $V_G$ plotted as a
function of $V_G$ and $V_P$. Note that here $V_P$ is negative, and
therefore the region of tunneling extends to more positive gate
voltage as $|V_P|$ is increased. The curve in (a) is taken at the
dotted line. Data adapted from ~\textcite{ElzermanAPL2004}.}
\label{fig:ZeemanCharge}
\end{figure}

In this situation, excited-state spectroscopy can be performed by
raising the pulse amplitude
$V_P$~\cite{ElzermanAPL2004,JohnsonPRB2005}. For small pulse
amplitudes, at most one level is available for tunneling on and
off the dot, as in Fig.~\ref{fig:ZeemanCharge}b. Whenever $V_P$ is
increased such that an extra transition becomes energetically
possible, the effective tunnel rate changes as in
Fig.\ref{fig:ZeemanCharge}c. This change is reflected in the
average value of the dot charge and can therefore be measured
using the charge sensor.

The signal-to-noise ratio is enhanced significantly by lock-in
detection of $G_{QPC}$ at the pulse frequency, thus measuring the
average \textit{change} in $G_{QPC}$ when a voltage pulse is
applied~\cite{Sprinzak2002}. We denote this quantity by $\delta
G_{QPC}$. Figure~\ref{fig:ZeemanCharge}a shows such a measurement
of $\delta G_{QPC}$, lock-in detected at the pulse frequency, as a
function of $V_G$ around the $0\leftrightarrow1$ electron
transition. The different sections of the dip correspond to
Figs.\ref{fig:ZeemanCharge}b and c as indicated, where $GS$ ($ES$)
is the electrochemical potential of the $0\leftrightarrow\ua$
($0\leftrightarrow\da$) transition. Figure~\ref{fig:ZeemanCharge}d
shows a plot of the derivative of $\delta G_{QPC}$ with respect to
$V_G$ in the ($V_P,V_G$)-plane, where the one-electron Zeeman
splitting is clearly resolved. This measurement is analogous to
increasing the source-drain bias $V_{SD}$ in a transport
measurement, and therefore leads to a similar plot as in
Fig.~\ref{fig:highbias2}, with $V_{SD}$ replaced by $V_P$, and
$dI_{DOT}/dV_{SD}$ replaced by $d (\delta
G_{QPC})/dV_G$~\cite{FujisawaPRL02,ElzermanAPL2004}.

\begin{figure}[htb]
\includegraphics[width=3.4in]{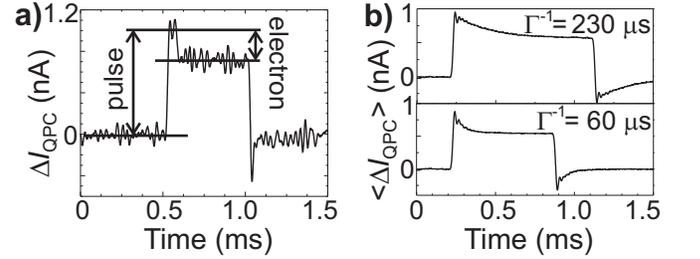}
\caption{(a) Measured changes in the QPC current, $\Delta
I_{QPC}$, when a pulse is applied to a gate, near the degeneracy
point between $0$ and $1$ electrons on the dot (bias voltage
across the QPC is 1 mV). The pulse of positive voltage increases
the QPC current due to the capacitive coupling between the pulsed
gate and the QPC. Shortly after the start of the pulse, an
electron tunnels onto the dot and the QPC current decreases. When
the pulse has ended, the electron tunnels off the dot again. (b)
Average of 286 traces as in (a). The top and bottom panel are
taken with a different gate settings, and therefore different
tunnel rates are observed. The damped oscillation following the
pulse edges is due to the 8th-order 40 kHz filter used. Data
adapted from ~\textcite{VandersypenAPL2004}.}
\label{fig:RealTimeTunn}
\end{figure}

The QPC response as a function of pulse length is a unique function of tunnel rate. Therefore, comparison of the obtained response function with the theoretical function yields an accurate value of the tunnel rate~\cite{ElzermanAPL2004,HansonThesis2005}. In a double dot, charge sensing can be used to quantitatively set the ratio of the tunnel rates (see~\textcite{JohnsonSpinBlock} for details), and also to observe the \textit{direction} of tunnel events~\cite{FujisawaScience2006}.

Electron tunneling can be observed in \textit{real time} if the
time between tunnel events is longer than the time needed to
determine the number of electrons on the dot -- or equivalently:
if the bandwidth of the charge detection exceeds the tunnel rate and the signal from a single electron charge exceeds the noise level over that bandwidth~\cite{Schoelkopf_rfSET,Lu03a}. Figure~\ref{fig:RealTimeTunn}a
shows gate-pulse-induced electron tunneling in real time. In
Fig.~\ref{fig:RealTimeTunn}b, the average of many such traces is
displayed; from the exponential decay of the signal the tunnel
rate can be accurately determined.

Optimized charge sensing setups typically have a bandwidth that
allows tunneling to be observed on a microsecond
timescale~\cite{Lu03a,Fujisawa04a,VandersypenAPL2004,Schleser04}. If the
relaxation of the electron spin occurs on a longer timescale,
single-shot readout of the spin state becomes possible. This is
the subject of the next section.

\section{Single-shot readout of electron spins}
\label{Section:Readout}
\subsection{Spin-to-charge conversion}

The ability to measure individual quantum states in a single-shot mode is important both for fundamental science and for possible applications in quantum information processing. Single-shot immediately implies that the measurement must have high fidelity (ideally $100 \%$) since only one copy of the state is available and no averaging is possible. 

\begin{figure}[htb]
\includegraphics[width=3.4in, clip=true]{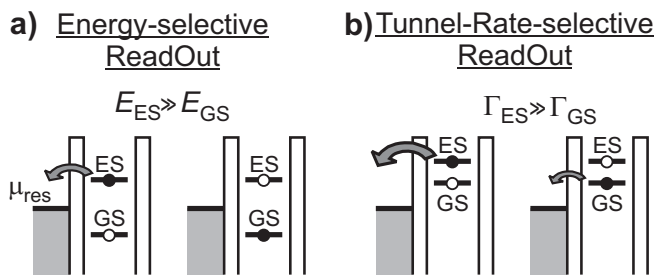}
\caption{Energy diagrams depicting two different methods for
spin-to-charge conversion: \textbf{(a)} Energy-selective readout
(E-RO) and \textbf{(b)} Tunnel-rate-selective readout (TR-RO).}
\label{Fig:ROdiagrams}
\end{figure}

Because of the tiny magnetic moment associated with the electron spin it is very challenging to measure it directly. However, by correlating the spin states to different charge states and subsequently measuring the charge on
the dot, the spin state can be determined \cite{LossDiVincenzo}.
This way, the measurement of a single spin is replaced by the
measurement of a single charge, which is a much easier task.
Several schemes for  such a \textit{spin-to-charge conversion}
have been
proposed~\cite{LossDiVincenzo,KaneNature1998,VandersypenMQC2002,FriesenPRL2004,EngelPRL2004,IonicioiuNJP2005,GreentreePRB2005}.
Two methods, both outlined in Fig.~\ref{Fig:ROdiagrams}, have been
experimentally demonstrated.

In one method, a difference in energy between the spin states is
used for spin-to-charge conversion. In this energy-selective
readout (E-RO), the spin levels are positioned around the
electrochemical potential of the reservoir $\mu_{res}$ (see
Fig.~\ref{Fig:ROdiagrams}a), such that one electron can tunnel off
the dot from the spin excited state, \ket{\:ES}, whereas tunneling
from the ground state, \ket{\:GS}, is energetically forbidden.
Therefore, if the charge measurement shows that one electron
tunnels off the dot, the state was \ket{\:ES}, while if no
electron tunnels the state was \ket{\:GS}. This readout concept was pioneered by Fujisawa \textit{et al.} in a series of transport experiments, where the measured current reflected the average state of the electron after a pulse sequence (see ~\textcite{FujisawaRPP2006} for a review). Using this pump-probe technique, the orbital relaxation time and a lower bound on the spin relaxation time in few-electron vertical and lateral dots was determined~\cite{FujisawaPRB2001,FujisawaPhysB2001,FujisawaNature2002,HansonPRL2003}. A variation of E-RO can be used for reading out the two-electron spin states in a double dot (see Section~\ref{Sec:DDspinstates}).

Alternatively, spin-to-charge conversion can be achieved by
exploiting the difference in \textit{tunnel rates} of the
different spin states to the reservoir. We outline the concept of
this tunnel-rate-selective readout (TR$-$RO) in
Fig.~\ref{Fig:ROdiagrams}b. Suppose that the tunnel rate from
\ket{\:ES} to the reservoir, $\Gamma_{ES}$, is much higher than
the tunnel rate from \ket{\:GS}, $\Gamma_{GS}$, i.e.
$\Gamma_{ES}\gg\Gamma_{GS}$. Then, the spin state can be read out
as follows. At time $t$=0, the levels of both \ket{\:ES} and
\ket{\:GS} are positioned far above $\mu_{res}$, so that one
electron is energetically allowed to tunnel off the dot regardless
of the spin state. Then, at a time $t=\tau$, where
$\Gamma_{GS}^{-1}\gg \tau \gg \Gamma_{ES}^{-1}$, an electron will
have tunneled off the dot with a very high probability if the
state was \ket{\:ES}, but most likely no tunneling will have
occurred if the state was \ket{\:GS}. Thus, the spin information
is converted to charge information, and a measurement of the
number of electrons on the dot reveals the original spin state.
The TR-RO can be used in a similar way if $\Gamma_{ES}$ is much
\textit{lower} than $\Gamma_{GS}$. A conceptually similar
scheme has allowed single-shot readout of a superconducting charge
qubit~\cite{Astafiev04}.

\subsection{Single-shot spin readout using a difference in energy}
\label{Subsection:ERO}

Single-shot readout of a single electron spin has first been
demonstrated using the E-RO technique~\cite{ElzermanNature2004}.
In this section we discuss this experiment in more detail.

A quantum dot containing zero or one electrons is tunnel coupled
to a single reservoir and electrostatically coupled to a QPC that
serves as an electrometer. The electrometer can determine the
number of electrons on the dot in about 10~$\mu$s. The Zeeman
splitting is much larger than the thermal broadening in the
reservoir. The readout configuration therefore is as in
Fig.~\ref{Fig:ROdiagrams}a, with the $0\leftrightarrow \ua$
transition as  \ket{\:GS} and the $0\leftrightarrow \da$
transition as \ket{\:ES}. In the following, we will also use just
$\ua$ and $\da$ to denote these transitions.

\begin{figure}[hbt]
\includegraphics[width=3.4in, clip=true]{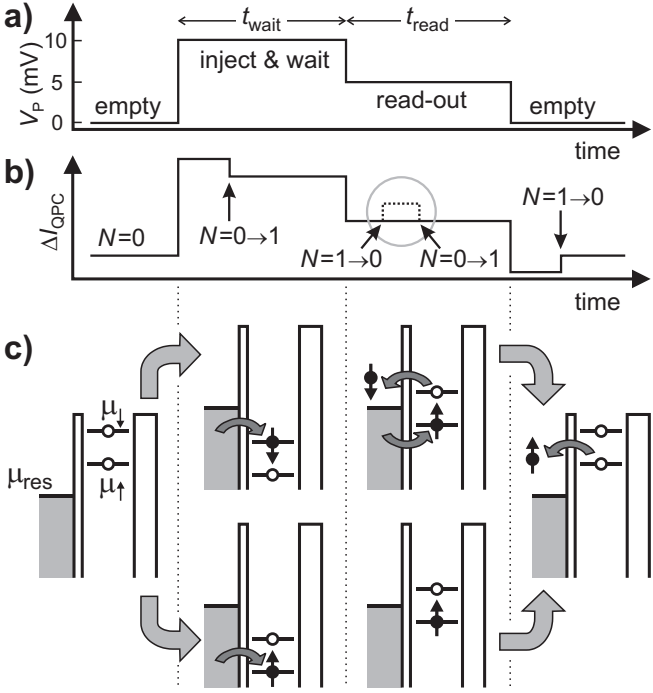}
\caption{Scheme for E-RO of a single electron spin. \textbf{(a)}
Two-level voltage pulse scheme. The pulse level is 10 mV during
$t_{wait}$ and 5 mV during $t_{read}$ (which is 0.5 ms for all
measurements). \textbf{(b)} Schematic response of the QPC if the
injected electron has spin-$\ua$ (solid line) or spin-$\da$
(dotted line; the difference with the solid line is only seen
during the read-out stage). Arrows indicate the moment an electron
tunnels into or out of the quantum dot. \textbf{(c)} Energy
diagrams for spin-up ($E_{\ua}$) and spin-down ($E_{\da}$) during
the different stages of the pulse. If the spin is up at the start
of the read-out stage, no change in the charge on the dot occurs
during $t_{read}$. In contrast, if the spin is down, the electron
can escape and be replaced by a spin-up electron.}
\label{Fig:EROpulse}
\end{figure}

\begin{figure*}[htb]
\includegraphics[width=12cm, clip=true]{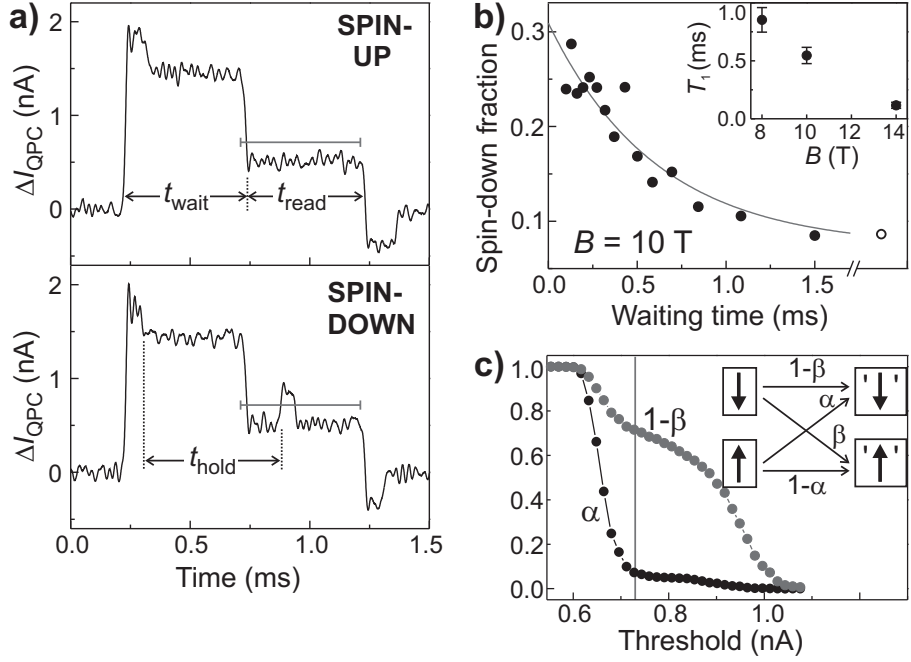}
\caption{Experimental results of single-shot readout of a single
electron spin. \textbf{(a)} The two types of measurement outcomes,
corresponding to a spin-up electron (upper panel) and spin-down
electron (lower panel); see Fig.~\ref{Fig:EROpulse}b for comparison. \textbf{(b)} Dependence of the fraction of
spin-down electrons on the waiting time, showing a clear
exponential decay. Red line is a fit to the data. Inset: the spin
relaxation time $T_1$ as a function of $B$. \textbf{(c)}
Determination of the readout fidelity. Inset: definition of the readout error probabilities $\alpha$ and $\beta$. Main figure: experimentally determined error probabilities at $B$=10~T. At the vertical line, the visibility 1-$\alpha$-$\beta$ reaches a maximum of
65\%. Data reproduced from ~\textcite{ElzermanNature2004}.} \label{Fig:EROresult}
\end{figure*}

To test the single-spin measurement technique, the following
three-stage procedure is used: 1) empty the dot, 2) inject one
electron with unknown spin, and 3) measure its spin state. The
different stages are controlled by gate voltage pulses as in
Fig.~\ref{Fig:EROpulse}a, which shift the dot's energy levels as
shown in Fig.~\ref{Fig:EROpulse}c. Before the pulse the dot is
empty, as both the spin-up and spin-down levels are above the
electrochemical potential of the reservoir $\mu_{res}$. Then a
voltage pulse pulls both levels below $\mu_{res}$. It is now
energetically allowed for one electron to tunnel onto the dot,
which will happen after a typical time $\approx \Gamma^{-1}$. That
particular electron can have spin-up or spin-down as shown in the
lower and upper diagram respectively. During this stage of the
pulse, lasting $t_{wait}$, the electron is trapped on the dot and
Coulomb blockade prevents a second electron to be added. After
$t_{wait}$ the voltage pulse is reduced, in order to position the
energy levels in the readout configuration. If the electron has
spin-up, its energy level is below $\mu_{res}$, so the electron
remains on the dot. If the electron has spin-down, its energy
level is above $\mu_{res}$, so the electron tunnels to the
reservoir after a typical time $\approx \Gamma^{-1}$. Now Coulomb
blockade is lifted and an electron with spin-up can tunnel onto
the dot. Effectively, the spin on the dot has been flipped by a
single electron exchange with the reservoir. After $t_{read}$, the
pulse ends and the dot is emptied again.

The expected QPC-response, $\Delta I_{QPC}$, to such a two-level
pulse is the sum of two contributions (Fig.~\ref{Fig:EROpulse}b).
First, due to a capacitive coupling between pulse-gate and QPC,
$\Delta I_{QPC}$ will change proportionally to the pulse
amplitude. Second, $\Delta I_{QPC}$ tracks the charge on the dot,
i.e. it goes up whenever an electron tunnels off the dot, and it
goes down by the same amount when an electron tunnels onto the
dot. Therefore, if the dot contains a spin-down electron at the
start of the readout stage, a characteristic step appears in
$\Delta I_{QPC}$ during $t_{read}$ for spin-down (dotted trace
inside grey circle). In contrast, $\Delta I_{QPC}$ is flat during
$t_{read}$ for a spin-up electron. Measuring whether a step is
present or absent during the readout stage constitutes the spin
measurement.

Fig.~\ref{Fig:EROresult}a shows experimental traces of the
pulse-response at an in-plane field of 10~T. The expected two
types of traces are indeed observed, corresponding to spin-up
electrons (as in the top panel of Fig.~\ref{Fig:EROresult}a), and
spin-down electrons (as in the bottom panel of
Fig.~\ref{Fig:EROresult}a). The spin state is assigned as follows:
if $\Delta I_{QPC}$ crosses a threshold value (grey line in
Fig.~\ref{Fig:EROresult}a), the electron is declared `spin-down';
otherwise it is declared `spin-up'.

As $t_{wait}$ is increased, the number of `spin-down' traces
decays exponentially (see Fig.~\ref{Fig:EROresult}b), precisely as expected because of spin relaxation to the ground state. This confirms
the validity of the spin readout procedure. The spin decay time
$T_1$ is plotted as a function of $B$ in the inset of
Fig.~\ref{Fig:EROresult}b. The processes underlying the spin
relaxation will be discussed in section~\ref{Spinenvironment}.

The fidelity of the spin measurement is characterized by two error
probabilities $\alpha$ and $\beta$ (see inset to
Fig.~\ref{Fig:EROresult}c). Starting with a spin-up electron, there is a probability $\alpha$ that the measurement yields the wrong outcome `$\da$'. Similarly, $\beta$ is the probability that a spin-down electron is mistakenly measured as `$\ua$'. These error probabilities can be determined from
complementary measurements~\cite{ElzermanNature2004}. Both
$\alpha$ and $\beta$ depend on the value of the threshold as shown
in Fig.~\ref{Fig:EROresult}c for data taken at 10 T. The optimal
value of the threshold is the one for which the visibility
$1-\alpha-\beta$ is maximal (vertical line in
Fig.~\ref{Fig:EROresult}c). For this setting, $\alpha$=0.07 and
$\beta$=0.28, so the measurement fidelity for the spin-up and the
spin-down state is ~0.93 and ~0.72 respectively. The measurement
visibility in a single-shot measurement is thus 65\%, and the
fidelity ($1-(\alpha+\beta)/2$) is 82\%. Significant improvements
in the spin measurement visibility can be made by lowering the
electron temperature (smaller $\alpha$) and by making the charge
measurement faster (smaller $\beta$).

The first all-electrical single-shot readout of an electron spin
has thus been performed using E-RO. However, this scheme has a few
drawbacks: (i) E-RO requires an energy splitting of the spin
states larger than the thermal energy of the electrons in the
reservoir. Thus, for a single spin the readout is only effective
at very low electron temperature and high magnetic fields ($k_B
T\ll \Delta E_Z$). Also, interesting effects occurring close to
degeneracy, e.g. near the singlet-triplet crossing for two
electrons, can not be probed. (ii) Since the E-RO relies on
precise positioning of the spin levels with respect to the
reservoir, it is very sensitive to fluctuations in the
electrostatic potential. Background charge
fluctuations~\cite{FujisawaChargeNoise} can easily push the levels
out of the readout configuration. (iii) High-frequency noise can
spoil the E-RO by inducing photon-assisted tunneling from the spin
ground state to the reservoir~\cite{OnacPRL2006}. Since the QPC is
a source of shot noise, this limits the current through the QPC
and thereby the bandwidth of the charge
detection~\cite{VandersypenAPL2004}. These constraints have
motivated the search for a different method for spin-to-charge
conversion, and have led to the demonstration of the
tunnel-rate-selective readout (TR-RO) which we treat in the next
section.

\subsection{Single-shot spin readout using a difference in tunnel rate}
\label{Subsection:TRRO}
The main ingredient necessary for TR-RO is a spin dependence in
the tunnel rates. To date, TR-RO has only been demonstrated for a two-electron dot, where the electrons are either
in the spin-singlet ground state, denoted by \ket{\:S}, or in a
spin-triplet state, denoted by \ket{\:T}. In \ket{\:S}, the two
electrons both occupy the lowest orbital, but in \ket{\:T} one
electron is in the first excited orbital. Since the wave function
in this excited orbital has more weight near the edge of the dot
\cite{kouwenhoven01}, the coupling to the reservoir is stronger than
for the lowest orbital. Therefore, the tunnel rate from a triplet
state to the reservoir $\Gamma_T$ is much larger than the rate
from the singlet state $\Gamma_S$, i.e. $\Gamma_T\gg\Gamma_S$
\cite{RonaldMoriond}.

\begin{figure}[htb]
\includegraphics[width=3.2in, clip=true]{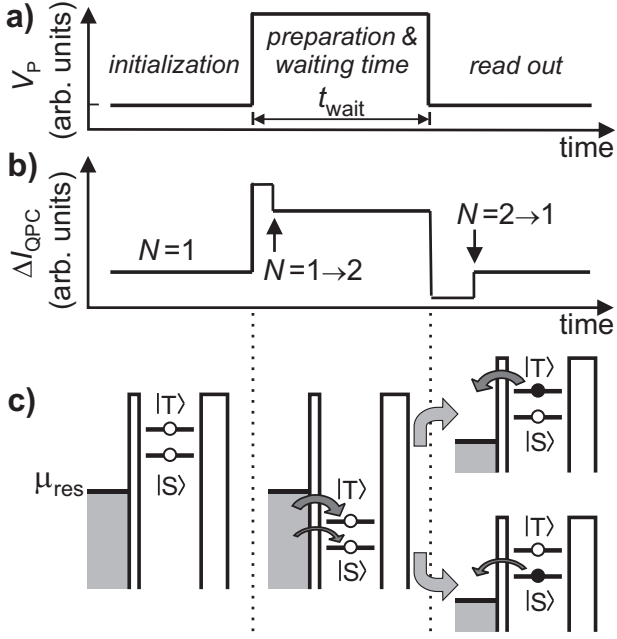}
\caption{Single-shot readout of two-electron spin states using
TR-RO. (a) Voltage pulse waveform applied to one of the gate
electrodes. (b) Response of the QPC current to the waveform of
(a). (c) Energy diagrams indicating the positions of the levels
during the three stages. In the final stage, spin is converted to
charge information due to the difference in tunnel rates for
states \ket{\:S} and \ket{\:T}.} \label{Fig:TRROpulse}
\end{figure}

The TR-RO is tested experimentally in ~\textcite{HansonPRL2005} by applying gate voltage pulses as depicted in Fig.~\ref{Fig:TRROpulse}a. Figure \ref{Fig:TRROpulse}b shows the expected response of $I_{QPC}$ to the pulse, and Fig.~\ref{Fig:TRROpulse}c depicts the level diagrams in the three different stages. Before the pulse starts, there is one electron on the dot. Then, the pulse pulls the levels down so that a second electron can tunnel onto the dot ($N\!=\!1\!\rightarrow\!2$), forming either a singlet or a triplet state with the first electron.
The probability that a triplet state is formed is given by
$3\Gamma_T/(\Gamma_S + 3\Gamma_T)$, where the factor of 3 is due
to the degeneracy of the triplets. After a variable waiting time
$t_{wait}$ the pulse ends and the readout process is initiated,
during which one electron can leave the dot again. The rate for
tunneling off depends on the two-electron state, resulting in the
desired spin-to-charge conversion. Due to the direct capacitive
coupling of the pulse gate to the QPC channel, $\Delta I_{QPC}$
follows the pulse shape. Tunneling of an electron on or off the
dot gives an additional step in $\Delta I_{QPC}$ as indicated by
the arrows in Fig.~\ref{Fig:TRROpulse}b.

\begin{figure}[htb]
\includegraphics[width=3.4in, clip=true]{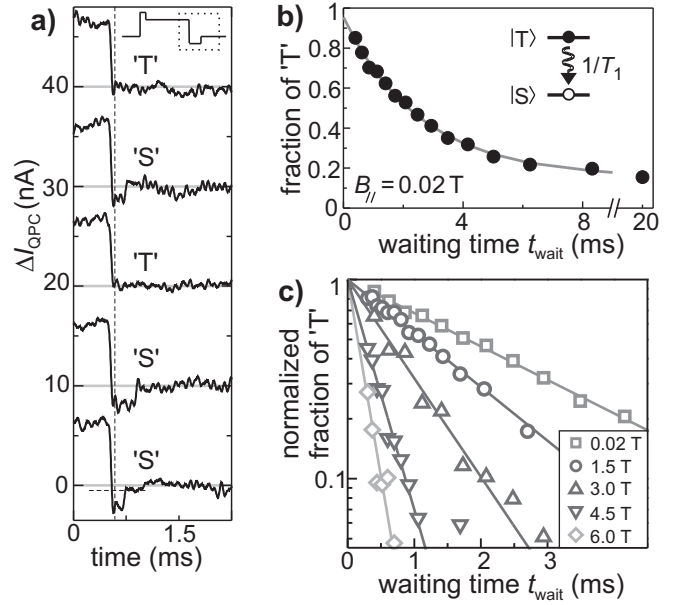}
\caption{(a) Real-time traces of $\Delta I_{QPC}$ during the last
part of the waveform (dashed box in the inset), for
$t_{wait}\!=0.8$~ms. At the vertical dashed line, $N$ is
determined by comparison with a threshold (horizontal dashed line
in bottom trace) and the spin state is declared $'T'$ or $'S'$
accordingly. (b) Fraction of $'T'$ as a function of waiting time
at $B_{/\!/}\!$~=~0.02~T, showing a single-exponential decay with
a time constant $T_1$ of 2.58~ms. (c) Normalized fraction of $'T'$
vs. $t_{wait}$ for different values of $B_{/\!/}$. The
singlet-triplet splitting $E_{ST}$ in this experiment is 1~meV. Data reproduced from ~\textcite{HansonPRL2005}.}
\label{Fig:TRROresult}
\end{figure}

In the experiment, $\Gamma_S$ is tuned to 2.5~kHz, and $\Gamma_T$
is $\approx\:$50~kHz. The filter bandwidth is 20~kHz, and
therefore many of the tunnel events from \ket{\:T} are not
resolved, but the tunneling from \ket{\:S} is clearly visible.
Figure~\ref{Fig:TRROresult}a shows several traces of $\Delta
I_{QPC}$, from the last part (0.3~ms) of the pulse to the end of
the readout stage (see inset), for a waiting time of 0.8~ms. In
some traces, there are clear steps in $\Delta I_{QPC}$, due to an
electron tunneling off the dot. In other traces, the tunneling
occurs faster than the filter bandwidth. In order to discriminate
between \ket{\:S} and \ket{\:T}, the number of electrons on the
dot is determined at the readout time (vertical dashed line in
Fig.~\ref{Fig:TRROresult}a) by comparing $\Delta I_{QPC}$ to a
threshold value (as indicated by the horizontal dashed line in the
bottom trace of Fig.~\ref{Fig:TRROresult}a). If $\Delta I_{QPC}$
is below the threshold, it means $N\!=\!2$ and the state is
declared $'S'$. If $\Delta I_{QPC}$ is above the threshold, it
follows that $N\!=\!1$ and the state is declared $'T'$.

To  verify that $'T'$ and $'S'$ indeed correspond to the spin
states \ket{\:T} and \ket{\:S}, the relative occupation
probabilities are changed by varying the waiting time. As shown in
Fig.~\ref{Fig:TRROresult}b, the fraction of $'T'$ indeed decays
exponentially as $t_{wait}$ is increased, due to relaxation, as before. The error probabilities are
found to be $\alpha\!~=\!~0.15$ and $\beta\!=\!~0.04$, where $\alpha$ ($\beta$) is the probability that a measurement on the state \ket{\:S} (\ket{\:T} yields the wrong outcome $'T'$ ($'S'$). The single-shot visibility is thus
81\% and the fidelity is 90\%. These numbers agree very well with
the values predicted by a simple rate-equation
model~\cite{HansonPRL2005}. Figure~~\ref{Fig:TRROresult}c shows
data at different values of the magnetic field. These results are
discussed in more detail in section~\ref{Spinenvironment}.

A major advantage of the TR-RO scheme is that it does not rely on
a large energy splitting between the spin states. Furthermore, it
is robust against background charge fluctuations, since these
cause only a small variation in the tunnel rates (of order
$10^{-3}$ in Ref. \cite{FujisawaChargeNoise}). Finally,
photon-assisted tunneling is not harmful since here tunneling is
energetically allowed regardless of the initial spin state. Thus,
TR-RO overcomes several constraints of E-RO. However, TR-RO can
only be used if there exist state-dependent tunnel rates. In
general, the best choice of readout method will depend on the
specific demands of the experiment and the nature of the states
involved.

It is interesting to think about a measurement protocol that would leave the spin state unaffected, a so-called Quantum Non-Demolition (QND) measurement. With readout schemes that make use of tunneling to a reservoir as the ones described in this section, QND measurements are not possible because the electron is lost after tunneling; the best one can do in this case is to re-initialize the dot electrons to the state they were in before the tunneling occured~\cite{Meunier06}. However, by making the electron tunnel not to a reservoir, but to a second dot~\cite{EngelPRL2004,EngelScience05}, the electron can be preserved and QND measurements are in principle possible. One important example of such a scheme is the readout of double-dot singlet and triplet states using Pauli blockade that we will discuss in Section~\ref{Sec:PauliBlockade}.

\section{Spin-interaction with the environment}
\label{Spinenvironment}

The magnetic moment of a single electron spin,
$\mu_B$=9.27$\times$10$^{-24}$ J/T, is very small. As a result,
electron spin states are only weakly perturbed by their magnetic
environment. Electric fields affect spins only indirectly, so generally spin states are only weakly influenced by their electric environment as well. One notable exception is the case of two-electron spin states -- since the singlet-triplet splitting directly depends on the Coulomb repulsion between the two electrons, it is very sensitive to electric field fluctuations~\cite{Hu06} -- but we will not discuss this further here.

For electron spins in semiconductor quantum dots, the most important interactions with the environment occur via the spin-orbit coupling, the hyperfine coupling with the nuclear spins of the host material and virtual exchange processes with electrons
in the reservoirs. This last process can be efficiently suppressed by
reducing the dot-reservoir tunnel coupling or creating a large gap between the eletrochemical potentials in the dot and in the lead~\cite{FujisawaNature2002}, and we will not
further consider it in this section. The effect of the spin-orbit
and hyperfine interactions can be observed in several ways. First,
the spin eigenstates are redefined and the energy splittings are
renormalized. A good example is the fact that the $g$-factor of
electrons in bulk semiconductors can be very different from $2$,
due to the spin-orbit interaction. In bulk GaAs, for instance, the
$g$-factor is $-0.44$. Second, fluctuations in the environment can
lead to phase randomization of the electron spin, by convention
characterized by a time scale $T_2$. Finally, electron spins can
also be flipped by fluctuations in the environment, thereby
exchanging energy with degrees of freedom in the environment. This
process is characterized by a timescale $T_1$.


\subsection{Spin-orbit interaction}
\label{Sec:spin-orbit}
\subsubsection{Origin}

The spin of an electron moving in an electric field $\vec{E}$
experiences an internal magnetic field, proportional to $\vec{E}
\times \vec{p}$, where $\vec{p}$ is the momentum of the electron.
This is the case, for instance, for an electron ``orbiting'' about
a positively charged nucleus. This internal magnetic field acting
on the spin depends on the orbital the electron occupies, i.e.
spin and orbit are coupled. An electron moving through a solid
also experiences electric fields, from the charged atoms in the
lattice. In crystals that exhibit bulk inversion asymmetry (BIA), such as in the zinc-blende structure of GaAs, the local electric fields lead to a net contribution to the spin-orbit interaction (which generally becomes stronger for heavier elements). This effect is  known as the Dresselhaus contribution to the spin-orbit interaction\cite{dresselhaus55,dyakonov86,wrinkler03}.

In addition, electric fields associated with asymmetric confining
potentials also give rise to a spin-orbit interaction (SIA or
structural inversion asymmetry). This occurs for instance in a
2DEG formed at a GaAs/AlGaAs heterointerface. It is at first sight surprising that there is a net spin-orbit interaction: since the state is bound along the growth direction, the average electric field in the conduction band must be zero (up to a correction due to the effective mass discontinuity at the interface, which results in a small force that is balanced by a small average electric field). The origin of the net spin-orbit interaction lies in mixing with other bands, mainly the valence band, which contribute a non-zero average electric field~\cite{wrinkler03,pfeffer99}.
Only in symmetric quantum wells with symmetric doping, these other contributions are zero as well. The spin-orbit contribution from SIA is known as the Rashba term~\cite{rashba60,bychkov84}.


\subsubsection{Spin-orbit interaction in bulk and 2D}
\label{sec:SO_bulk_2D}

In order to get insight in the effect of the Dresselhaus
spin-orbit interaction in zinc-blende crystals, we start from the
bulk Hamiltonian~\cite{dyakonov71,wrinkler03}, 
\be 
{\cal H}_D^{3D} \propto
[p_x (p_y^2-p_z^2) \sigma_x + p_y (p_z^2-p_x^2) \sigma_y +
p_z (p_x^2-p_y^2) \sigma_z] 
\ee 
where $x, y$ and $z$ point along
the main crystallographic directions, $(100)$, $(010)$ and
$(001)$.

In order to obtain the spin-orbit Hamiltonian in 2D systems, we
integrate over the growth direction. For 2DEGs grown along the
$(001)$ direction, $\langle p_z \rangle = 0$, and $\langle p_z^2
\rangle$ is a heterostructure dependent but fixed number. The
Dresselhaus Hamiltonian then reduces to 
\be 
{\cal H}_D^{2D,(001)}
\propto [- p_x \langle p_z^2 \rangle \sigma_x + p_y
\langle p_z^2 \rangle \sigma_y +  p_x p_y^2 \sigma_x - p_y p_x^2 \sigma_y]
\ee 
The first two terms are the linear Dresselhaus terms and the
last two are the cubic terms. Usually the cubic terms are much
smaller than the linear terms, since $\langle p_z^2 \rangle \gg
p_x^2, p_y^2$ due to the strong confinement along $z$. We then
retain~\cite{dresselhaus55} 
\be 
{\cal H}_D^{2D,(001)} = \beta [- p_x \sigma_x +  p_y \sigma_y] \;, \label{eq:H_D} 
\ee 
where $\beta$
depends on material properties and on $\langle p_z^2 \rangle$. It
follows from Eq.~\ref{eq:H_D} that the internal magnetic field is
aligned with the momentum for motion along $(010)$, but is
opposite to the momentum for motion along $(100)$ (see
Fig.~\ref{fig:spinorbit}a).

\bfig
\begin{center}
\includegraphics[width=7cm]{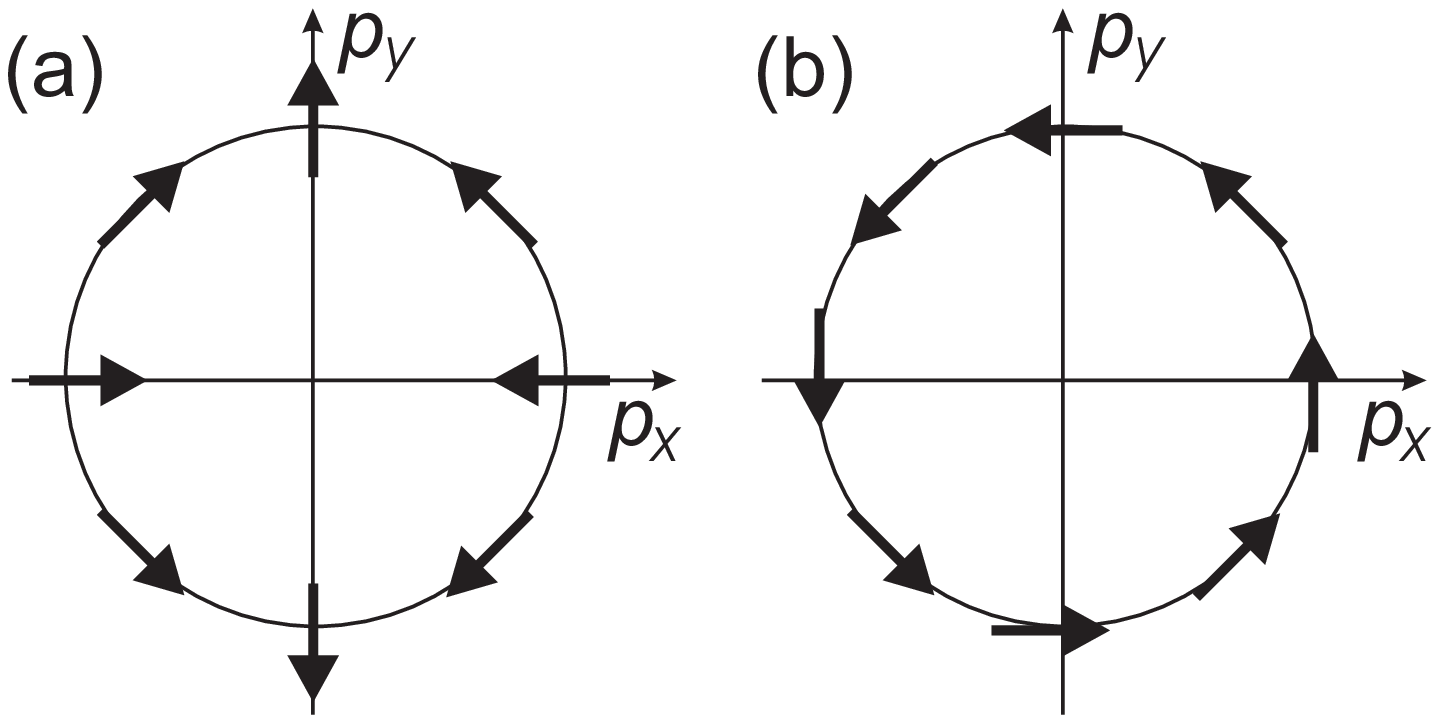}
\end{center}
\caption{The small arrows indicate the orientation of the apparent
magnetic field acting on the electron spin as a result of (a) the
Dresselhaus and (b) the Rashba spin-orbit interaction when the
electron travels through a GaAs crystal with momentum $\vec{p}$.}
\label{fig:spinorbit} \efig

Similarly, we now write down the spin-orbit Hamiltonian for the
Rashba contribution. Assuming that the confining electric field is
along the $z$-axis, we have 
\be 
{\cal H}_R \propto [ \vec{E}
\times \vec{p}\;]\; \vec{\sigma} = E_z (- p_y  \sigma_x + p_x \sigma_y) \;, 
\ee
or 
\be 
{\cal H}_R = \alpha (- p_y \sigma_x + p_x \sigma_y) \;, 
\label{eq:H_R} 
\ee 
with $\alpha$ a number that is
material-specific and also depends on the confining potential.
Here the internal magnetic field is always orthogonal to the
momentum (see Fig.~\ref{fig:spinorbit}b).

We point out that as an electron moves ballistically over some
distance, $l$, the angle by which the spin is rotated, whether
through Rashba or linear Dresselhaus spin-orbit interaction, is
independent of the velocity with which the electron moves. The
faster the electron moves, the faster the spin rotates, but the
faster the electron travels over the distance $l$ as well. In the
end, the rotation angle is determined by $l$ and the spin-orbit
strength only. A useful quantity is the distance associated with a
$\pi$ rotation, known as the spin-orbit length, $l_{SO}$. In GaAs,
estimates for $\beta$ vary from $10^3$ m/s to $3\times 10^3$ m/s,
and it follows that the spin-orbit length, $l_{SO} = \hbar /
(\beta m^*)$ is $1-10 \mu$m, in agreement with experimentally
measured values~\cite{zumbuhl02}. The Rashba contribution can be
smaller or larger than the Dresselhaus contribution, depending on
the structure. From Fig.~\ref{fig:spinorbit}, we see that the
Rashba and Dresselhaus contributions add up for motion along
the $(110)$ direction and oppose each other along $(\bar{1}10)$, i.e. 
the spin-orbit interaction is anisotropic~\cite{Haug05}.

In 2DEGs, spin-orbit coupling (whether Rashba or Dresselhaus) can lead to spin relaxation via several mechanisms~\cite{zutic04}. The D'yakonov-Perel mechanism~\cite{dyakonov71,wrinkler03} refers to spin randomization that occurs when the electron follows randomly oriented ballistic trajectories \emph{between} scattering events (for each trajectory, the internal magnetic field is differently oriented). In addition, spins can be flipped \emph{upon} scattering, via the Elliot-Yafet mechanism~\cite{elliott54,yafet63} or the Bir-Aronov-Pikus mechanism~\cite{bir75}.


\subsubsection{Spin-orbit interaction in quantum dots}

From the semi-classical picture of the spin-orbit interaction, we
expect that in 2D quantum dots with dimensions much smaller than
the spin-orbit length $l_{SO}$, the electron spin states will be
hardly affected by the spin-orbit interaction. We now show that the same result follows from the quantum-mechanical description, where the
spin-orbit coupling can be treated as a small perturbation to the
discrete orbital energy level spectrum in the quantum dot.

First, we note that stationary states in a quantum dot are bound
states, for which $\langle p_x \rangle = \langle p_y \rangle = 0$.
This leads to the important result that 
\be 
\bra{n l \down} H_{SO} \ket{\,n l \up} 
\propto \bra{n l} p_{x,y} \ket{\,n l} \bra{\down} \sigma_{x,y} \ket{\up} = 0 \;, 
\ee 
where $n$ and $l$ label the orbitals in
the quantum dot, and 
$H_{SO}$ stands for the spin-orbit Hamiltonian, which consists of terms of the form $p_{x,y} \sigma_{x,y}$ both for the Dresselhaus and Rashba contributions. Thus, the spin-orbit
interaction does not directly couple the Zeeman-split
sublevels of a quantum dot orbital. However, the spin-orbit 
Hamiltonian does couple states that contain both different orbital 
and different spin parts\cite{khaetskii00}. As a result, what we
usually call the electron spin states `spin-up' and `spin-down' in
a quantum dot, are in reality admixtures of spin and orbital
states~\cite{khaetskii01}. When the Zeeman splitting is well below
the orbital level spacing, the perturbed eigenstates can be
approximated as \bea
\ket{\,n l\up}^{(1)} = \ket{n l\up} \hspace*{3.5cm} \nonumber \\
 + \sum_{n' l'\neq n l}
\frac{\bra{n' l'\down}H_{SO} \ket{\,n l\up}}{E_{n l} - E_{n' l'} -
\Delta E_Z} \ket{n' l'\down} \;, \hspace*{-.5cm}
\label{eq:SO_zeeman}
\\
\ket{\,nl\down}^{(1)} =  \ket{nl\down} \hspace*{3.5cm} \nonumber\\
+ \sum_{n'l'\neq nl} \frac{\bra{n'l'\up}H_{SO}
\ket{\,nl\down}}{E_{nl} - E_{n'l'} + \Delta E_Z} \ket{n'l'\up}
\hspace*{-.5cm} \label{eq:SO_zeeman2} 
\eea 
(the true eigenstates can be obtained via exact
diagonalization~\cite{cheng04}). Here $\Delta E_Z$ refers to the unperturbed spin splitting (in the remainder of the review, $\Delta E_Z$ refers to the actual spin splitting, including all perturbations). The energy splitting between the spin-up and spin-down states will be renormalized accordingly, $\Delta E_Z^{(1)} = E_\down^{(1)} - E_\up^{(1)}$~\cite{sousa03b,stano05} (see also Fig.~\ref{fig:relaxation}(a)). In GaAs few-electron quantum dots, the measured $g$-factor in absolute value is usually in the range of $0.2 - 0.4$, and is sometimes magnetic field dependent (see Fig.~\ref{Fig:Zeeman01}). A similar behavior of the $g$-factor was found in GaAs/AlGaAs 2DEGs~\cite{dobers88}.

In contrast to single-electron spin states in a quantum dot, the
lowest two-electron spin states, singlet and triplet, are coupled
directly by the spin-orbit interaction (except for $T_0$ and $S$,
which are not coupled to lowest order in the spin-orbit
interaction, due to spin selection
rules~\cite{dickmann03,sasaki05,golovach07,florescu06,climente06}). This is not so surprising since the singlet and triplet states
themselves involve different orbitals. Nevertheless, coupling to
two-electron spin states composed of higher orbitals needs to be
included as well, as their effect is generally not 
negligible~\cite{golovach07,climente06}. The leading order correction
to the two-electron wavefunction is then given by 
\bea
\ket{\,S q}^{(1)} = \ket{\,S q} \hspace*{3.5cm} \nonumber \\
+ \sum_{q'\neq q} \frac{\langle T_\pm q' |H_{SO}|S q\rangle}{E_{T_\pm q'} - E_{S q}} \ket{\,T_\pm q'}\\
\ket{\,T_0 q}^{(1)} = \ket{\,T_0 q} \hspace*{3.5cm} \nonumber \\
+ \sum_{q'\neq q} \frac{\langle T_0 q' |H_{SO}|T_\pm q\rangle}{E_{T_\pm q'} - E_{T_0 q}} \ket{\,T_\pm q'}\\
\ket{\,T_\pm q}^{(1)} = \ket{\,T_\pm q}
+ \sum_{q'\neq q} \frac{\langle T_\pm q' |H_{SO}|S q\rangle}{E_{S q'} - E_{T_\pm q}} \ket{\,S q'} \nonumber \\
+ \sum_{q'\neq q} \frac{\langle T_\pm q' |H_{SO}|T_0
q\rangle}{E_{T_0 q'} - E_{T_\pm q}} \ket{T_0 q'}
\label{eq:SO_ST} 
\eea 
where $q$ is shorthand for the quantum
numbers $n_1 l_1 n_2 l_2$ that label the orbital for each of the
two electrons. It can be seen from inspection of the spin-orbit
Hamiltonian and the form of the wavefunctions that many of the
matrix elements in these expressions are zero. A more detailed
discussion is beyond the scope of this review but can be found
in~\textcite{golovach07,climente06}.


\subsubsection{Relaxation via the phonon bath}
\label{Sec:RelaxationviaPhononBath}
Electric fields cannot cause transitions between pure spin states.
However, we have seen that the spin-orbit interaction perturbs the
spin states and the eigenstates become admixtures of spin and
orbital states, see Eqs.~\ref{eq:SO_zeeman}-\ref{eq:SO_ST}. These
new eigenstates can be coupled by electric fields (see
Fig.~\ref{fig:relaxation}), and electric field fluctuations can
lead to spin relaxation~\cite{khaetskii00,khaetskii01,woods02}. As
we will see, this indirect mechanism is not very efficient, and
accordingly, very long spin relaxation times have been observed
experimentally~\cite{FujisawaPRB2001,FujisawaNature2002,HansonPRL2003,ElzermanNature2004,kroutvar04,HansonPRL2005,sasaki05,meunier07,amasha06}.

In general, fluctuating electric fields could arise from many sources, including fluctuations in the gate potentials, background charge
fluctuations or other electrical noise
sources~\cite{marquardt05,borhani06}. However, as we shall see, it appears that in carefully designed measurement systems, the electric field fluctuations of these extraneous noise sources is less important than those caused by
the phonon bath. Phonons can produce electric field fluctuations
in two ways. First, so-called deformation potential phonons
\emph{inhomogeneously} deform the crystal lattice, thereby
altering the bandgap in space, which gives rise to fluctuating
electric fields. This mechanism occurs in all semiconductors.
Second, in polar crystals such as GaAs, also \emph{homogeneous}
strain leads to electric fields, through the piezo-electric effect
(piezo-electric phonons).

The phonon-induced transition rate between the renormalized states $\ket{\,n,l,\up}^{(1)}$
and $\ket{\,n,l,\down}^{(1)}$ is given by Fermi's golden rule (an analogous expression can be derived for relaxation from triplet to singlet states, or between other spin states): 
\be
\Gamma = \frac{2\pi}{\hbar} \sum_{n,l} |{^{(1)}\bra{nl\up}} {\cal
H}_{e,ph} \ket{\,nl\down}^{(1)}|^2 D(\Delta E_Z^{(1)})
\label{eq:golden_rule} 
\ee 
where $D(E)$ is the phonon density of
states at energy $E$. ${\cal H}_{e,ph}$ is the electron-phonon coupling  Hamiltonian, given by
\be 
{\cal
H}_{e,ph}^{\vec{q}j} = M_{\vec{q}j} \, e^{i\vec{q}\vec{r}} \; (b_{\vec{q}j}^\dagger + b_{\vec{q}j}) \;, 
\ee
where $M_{\vec{q}j}$ is a measure of the electric field strength of a phonon with wavevector $\vec{q}$ and phonon branch $j$ (one longitudinal and two transverse modes), $\vec{r}$ is the position vector of the electron and $b_{\vec{q}j}^\dagger$ and $b_{\vec{q}j}$ are the phonon creation and annihilation operators respectively.

The relaxation rate thus depends on the phonon density of states
at the spin-flip energy (the phonons have to carry away the energy), and on how strongly the electron-phonon coupling connects the spin-orbit perturbed spin states (see Eq.~\ref{eq:golden_rule}). The latter in turn depends on (i) the degree of admixing between spin and orbital states (see Eqs.~\ref{eq:SO_zeeman}-~\ref{eq:SO_zeeman2}), (ii) the electric field strength of a single phonon ($\propto M_{\vec{q}j}$), (iii) the effectiveness of phonons at coupling different dot orbitals, via $e^{i\vec{q}\vec{r}}$, and (iv) the phonon occupation, via $(b_{\vec{q}j}^\dagger + b_{\vec{q}j})$. In addition, (v), we will see that an external magnetic field is necessary for spin relaxation to occur. We next discuss each of these elements separately, starting with the density of states.

\bfig
\includegraphics[width=8cm]{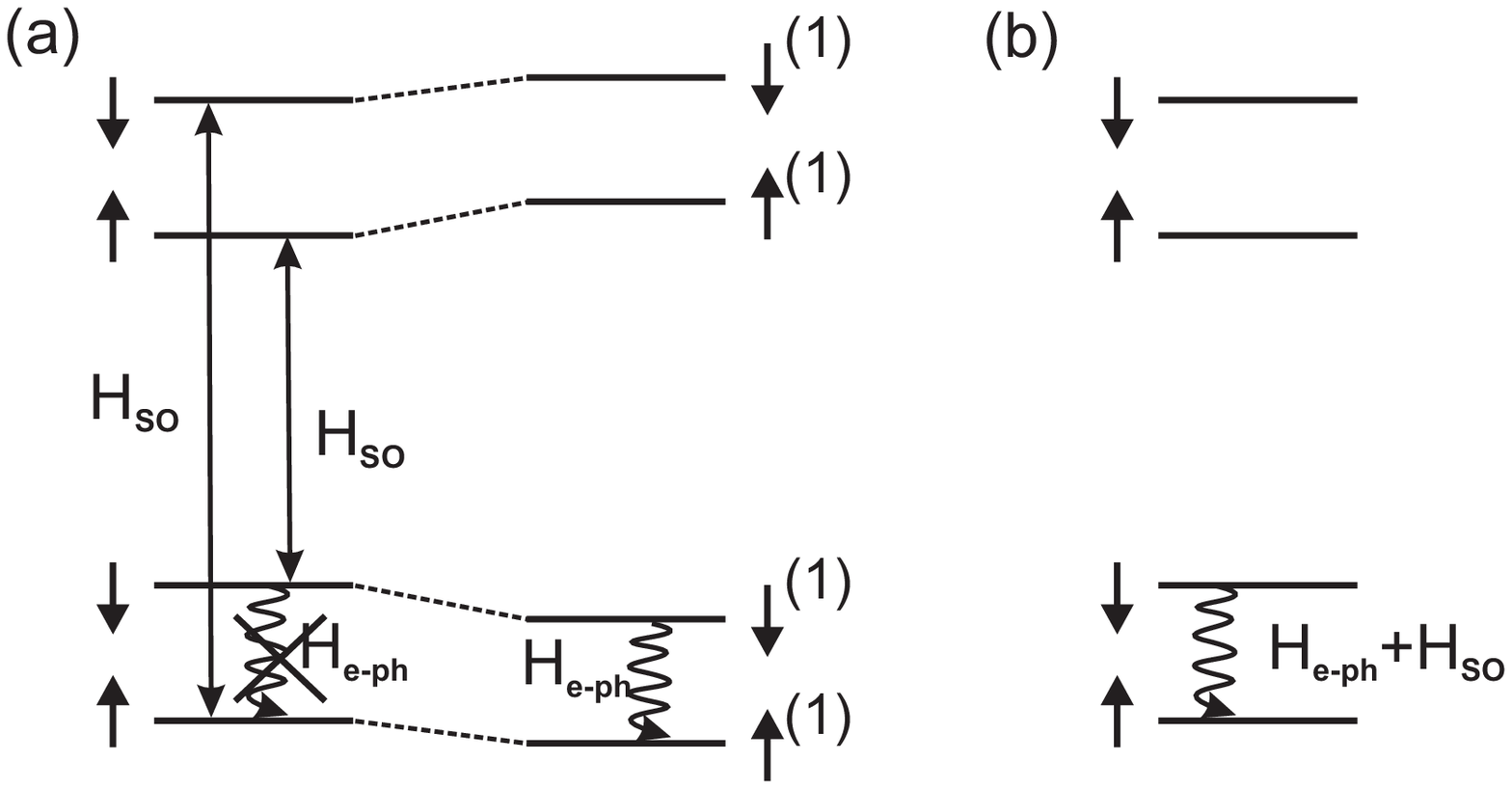}
\caption{Two views on spin relaxation due to the spin-orbit
interaction and the phonon bath. (a) The electron phonon
interaction doesn't couple pure spin states, but it does couple
the spin-orbit perturbed spin states, labeled with superscripts $^{(1)}$. 
Here ${\cal H}_{SO}$ and ${\cal H}_{e,ph}$ are treated sequentially 
(although they don't commute). (b) The combined
electron-phonon and spin-orbit Hamiltonian couples pure spin
states.} \label{fig:relaxation} \efig

The phonons are taken to be bulk phonons in most discussions of relaxation in GaAs quantum dots. This may not be fully accurate since the dot is formed at a hetero-interface, and furthermore is in close vicinity to the surface (e.g. surface acoustic waves have also been observed to couple to dots, when explicitly excited~\cite{naber06}). Nevertheless, this simplification has worked reasonably well so far for explaining observations of relaxation in dots~\cite{fujisawa98,kroutvar04,amasha06,meunier07}. 
In what follows, we therefore consider bulk phonons only. Furthermore, we only include acoustic phonons, as the energies for optical phonons are much higher than typical spin-flip energies~\cite{ashcroft}. Since bulk acoustic phonons have a linear dispersion relation at low energies~\cite{ashcroft}, the phonon density of states increases quadratically with energy.

(i) The degree of admixing between spin and orbitals obviously scales with the spin-orbit coupling parameters, $\alpha$ and $\beta$. Since the spin-orbit interaction is anisotropic ($\alpha$ and $\beta$ can add up or cancel out depending on the magnetic field orientation with respect to the crystal axis), the admixing and hence the relaxation rate are anisotropic as well~\cite{falko05}. Furthermore, the admixing depends on how close together in energy the relevant orbitals are (see Eqs.~\ref{eq:SO_zeeman}-~\ref{eq:SO_ST}). At an avoided crossing of two levels caused by the spin-orbit interaction, the admixing will be complete~\cite{bulaev05,stano05,stano06}.

(ii) The electric field associated with a single phonon scales as
$1/\sqrt{q}$ for piezo-electric phonons and as $\sqrt{q}$ for
deformation potential phonons, where $q$ is the phonon wavenumber.
This difference can be understood from the fact that small phonon
energies correspond to long wavelengths, and therefore nearly
homogeneous crystal strain, which can only create electric fields
through the piezo-electric effect. At sufficiently small energies
(below $\sim 0.6$ meV in GaAs), the effect of piezo-electric
phonons thus dominates over the effect of deformation potential
phonons. As the phonon energy increases, deformation potential
phonons become more important than piezo-electric phonons.

(iii) How effectively different orbitals are coupled by phonons, i.e. the size of the matrix element $\langle n l \uparrow | e^{i\vec{q}\vec{r}} | n' l' \uparrow \rangle$, depends on the phonon wavelength and the dot size~\cite{bockelmann94} (this matrix element is obtained when substituting Eqs.~\ref{eq:SO_zeeman}-~\ref{eq:SO_zeeman2} into
Eq.~\ref{eq:golden_rule}). In GaAs, the speed of sound $c_{ph}$ is
of the order of 4000 m/s, so the phonon wavelength is $h
c_{ph}/E_{ph}$, which gives $\sim 16$ nm for a 1 meV phonon. For
phonon wavelengths much shorter than the dot size (phonon energies
much larger than a few hundred $\mu$eV), the electron-phonon
interaction is averaged away (the matrix element vanishes). Also for phonon wavelengths much longer than the dot size, the electron-phonon coupling becomes inefficient, as it just shifts the entire dot potential uniformly up and down, and no longer couples different dot orbitals to each other (this is the regime where the often-used dipole approximation applies, where the matrix element $\langle n l \uparrow | e^{i\vec{q}\vec{r}} | n' l' \uparrow \rangle$ is taken to be $\propto q$). When the phonon wavelength is comparable to the dot
size, the phonons can most efficiently couple the orbitals, and
spin relaxation is fastest~\cite{woods02,golovach04,bulaev05}. This role of the phonon wavelength (convoluted with the other effects discussed in this section) has been clearly observed experimentally, see Fig.~\ref{fig:meunier_T1}. 

\bfig
\includegraphics[width=7cm]{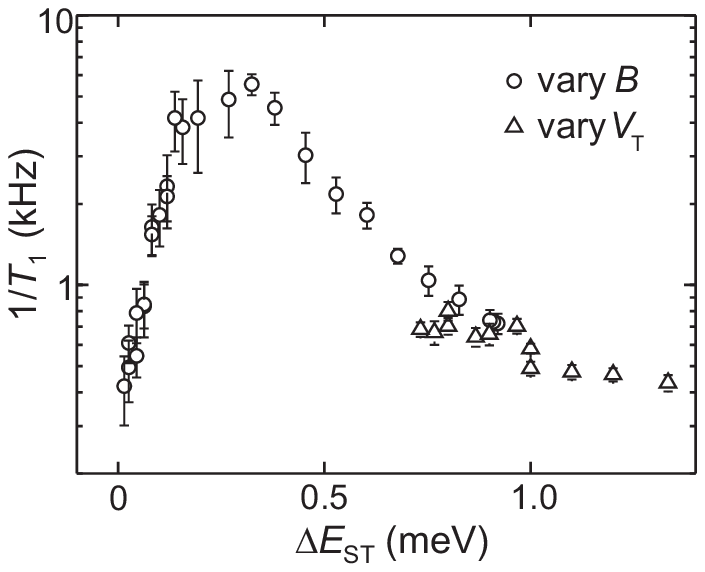}
\caption{The relaxation rate from two-electron triplet to singlet states, as a function of the singlet-triplet energy splitting (measured with TR-RO, see section~\ref{Subsection:TRRO}). The relaxation rate shows a maximum when the wavelength of the phonons with the right energy matches the size of the dot. The energy splitting was varied by a magnetic field with a component perpendicular to the 2DEG (circles) and via the gate voltages that control the dot potential landscape (triangles). We note that the relaxation rate goes down near the singlet-triplet crossing (because of the long phonon wavelength and vanishing phonon density of states), even though the spin-orbit admixing of singlet and triplet is maximum here. Data reproduced from~\textcite{meunier07}.} \label{fig:meunier_T1} \efig

(iv) A finite phonon occupation $N_{ph}$ leads to stimulated emission. This is accounted for by multiplying the relaxation rate by a factor $(1+N_{ph})$. $N_{ph}$ is given by the Bose-Einstein distribution, and can be approximated by $k_B T / E_Z$ when $k_B T \gg E_Z$.

(v) The last necessary ingredient for spin-orbit induced spin
relaxation is a finite Zeeman splitting. Without Zeeman splitting,
the various terms that are obtained when expanding Eq.~\ref{eq:golden_rule} using Eqs.~\ref{eq:SO_zeeman}-~\ref{eq:SO_zeeman2} cancel out~\cite{khaetskii01} (this is known as van Vleck cancellation; it is a consequence of Kramer's theorem). A similar cancellation occurs
for spin states of two or more electrons. We can understand the need for a magnetic field intuitively from the semiclassical discussion of the spin-orbit interaction in Section~\ref{sec:SO_bulk_2D}. A phonon produces an electric field that oscillates along a certain axis, and this electric field will cause an electron in a quantum dot to
oscillate along the same axis. In the absence of any other terms
in the Hamiltonian acting on the electron spin, the spin-orbit
induced rotation that takes place during half a cycle of the 
electric field oscillation will be reversed in the next half cycle, so no net spin rotation takes place. This is
directly connected to the fact that the spin-orbit interaction
obeys time-reversal symmetry. In contrast, in the presence of an
external magnetic field, the spin rotation (about the sum of the
external and spin-orbit induced magnetic field) during the first
half period doesn't commute with the spin rotation during the
second half period, so that a net spin rotation results. 
Theory predicts that this effect leads to a $B_0^2$ dependence of the relaxation rate $1/T_1$~\cite{khaetskii01,bulaev05,golovach07}. A clear $B_0$ dependence was indeed seen experimentally, see Fig.~\ref{fig:T1_vs_zeeman}.

When two phonons are involved, a net spin rotation can be obtained even at zero field. Here, the electron will in general not just oscillate back and forth along one line, but instead describe a closed trajectory in two dimensions. Since the spin rotations induced during the various legs along this trajectory generally don't commute, a net rotation results~\cite{sanjose06}.
Such two-phonon relaxation processes become relatively more imporant at very low magnetic fields, where single-phonon relaxation becomes very inefficient~\cite{khaetskii01,sanjose06} (Fig.~\ref{fig:T1_vs_zeeman}).

\bfig
\begin{center}
\includegraphics[width=4.5cm]{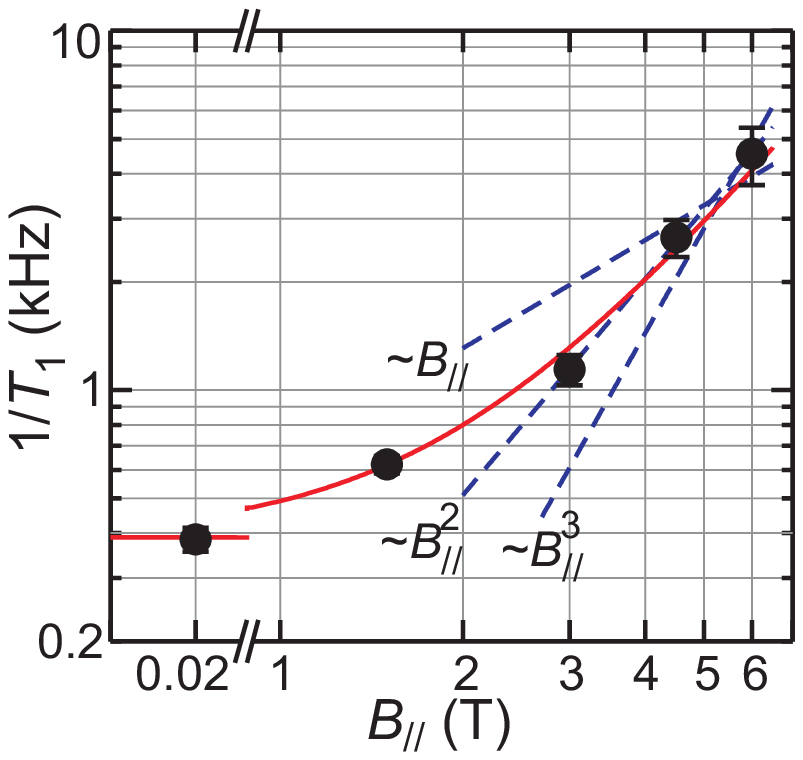}
\end{center}
\caption{(Color in online edition) Relaxation rate between the two-electron triplet and
singlet states in a single dot, as a function of in-plane magnetic
field $B_{//}$. The in-plane magnetic field doesn't couple to the orbitals and therefore hardly modifies the triplet-singlet energy splitting ($\Delta E_{ST} \sim 1$ meV, whereas the Zeeman splitting is only $\sim 20 \mu$eV/T in GaAs quantum dots). Nevertheless, and as expected, the experimentally measured rate $1/T_1$ at first markedly decreases as $B_{//}$ decreases, before saturating as $B_{//}$ approaches zero and two-phonon relaxation mechanisms set in. The solid line is a second-order polynomial fit to the data. For comparison, lines with linear, quadratic, and cubic $B_{//}$ dependences are shown. The data are extracted from Fig.~\ref{Fig:TRROresult}c and are reproduced from~\cite{HansonPRL2005}}
\label{fig:T1_vs_zeeman} 
\efig

Putting all these elements together, we can predict the $B_0$ dependence
of the relaxation rate $1/T_1$ beween Zeeman split sublevels of a
single electron as follows. First, the phonon density of states
increases with $\Delta E_{Z}^2$. Next, the electric field amplitude from
a single phonon scales as  $\sqrt{q}\propto\sqrt{\Delta E_Z}$ for
deformation potential phonons and as $1/\sqrt{q} \propto
1/\sqrt{\Delta E_Z}$ for piezo-electric phonons. Furthermore, for $B_0$
up to a few Tesla, $\Delta E_Z$ is well below the cross-over point
where the dot size matches the phonon energy (several 100
$\mu$eV), so we are in the long-wavelength limit, where the
matrix element $\langle n l \uparrow | e^{i\vec{q}\vec{r}} | n' l' \uparrow \rangle$ scales as $q\propto \Delta E_Z$.
Finally, due to the effect of the Zeeman splitting, the matrix
element in Eq.~\ref{eq:golden_rule} picks up another factor of
$\Delta E_Z$ (assuming only single-phonon processes are relevant). Altogether, and taking into account that the rate is proportional to the matrix element squared, $T_1^{-1}$ is predicted (at low temperature) to vary with $\Delta E_Z^5$ for coupling to piezo-electric phonons~\cite{khaetskii01}, and as $\Delta E_Z^7$ for coupling to deformation potential phonons. At high temperature, there is an extra factor of $\Delta E_Z^{-1}$.

We can similarly work out the $1/T_1$ dependence on the dot size, $l$, or equivalently, on the orbital level spacing, $\Delta E_{orb} \propto l^{-2}$ (in single dots, $\Delta E_{orb}$ can only be tuned over a small range, but in double dots, the splitting between bonding and antibonding orbitals can be modified over several orders of magnitudes~\cite{wang06}). The degree of admixing of spin and orbital states by ${\cal H}_{SO}$ contributes a factor $l^2$ to the rate via the numerator in Eqs.~\ref{eq:SO_zeeman}-~\ref{eq:SO_zeeman2} and another factor of $\sim l^4$ via $E_{nl} - E_{n'l'}$ (the dominant part in the denominator in Eqs.~\ref{eq:SO_zeeman}-~\ref{eq:SO_zeeman2}). Taking the long-wavelength limit as before, $|\langle n l \uparrow | e^{i\vec{q}\vec{r}} | n' l' \uparrow \rangle|^2$ contributes a factor $l^2$. We thus arrive at $1/T_1 \propto l^8 \propto \Delta E_{orb}^{-4}$.

In summary, the relaxation rate from spin down to spin up (for an electron in the ground state orbital of a quantum dot) scales as
\be
1/T_1 \propto \frac{\Delta E_Z^5 }{\Delta E_{orb}^4}
\ee
at temperatures low compared to $\Delta E_Z /k_B$, and as
\be
1/T_1 \propto \frac{\Delta E_Z^4 k_B T}{\Delta E_{orb}^4}
\ee
at temperatures much higher than $\Delta E_Z /k_B$.

Experimentally measured values for $T_1$ between Zeeman sublevels in a one-electron GaAs quantum dot are shown in Fig.~\ref{fig:amasha_elzerman}. The relaxation times range from 120 $\mu$s at 14 T to 170 ms at 1.75 T, about 7 orders of magnitude longer than the relaxation rate between dot orbitals~\cite{FujisawaNature2002}. The expected $B^5$ dependence of $T_1^{-1}$ is nicely observed over the applicable magnetic field range. A similar dependence was observed in optically measured quantum dots~\cite{kroutvar04}. In that system, the $1/T$ temperature dependence of $T_1$ was also verified~\cite{heiss05}. There are no systematic experimental studies yet of the dependence on dot size.

\bfig
\begin{center}
\includegraphics[width=8cm]{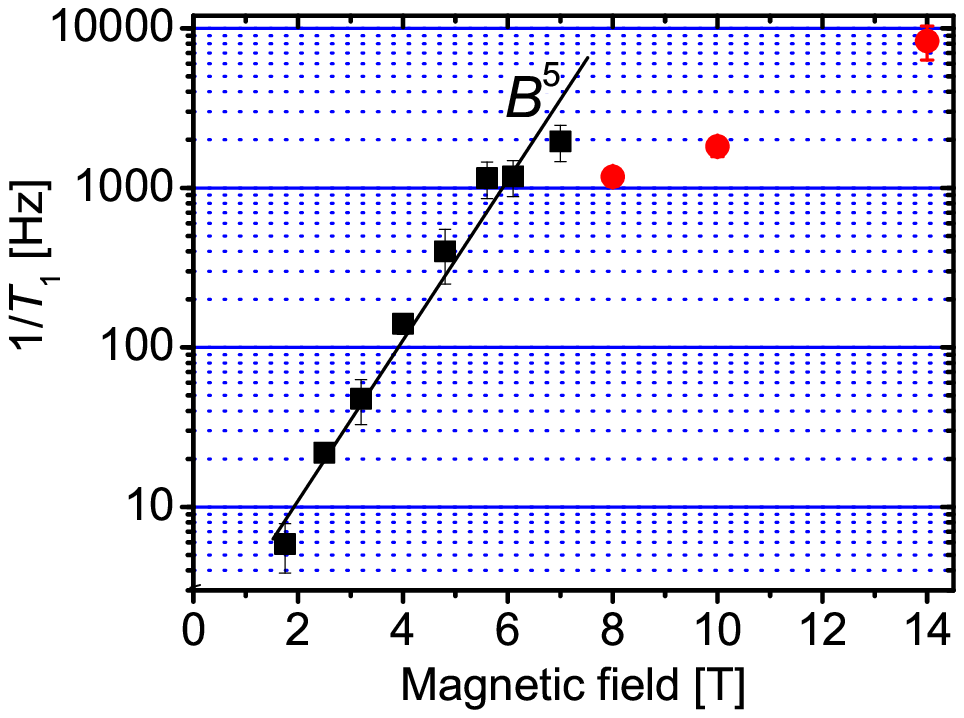}
\end{center}
\caption{(Color in online edition) Relaxation rate between the Zeeman split sublevels of the
ground state orbital in a quantum dot (measured with E-RO, see section~\ref{Subsection:ERO}). The square data points are taken from~\textcite{amasha06}; the round datapoints are reproduced from~\textcite{ElzermanNature2004}. The fact that the two datasets don't connect is explained by a possible difference in orbital spacing, crystal orientation etc. For
comparison, a solid line with a $B^5$ dependence is shown.}
\label{fig:amasha_elzerman} \efig

So far we have first considered the effect of $H_{SO}$ on the
eigenstates and then looked at transitions between these new
eigenstates, induced by $H_{e,ph}$ (Fig.~\ref{fig:relaxation}(a)).
We point out that it is also possible to calculate the matrix
element between the unperturbed spin states directly, for $H_{SO}$
and $H_{e,ph}$ together, for instance as
$$
\bra{nl\down}(H_{SO} + H_{e,ph}) \ket{nl\up}
$$
for Zeeman split states of a single orbital
(Fig.~\ref{fig:relaxation}(b)).

Finally, we remark that whereas at first sight phonons cannot flip
spins by themselves as there are no spin operators in the phonon
Hamiltonian, ${\cal H}_{e,ph}$, this is not strictly true.
Since phonons deform the crystal lattice, the $g$-tensor may be
modulated, and this can in fact lead to electron spin flips
directly (when phonons modulate only the magnitude of the
$g$-factor but not the anisotropy of the $g$-tensor, the electron
spin phase gets randomized without energy exchange with the
bath~\cite{semenov04}). Furthermore, the electron spin could flip
due to the direct relativistic coupling of the electron spin to
the electric field of the emitted phonon. However, both mechanisms
have been estimated to be much less efficient than the mechanism
via admixing of spin and orbitals by the spin-orbit
interaction~\cite{khaetskii00,khaetskii01}.


\subsubsection{Phase randomization due to the spin-orbit interaction}

We have seen that the phonon bath can induce transitions between
different spin-orbit admixed spin states, and absorb the spin flip
energy. Such energy relaxation processes (described by a time constant $T_1$) unavoidably also lead to the loss of quantum coherence
(described by a time constant $T_2$). In fact, by definition $T_2
\le 2 T_1$.

Remarkably, in leading order in the spin-orbit interaction, there
is no pure phase randomization of the electron spin, such that in
fact $T_2 = 2 T_1$~\cite{golovach04}. For a magnetic field
perpendicular to the plane of the 2DEG, this can be understood
from the form of the spin-orbit Hamiltonian. Both the Dresselhaus
contribution, Eq.~\ref{eq:H_D}, and the Rashba contribution,
Eq.~\ref{eq:H_R}, only contain $\sigma_x$ and $\sigma_y$ terms.
With $B$ along $\hat{z}$, these terms lead to spin flips but not
to pure phase randomization. However, this intuitive argument doesn't capture the full story:
for $B_0$ along $\hat{x}$, one
would expect the $\sigma_x$ term to contribute to pure phase
randomization, but surprisingly, in leading order in the spin-orbit
interaction, there is still no pure randomization even with an
in-plane magnetic field~\cite{golovach04}.


\subsection{Hyperfine interaction}
\label{Section:Hyperfine}
\subsubsection{Origin}

The spin of an electron in an atom can interact with the spin of
``its" atomic nucleus through the hyperfine coupling. An electron
spin in a quantum dot, in contrast, may interact with many nuclear
spins in the host material (Fig.~\ref{fig:hyperfine}). The Hamiltonian for the Fermi contact hyperfine interaction is then given by 
\be 
{\cal H}_{HF} = \sum_k^N A_k \vec{I_k} \vec{S} \;, 
\label{eq:H_HF} 
\ee
where $\vec{I_k}$ and $\vec{S}$ are the spin operator for nucleus
$k$ and the electron spin respectively~\cite{Abragam,Slichter,Abragam-Bleaney,BookOptical}. Since the electron wavefunction is inhomogeneous, the coupling strength, $A_k$, between each nucleus $k$ and the electron spin varies, as it is proportional to the overlap squared between the nucleus and the electron wavefunction.  

This asymmetric situation combined with fast electron spin dynamics and slow nuclear spin dynamics, gives rise to a subtle and complex many-body quantum mechanical behavior, whereby the nuclear spins affect the electron spin time evolution, and the electron spin in turn acts back on the dynamics of each of the nuclei.

\bfig
\begin{center}
\includegraphics[width=7cm]{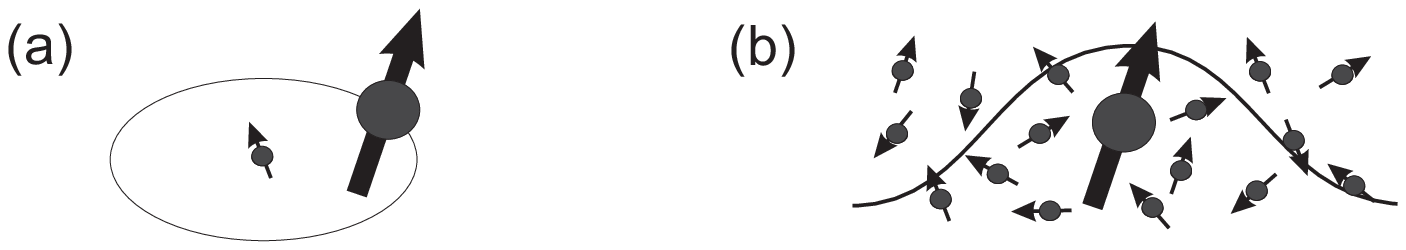}
\end{center}
\caption{One electron spin interacts with (a) a single nuclear
spin in an atom, versus (b) many nuclear spins in a semiconductor
quantum dot.} \label{fig:hyperfine} \efig

Since both the nuclear spins and the localized electron spin are quantum objects, the hyperfine coupling could in principle create entanglement between them (if both the electron spin and the nuclear spins had a sufficiently pure initial state, see~\textcite{braunstein99}. For the electron spin, this interaction with
uncontrolled degrees of freedom in the environment leads to decoherence~\cite{merkulov02,khaetskii02,khaetskii03,coish04}. This implies that an electron spin starting off in a pure state will
evolve to a statistical mixture of several states, i.e. to one of
several states, each with some probability~\cite{nielsen00}.

An alternative and very useful description of the effect of the
nuclei on the electron spin, is to treat the ensemble of nuclear
spins as an apparent magnetic field, $B_N$. This nuclear
field, also known as the Overhauser field, acts on the electron
spin much like an external magnetic field: 
\be 
\left( \sum_k^N A_k \vec{I_k}\right) \vec{S}
= g \mu_B \vec{B_N} \vec{S} \;.
\label{eq:H_HF_class} 
\ee 
When this nuclear field assumes a random, unknown value, the electron spin will subsequently evolve in a random way and thus end up in a statistical mixture of states as well, just like in the quantum mechanical description.

The semiclassical description of the nuclear spins yields an intuitive
picture of the electron-nuclear dynamics and is sufficient to explain all the experimental observations discussed in this review. However, we note that the full quantum description is required to analyze the correlations between the microscopic nuclear spin states and the single electron spin state, as e.g. in a study of the entanglement between electron and nuclear spins.

The magnitude of the nuclear field, $B_N = \sum_k^N A_k \vec{I_k} / (g \mu_B)$ is maximum when all nuclear spins are fully polarized. In GaAs, $B_{N,max}$ is about $5$ T~\cite{paget77}. For any given host material, this value is \emph{independent} of the number of nuclei, $N$, that the electron overlaps with -- for larger numbers of nuclei, the contribution from each nuclear spin to $B_N$ is smaller (the typical value for $A_k$ is proportional to $1/N$).

This is distinctly different in the case of (nearly) unpolarized nuclear spins, for instance nuclear spins in thermodynamic equilibrium under typical experimental conditions. First there is a small \emph{average} nuclear polarization, oriented along the external magnetic field and with an amplitude given by the Boltzman distribution (see Appendix~\ref{App:nuclearfield}). In addition, there is a \emph{statistical fluctuation} about the average, analogous to the case of $N$ coin tosses. For an electron spin interacting with $N$ nuclear spins $1/2$, the root mean square value of the statistical fluctuation will be $B_{N,max} /\sqrt{N}$ T \cite{merkulov02,khaetskii02}. This quantity has recently been measured in various semiconductor quantum dots, both optically~\cite{braun05,dutt05} and electrically~\cite{johnson05,koppens05}, giving values in the range
of a few mT, as expected since $N\approx 10^6$ in these dots. Similar values were obtained earlier for electrons bound to shallow donors in GaAs~\cite{dzhioev02}.

A few comments on the importance of the host material are in order. 
First, the value of $A_k$ is typically smaller for lighter nuclei. Second, for particles in $p$-like orbitals, such as holes in GaAs, the wavefunction has almost no overlap with the nuclei (only $s$ orbitals have a finite amplitude at the nucleus), so the Fermi contact hyperfine coupling constant, $A_k$, will be very small. Third, if a fraction $x$ of the nuclei in the host material has zero nuclear spin, $B_{N,max}$ is scaled down with a factor $1-x$. The number of nuclei contributing to the statistical fluctuations in the nuclear field also scales down by $1-x$. As a result, the r.m.s. value of the nuclear field scales with $\sqrt{1-x}$. While $x=0$ in GaAs, a fraction $x\approx0.95$ of the nuclei ($^{28}$Si) is non-magnetic in natural silicon, and $x\approx0.99$ in carbon ($^{12}$C). Furthermore, both for silicon and carbon, purification to nearly 100$\%$ zero-spin isotopes is possible, so really small nuclear fields can be obtained

Finally, we point out that since the r.m.s. value of the statistically fluctuating hyperfine field scales with $1/\sqrt{N}$, it is much stronger for electrons localized in dots or bound to impurities than for electrons with extended wavefunctions, for instance in 2DEGs, where the electron wavefunction overlaps with a very large number of nuclei. This is in sharp contrast to the effect of the spin-orbit interaction, which becomes suppressed when the electron is confined to dimensions shorter than the spin-orbit length, such as in small quantum dots.


\subsubsection{Effect of the Overhauser field on the electron spin time evolution}
\label{sec:effect_BN}

The electron spin will precess about the vector of the total
magnetic field it experiences, here the vector sum of the
externally applied magnetic field $\vec{B_0}$ and the nuclear
field $\vec{B_N}$. The \emph{longitudinal} component of
$\vec{B_N}$, i.e. the component oriented parallel or opposed to
$\vec{B_0}$, directly changes the precession frequency by $g \mu_B
B_N$, irrespective of the strength $B_0$
(Fig.~\ref{fig:B0_plus_Bn}a). Throughout this section, we shall
call the longitudinal component $B_N^z$. For $B_N^z = 1$ mT, the
precession rate is increased by about 6 MHz (taking $g=-0.44$), and the electron spin picks up an extra phase of $180^\circ$ in just 83 ns. The effect of the \emph{transverse} components of the nuclear field,
$B_N^{x,y}$, strongly depends on the strength of $B_0$. For $B_0
\ll B_N^{x,y}$, the electron spin will precess about an axis
very close to $B_N^{x,y}$ (Fig.~\ref{fig:B0_plus_Bn}b). For $B_0
\gg B_N^{x,y}$, in contrast, the transverse components of the
nuclear field only have a small effect: a change in the electron
spin precession rate by $\approx g \mu_B B_N^2/(2 B_0)$, and a
tilt of the rotation axis by arctan($B_N/B_0)$
(Fig.~\ref{fig:B0_plus_Bn}c). Taking $B_0 = 1$ T and $B_N^x = 1$
mT, the precession frequency is shifted by just 3 kHz, causing an
extra phase of $180^\circ$ only after 166 ms; the rotation axis is
then tilted by $\approx 0.06^\circ$. For external magnetic fields
above say 100 mT, we are therefore mainly concerned with the
longitudinal nuclear field.

\bfig
\begin{center}
\includegraphics[width=5cm]{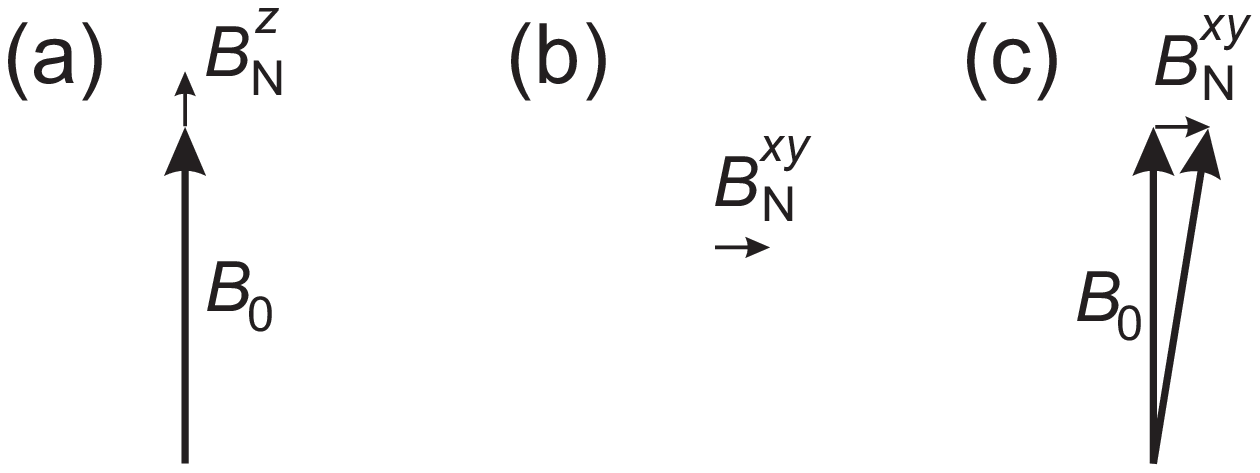}
\end{center}
\caption{Longitudinal magnetic field fluctuations, $B_N^z$, add
directly to the external field, $B_0$, whereas transverse
fluctuations, $B_N^{x,y}$ , change the total field only in second
order when $B_0 \gg B_N^{x,y}$.} 
\label{fig:B0_plus_Bn} 
\efig

If the nuclear field, $B_N$, were fixed and precisely
known, it would affect the electron spin dynamics in a systematic
and known way. In this case, there would be no contribution to
decoherence. However, the orientation and magnitude of the nuclear
field change over time. First, the hyperfine field or Overhauser
field, $B_N$, will change if the local nuclear polarization,
$\sum_k I_k$, changes. This can occur for instance through dynamic 
nuclear polarization. Second, $B_N$ can also change while the net 
nuclear polarization remains
constant. This happens when two nuclei with different $A_k$
flip-flop with each other, such that $\sum_k A_k \vec{I_k}$
changes.

\emph{At any given time, the nuclear field thus assumes a random
and unknown value and orientation, and this randomness in the
nuclear field directly leads to a randomness in the electron spin
time evolution}. During free evolution, the electron spin will
thus pick up a random phase, depending on the value of the nuclear
field, i.e. the single-spin coherence decays. The shape of the decay (exponential, power-law, etc.) is determined by the distribution of nuclear field values. For a longitudinal nuclear field, $B_N^z$ that is randomly
drawn from a Gaussian distribution of nuclear fields with standard
deviation $\sqrt{\langle(B_N^z)^2\rangle}$ (e.g. when every
nuclear spin had equal probabilities for being up or down), the decay would be Gaussian as well, i.e. of the form $\mbox{exp}(-t^2/(T_2^*)^2)$, where~\cite{merkulov02} 
\be
T_2^* = \frac{\hbar}{g \mu_B \sqrt{2\langle(B_N^z)^2\rangle}} \;.
\ee 
For $\sqrt{\langle(B_N^z)^2\rangle} = 1$ mT, $T_2^*$ would be as short as $30$ ns. 

The timescale $T_2^*$ can be measured as the decay time of the electron spin signal during free evolution, averaged over the nuclear field distribution(Fig.~\ref{fig:nuclear_dephasing}a). The free evolution of the spin can be measured by tipping the spin into the $x-y$ plane, subsequently allowing the spin to freely evolve about $\vec{B_0}$ (assumed to be along $\hat{z}$), and recording the magnetization in this plane as a function of the free evolution time interval (in NMR, this is known as the free induction decay or FID).  If the spin can only be measured in the $\pm\hat{z}$ basis, a so-called Ramsey experiment should be performed instead. It starts off like an FID but the magnetization is rotated back to the $\hat{z}$ axis after the free evolution time interval, so it can be measured along $\hat{z}$. Averaging each data point in an FID or Ramsey experiment over a sufficiently long time, is equivalent to averaging over a large number of uncorrelated nuclear field values (assuming the system is ergodic). Such experiments have recently been performed (see Section~\ref{Section:Coherent}), and gave the expected short timescales for $T_2^*$.

\bfig
\begin{center}
\includegraphics[width=8.5cm]{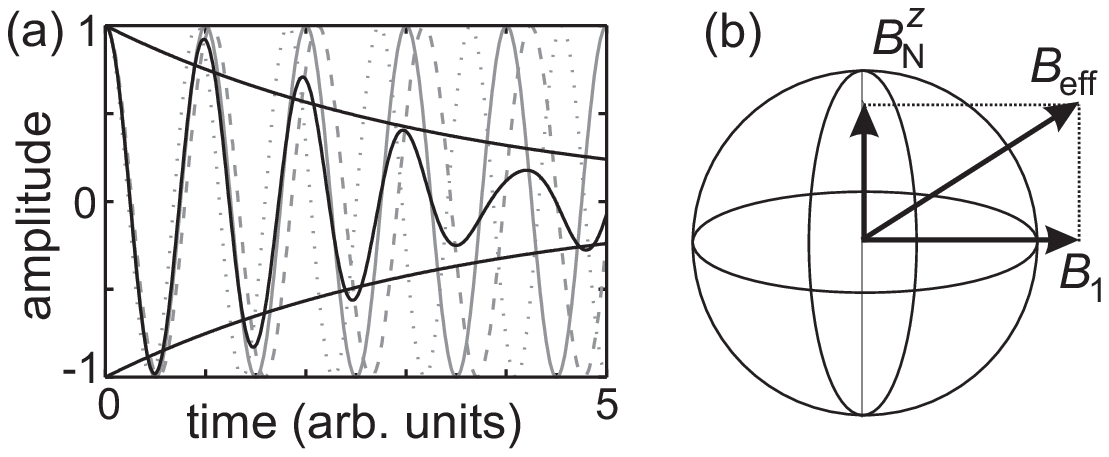}
\end{center}
\caption{(a) Amplitude of the $x$-component of the electron spin
as a function of time, under free precession about $(B_0 + B_N^z)$
for three different values of $B_N^z$ (dotted and dashed lines).
Also shown is the average of the three oscillations, which is seen
to rapidly decay (solid line). If an average was taken over many
more values of $B_N^z$ taken from a Lorentzian distribution in the same range, the envelope of the
oscillation would decay with a single exponent (solid lines). A Gaussian distribution would yield a Gaussian decay (see text). (b)
Representation of the rotating magnetic field, $\vec{B_1}$, and
the longitudinal nuclear field, $\vec{B_N^z}$, in a reference
frame rotating about $\hat{z}$ at the same rate as $B_1$. The
electron spin will precess about the vector sum of these two
fields, rather than about the axis defined by $\vec{B_1}$.}
\label{fig:nuclear_dephasing} \efig

Also in the case of driven evolution, the nuclear field will
affect the time evolution. In a spin resonance experiment, for
instance, a rotating magnetic field $B_1$ is applied with
frequency $g \mu_B B_0/h$ (on-resonance with the Zeeman
splitting), and perpendicular to $\vec{B_0}$ (see also
Section~\ref{Sec:esr}). In the usual rotating reference frame,
this corresponds to a rotation about the $B_1$-axis. However, a
longitudinal nuclear field will shift the electron spin resonance
frequency, so that $B_1$ is no longer on-resonance with the
electron spin precession. In the rotating frame, the spin will
then rotate about the vector sum of $B_1$ and $B_N^z$
(Fig.~\ref{fig:nuclear_dephasing}b), which may be a rather
different rotation than intended. In fact, the nuclear field has
been the main limitation on the fidelity of spin rotations in
recent electron spin resonance experiments in a quantum dot (see
Section~\ref{Sec:esr}).

We have so far focussed on the effect of the nuclear field (mainly $B_N^z$) on the electron spin phase. We now briefly turn to electron spin flips caused by the nuclear field (mainly by $B_N^{x,y}$). The semi-classical picture says that for $B_0 \ll B_N^{x,y}$, the electron spin will rotate about $B_N^{x,y}$, i.e. it is changed from spin-up to spin-down and back, while for $B_0 \gg B_N^{x,y}$, this hardly occurs (see Fig.~\ref{fig:B0_plus_Bn}). We arrive at a similar conclusion from the quantum mechanically picture: the hyperfine Hamiltonian, Eq.\ref{eq:H_HF}, permits direct electron-nuclear flip-flops only when the two relevant electron spin states are very close in energy (since a nuclear spin flip can absorb only a small amount of energy). This effect has been observed in recent experiments on two-electron singlet and triplet states in a double quantum dot\cite{johnson05} (Section~\ref{Section:STmixing}).

There still is another contribution from $B_N^{x,y}$ to spin flips (i.e. to $T_1$). Since the nuclear field strength and orientation depend on the collection of nuclear spins that the electron wavefunction overlaps with, $B_N$ depends on the orbital the electron occupies. As a result, like the spin-orbit interaction (see section~\ref{Sec:spin-orbit}), the hyperfine interaction also leads to admixing of spin and orbital states. Here too, phonons can induce transitions between the perturbed spin states, and absorb the spin-flip energy~\cite{erlingsson01,erlingsson02,erlingsson04,abalmassov04}.
Whereas the transition amplitude due to the spin-orbit interaction
vanishes in lowest order at $B_0=0$ (see discussion in section~\ref{Sec:RelaxationviaPhononBath}), this is not the case for hyperfine mediated transitions, so the hyperfine mechanism will
be relatively more important at low magnetic fields.


\subsubsection{Mechanisms and timescales of nuclear field fluctuations}
\label{sec:BN_fluct}

We have seen that the nuclear field only leads to a loss of spin coherence
because it is random and unknown -- if $\vec{B}_N$ were fixed in
time, we could simply determine its value and the uncertainty
would be removed. We here discuss on what timescale the
nuclear field actually fluctuates. This timescale is denoted by $t_{nuc}$.

The study of nuclear dynamics due to internuclear and electron-nuclear interactions has a long and rich history~\cite{Abragam,Slichter,Abragam-Bleaney,BookOptical}. When applied to quantum dots, theory predicts that the two most important
mechanisms responsible for fluctuations in the nuclear field are
the internuclear magnetic dipole-dipole interaction~\cite{sousa03a,sousa03b,witzel05,yao05} and the
electron-nuclear hyperfine
interaction~\cite{khaetskii02,coish04,shenvi05}. 

The Hamiltonian describing magnetic dipole-dipole interactions between
neighbouring nuclei is of the form
\begin{equation}
{\cal H}_{DD} = \sum_{i<j} \frac{\mu_0 g_i g_j \mu_N^2 \hbar}{4\pi
|\vec{r_{ij}}|^3} \left[\vec{I_i} \cdot \vec{I_j} -
\frac{3}{|\vec{r_{ij}}|^2} (\vec{I_i} \cdot \vec{r_{ij}})
(\vec{I_j} \cdot \vec{r_{ij}}) \right] \;, \label{eq:H_dipdip}
\end{equation}
where $\mu_0$ is the permeability of free space, $g_i$ is the
g-factor of nucleus $i$, $\mu_N$ the nuclear magneton, and $\vec{r_{ij}}$ is the vector connecting the two nuclei. 
The strength of the effective magnetic dipole-dipole interaction between neighbouring nuclei in GaAs is about (100 $\mu$s)$^{-1}$\cite{shulman58}. In strong magnetic fields, we can discard the non-secular part of this Hamiltonian, and retain
\bea
{\cal H}_{DD,ij}^{sec} &\propto& 
\vec{I_i} \cdot \vec{I_j} - 3 I_i^z I_j^z \nonumber \\
&=& 
I_i^x I_j^x + I_i^y I_j^y - 2 I_i^z I_j^z \nonumber \\
&=& 
(I_i^+ I_j^- + I_i^- I_j^+ - 4 I_i^z I_j^z)/2 
\label{eq:H_dipdip_sec}
\eea
for the coupling term between nuclear spins $i$ and $j$ of the same species (for coupling between spins of different isotopes, only the $I_i^z I_j^z$ term survives at high field). Here $I^\pm$ are the nuclear spin raising and lowering operators.
The $I_i^z I_j^z$ terms in Eq.~\ref{eq:H_dipdip_sec} are responsible for changing $B_N^x$ and $B_N^y$. This may occur on the $100 \mu$s timescale. The flip-flop terms, $I_i^x I_j^x + I_i^y I_j^y$, affect $B_N^z$, but the flip-flop rate between nuclei $i$ and $i+1$ may be suppressed, namely when $|A_i - A_{i+1}|$ is greater than the internuclear coupling strength~\cite{deng06} (as this causes an energy mismatch). Thus, when we consider the dipolar interaction only, $B_N^{x,y}$ evolves on a 100 $\mu$s timescale, but $B_N^z$ may evolve more slowly.

We now turn to the hyperfine interaction. So far we only considered its effect on the localized electron spin, but naturally the interaction (Eq.\ ~\ref{eq:H_HF}) works both ways and the nuclei evolve about the electron spin, just like the electron spin evolves about the nuclear field. The apparent field experienced by the nuclei is called the Knight shift, and it has a strength $A_k \approx (10
\mu$s$)^{-1}$~\cite{merkulov02,khaetskii02,coish04} (the $N$
nuclei with which the electron wavefunction overlaps ``share" the
total coupling strength $A$). When we look at the hyperfine interaction Hamiltonian, Eq.~\ref{eq:H_HF}, we see that similar to the internuclear
dipole-dipole interaction, it contains $I_i^z S^z$ terms as well as flip-flop terms:
\bea
{\cal H}_{HF} &=& \sum_k^N A_k \left(I_k^x S^x + I_k^y S^y + I_k^z S^z\right) \nonumber \\
&=& \sum_k^N A_k \left(I_k^+ S^+ + I_k^- S^- + 2I_k^z S^z\right)/2 
\label{eq:H_HF_detailed}
\eea
where $S^\pm$ are the electron spin raising and lowering operators.
The transverse component of the nuclear field, $B_N^{x,y}$, will
evolve due to the $I_i^z S^z$ terms, on a $10 \mu$s timescale.
$B_N^z$ will change on the same timescale only near $B_0=0$, due to
the electron-nuclear flip-flop components in Eq.~\ref{eq:H_HF_detailed}.
At finite $B_0$, the energy mismatch between the electron and
nuclear Zeeman energies suppresses electron-nuclear flip-flops, so here
$B_N^z$ cannot change by direct electron-nuclear flip-flops. 

The hyperfine interaction can also affect $B_N^z$ indirectly. Two \emph{virtual} electron-nuclear flip-flops (between one nucleus and the electron and between the electron and another nucleus) can together lead to a nuclear-nuclear flip-flop~\cite{yao05,shenvi05}. Such a flip-flop process between two nuclei $i$ and $j$ modifies $B_N^z$ whenever $A_i \neq A_j$. This virtual electron-nuclear flip-flop process continues to be effective up to much higher $B_0$ than real electron-nuclear flip-flops. Eventually it is suppressed at high $B_0$ as well.

Altogether the dipole-dipole and hyperfine interactions are expected to lead to moderate timescales (10-100 $\mu$s) for $B_N^{x,y}$ fluctuations. At low $B_0$, the timescale for $B_N^z$ fluctuations is similar, but at high $B_0$, $B_N^z$ fluctuations are very slow. Here $t_{nuc}$ is certainly longer than $10-100 \mu$s and perhaps longer than a second. This still needs to be confirmed experimentally, but an indication that $t_{nuc}$ may indeed be very long is that the decay (due to spin diffusion) of nuclear polarization built up locally at a quantum dot or impurity, occurs on a timescale of seconds to minutes~\cite{koppens05,paget82,huettel04} (see
Section~\ref{Section:STmixing}).


\subsubsection{Electron spin decoherence in a fluctuating nuclear field}
\label{sec:T2_BN}

In section~\ref{sec:effect_BN}, we saw that we lose our knowledge of the electron spin phase after a time $T_2^*$, in case the nuclear field orientation and strength are unknown. Now suppose that we do know the orientation and strength of the nuclear field exactly at time $t=0$ , but that the nuclear spin bath subsequently evolves in a random fashion on a timescale $t_{nuc}$, as described in the previous subsection. On what timescale, $T_2$, will the phase of the electron spin then be randomized?

It may come as a surprise at first that $T_2$ is not simply the same or even of the same order as $t_{nuc}$. The reason is that $T_2$ depends not only on the \emph{timescale} of the nuclear field fluctuations ($t_{nuc}$), but also on the \emph{amplitude} and \emph{stochastics} of the fluctuations. The typical amplitude is given by the width of the nuclear field distribution, which can be expressed in terms of $1/T_2^*$. Examples of different stochastic models include Gaussian noise and Lorentzian noise, which lead to distinct decoherence characteristics~\cite{klauder62,sousa06}.  The actual value of $T_2$ is difficult to calculate exactly, but can be estimated in various regimes to be $1 - 100 \mu$s. 

The timescale $T_2$ is also hard to obtain experimentally. In principle $T_2$ could be determined by recording an FID or Ramsey decay, whereby $B_N^z$ is reset to the same initial value for every datapoint. This may require measuring $B_N^z$ accurately and quickly, i.e. better than the initial uncertainty in $B_N^z$ and in a time much shorter than $t_{nuc}$~\cite{klauser06,stepanenko06,giedke06}. Alternatively, $T_2$ could be obtained by recording all the datapoints needed to construct an FID or Ramsey experiment within a time short compared to $t_{nuc}$.

Experimentally, it may be much easier to obtain a spin-echo decay time, $T_{echo}$, well-known from NMR~\cite{vandersypen04c,freeman97}. In its simplest form, the Hahn-echo, the random time evolution that takes place during a certain time interval $\tau$ is reversed during a second time interval of the same duration, by applying a so-called echo pulse ($180^\circ$ rotation) in between the two time intervals. Importantly, this unwinding of random dephasing only takes place to the extent that the random field causing it is constant for the duration of the entire echo sequence. Thus, the slow time evolution of the nuclear field implies that the echo will not be complete. We call the timescale of the remaining loss of phase coherence $T_{echo}$.

Like $T_2$, $T_{echo}$ is much longer than $T_2^*$ but also much shorter than $t_{nuc}$. For example, if the nuclear field fluctuations had Gaussian noise characteristics, the electron spin coherence in a Hahn echo experiment would decay as $\exp[- t^3 / (t_{nuc} {T_2^*}^2)]$\cite{hahn56}. Taking $T_2^* = 10$ ns and $t_{nuc} = 10$ s, we would obtain a $T_2$ of $10 \mu$s, much faster than $t_{nuc}$ itself. The nuclear field fluctuations may not be characterized exactly by Gaussian noise, but nevertheless, predictions for $T_2$ still range from $1 \mu$s to $100 \mu$s, with contributions from the internuclear dipole-dipole interaction\cite{sousa03c,witzel05,yao05}, the electron-nuclear hyperfine interaction~\cite{khaetskii02,coish04}, and indirect nuclear-nuclear interactions, mediated by the hyperfine coupling~\cite{yao05,shenvi05}. Also, contrary to the usual case, the echo decay is not well described by a single exponential. The predicted form for the echo decay depends on the magnetic field strength, and on what terms in the Hamiltonian are then important, but can be very complex~\cite{coish04,yao05,sousa06}.

The usefulness of the echo technique for effectively obtaining an extended coherence time has been demonstrated experimentally with two electron spins in a double quantum dot, whereby a lower bound on $T_{echo}$ of $1 \mu$s was obtained at 100 mT~\cite{petta05} (Section~\ref{Section:Coherent:TwoSpin}). Similar echo-like decay times were observed in optical measurements on an ensemble of quantum dots that each contain a single electron spin~\cite{greilich06b}. 

We note that sometimes a distinction is made between ``dephasing", referring to a loss of phase coherence that can be reversed with echo techniques, and ``decoherence", referring to a loss of phase coherence that cannot be reversed. While this distinction is useful in practice, we note that it is also somewhat artificial, in the sense that \emph{any} time evolution can in principle be reversed by a sufficiently rapid sequence of multiple generalized echo pulses~\cite{augustine97,viola98a,viola98b}.

Finally, we point out that it may be possible to extend $t_{nuc}$, i.e. to (almost) freeze the nuclear field fluctuations. One possibility to do this is to fully polarize the nuclear spins: if all nuclear spins point the same way, nuclear-nuclear flip-flop processes can no longer take place and also electron-nuclear flip-flops can only have a very small effect~\cite{khaetskii02,khaetskii03,schliemann02}. For this approach to be effective, the nuclear spin polarization must be really very close to $100\%$~\cite{schliemann02}; just $90\%$ polarization hardly helps. At present the highest nuclear spin polarizations reached in quantum dots are 60 $\%$, via optical pumping~\cite{bracker05a}. Certainly other mechanisms for freezing the nuclear spin fluctuations could be considered, but no such effect has been demonstrated to date.\\

\subsection{Summary of mechanisms and timescales}

Our present understanding of the mechanisms and timescales for
energy relaxation and phase randomization of electron spins in
few-electron quantum dots is summarized as follows (as before, most numbers are specific to GaAs dots, but the underlying physics is similar in other dot systems).

Energy relaxation is dominated by direct electron-nuclear
flip-flops near zero field (or whenever the relevant electron spin
states are degenerate). In this case, $T_1$ is as low as 10-100
ns. As $B_0$ increases, electron-nuclear flip-flops become
suppressed, and energy must be dissipated in the phonon bath. Spin-phonon coupling is inefficient, and occurs mostly indirectly, either mediated by the hyperfine interaction or by spin-orbit interaction. As a result $T_1$ rapidly increases with $B_0$, and at 1.75 T, $T_1$ has been measured to be 170 ms. As $B_0$ further increases, the phonon density of states
increases and the phonons couple more efficiently to the dot
orbitals (the phonon wavelength gets closer to the dot size), so
at some point relaxation becomes faster again and $T_1$ decreases with field. At 14 T, a $120 \mu$s $T_1$ has been observed. At still higher fields, the phonon wavelength would become shorter than the dot size, and $T_1$ is once more expected to go up with field.

Phase coherence is lost on much shorter timescales. A rapid dephasing of the electron spin results from the
uncertainty in the nuclear field, $T_2^*\approx 10$ ns, irrespective of
$B_0$. If the uncertainty in the nuclear field is removed or if
the resulting unknown time evolution is unwound, we recover $T_2$ or $T_{echo}$ respectively, which are much longer. Phase randomization of the electron spin then results from the (slow) fluctuations in the nuclear field, which occur on a timescale of 100 $\mu$s to perhaps seconds, and should lead to a $T_2$ or $T_{echo}$ of $1-100 \mu$s. Indeed, a lower bound on $T_{echo}$ of $1 \mu$s was experimentally observed at 100 mT. If the effect of the nuclear field on the electron spin coherence could be suppressed, the spin-orbit interaction would limit $T_2$, to a value of $2T_1$ (to first order in the spin-orbit interaction), which is as we have seen a very long time.


\section{Spin states in double quantum dots}
\label{Section:DoubleDotSpin} In this section, we discuss the spin
physics of double quantum dots. We start by describing the
properties of ``spinless'' electrons. Then, we show how the spin
selection rules can lead to a blockade in electron transport
through the double dot. Finally, we describe how this spin
blockade is influenced by the hyperfine interaction with the
nuclear spins, and discuss the resulting dynamics.

\subsection{Electronic properties of electrons in double dots}
\label{DDspinless} We first ignore the spin of the electrons and
describe the basic electronic properties of double quantum dots.
The properties of ``spinless'' electrons in double dots are
treated in detail by Van der Wiel \textit{et
al.}~\cite{WielRMP2003}. Here, we give all the theory relevant for
electron spins in double dots without going into the details of
the derivations.

\subsubsection{Charge stability diagram}
\begin{figure}[htb]
\includegraphics[width=3.4in, clip=true]{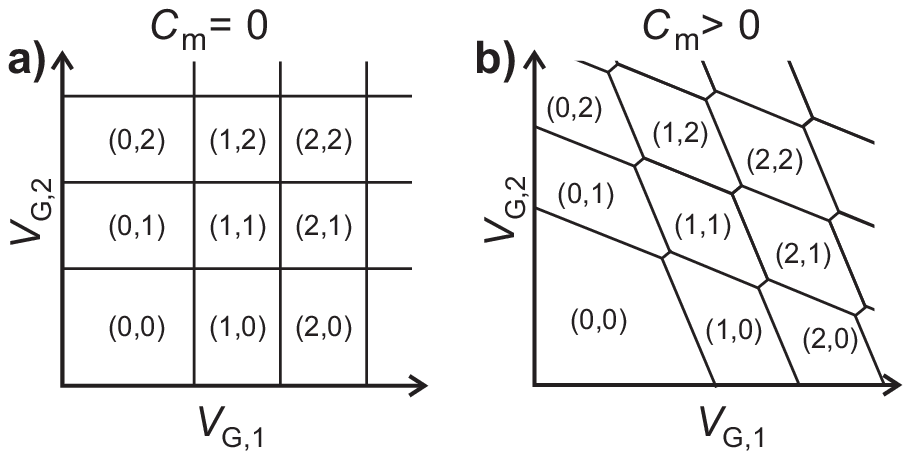}
\caption{Charge stability diagrams for (a) uncoupled and (b)
coupled double dots, depicting the equilibrium electron numbers
($N_1,N_2$) in dot 1 and 2 respectively. The lines indicate the
gate voltage values at which the electron number changes. In (b),
a finite cross-capacitance between gate 1 (2) and dot 2 (1) is
taken into account.} \label{Fig:DDChargDiagr}
\end{figure}

Consider two quantum dots, labelled 1 and 2, whose electrochemical
potentials are controlled independently by the gate voltages
$V_{G,1}$ and $V_{G,2}$, respectively.
Figure~\ref{Fig:DDChargDiagr}a shows the equilibrium electron
numbers ($N_1$,$N_2$) of the quantum dots as a function of
$V_{G,1}$ and $V_{G,2}$, for the case that the dots are completely
uncoupled. Such a plot is called a charge stability diagram. The
lines indicate the values of the gate voltages at which the number
of electrons in the ground state changes. Note that the lines are
exactly horizontal and vertical, since the electrochemical
potential in either dot is independent of the charge on the other
dot, and each gate voltage only affects one of the dots.

When the dots are capacitively coupled, addition of an electron on
one dot changes the electrostatic energy of the other dot. Also,
the gate voltage $V_{G,1}$ ($V_{G,2}$) generally has a direct
capacitive coupling to quantum dot 2 (1). The resulting charge
stability diagram is sketched in Fig.~\ref{Fig:DDChargDiagr}b.
Each crosspoint is split into two so-called \textit{triple
points}. The triple points together form a hexagonal or
``honeycomb'' lattice. At a triple point, three different charge
states are energetically degenerate. The distance between the
triple points is set by the capacitance between the dots (the
interdot capacitance) $C_{m}$. At low source-drain bias voltage,
electron transport through the double dot is possible only at
these triple points. In contrast, a charge sensing measurement
will detect any change in the electron configuration and therefore
map out all the transitions, including those where an electron
moves between the dots (e.g. from (0,1) to (1,0)).

\begin{figure}[hbt]
\includegraphics[width=3.4in, clip=true]{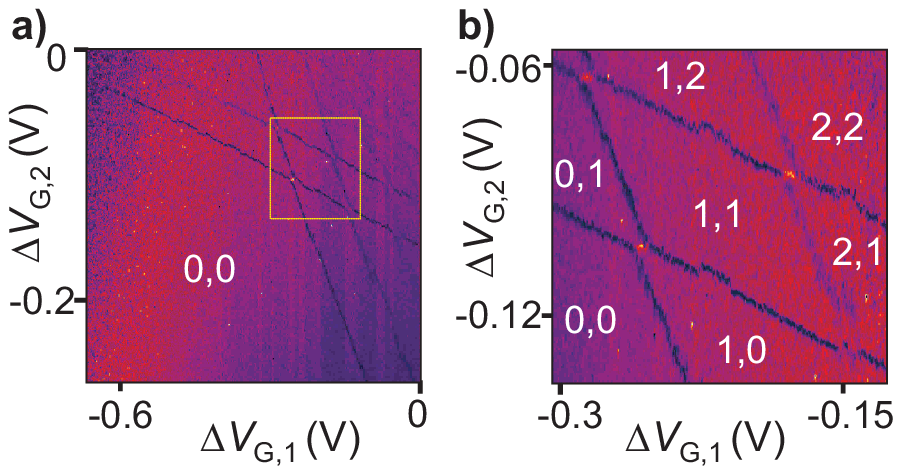}
\caption{(Color in online edition) Charge sensing data on a double dot in the few-electron
regime. The dark lines signal the addition of a single electron to
the double dot sytem. The absolute number of electrons in dot 1
and 2 is indicated in each region as ``$N_1,N_2$''. \textbf{(a)}
The absence of dark lines in the lower left region indicates that
the dot is empty there. \textbf{(b)} Zoom-in of the boxed region
of (a). Data adapted from ~\textcite{ElzermanPRB2003}.}
\label{fig:DDFewEl}
\end{figure}

\begin{figure*}[hbt]
\includegraphics[width=6.8in, clip=true]{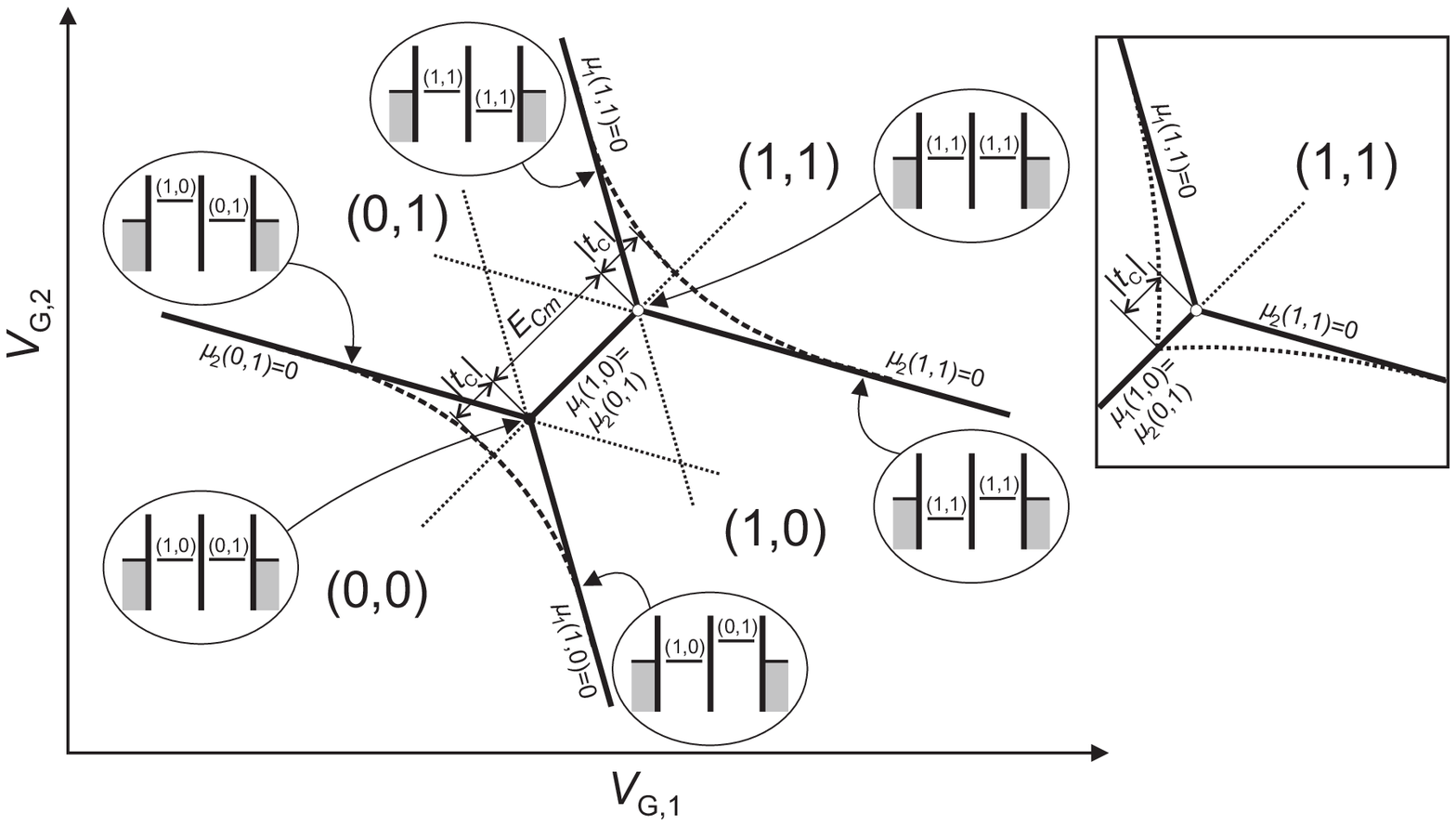}
\caption{Electrochemical potential lines around the triple points
for weak tunnel coupling (solid and dotted lines) and strong tunnel coupling
(dashed lines). Each line outlines the gate voltages where the corresponding electrochemical potential in the dots is equal to the electrochemical potential in the reservoirs (which is defined as zero). The level diagrams indicate the positions of the electrochemical potentials at several points in gate space. Inset: dotted line indicates the ``naive'' expectation for the electrochemical potentials when spin is included.} \label{Fig:DDelchem}
\end{figure*}

Figure~\ref{fig:DDFewEl} shows charge sensing data in the
few-electron regime. The absence of charge transitions in the
lower left corner of Fig.~\ref{fig:DDFewEl}a indicates that here
the double dot structure is completely depleted of electrons. This
allows the absolute number of electrons to be determined
unambiguously in any region of gate voltage space, by simply
counting the number of charge transition lines from the (0,0)
region to the region of interest. Figure~\ref{fig:DDFewEl}b
displays a zoom-in of the boxed region in Fig.~\ref{fig:DDFewEl}a.
The bright yellow lines in between the triple points in
Fig.~\ref{fig:DDFewEl}b are due to an electron moving from one dot
to the other. This changes the number of electrons on each
individual dot, while keeping the total number of electrons on the
double dot system constant.

From now on, we assume the dots are in series, such that dot 1 is
connected to the source and dot 2 to the drain reservoir. From a
similar analysis as in Section~\ref{CImodel}, the electrochemical
potential of dot 1 is found to be:
\begin{eqnarray}
&&\mu _{1}(N_{1},N_{2}) \equiv  U(N_{1},N_{2})-U(N_{1}-1,N_{2}) \nonumber \\
&\!\!=&\!\!\!(N_{1}\!\!-\!\!\frac{1}{2})E_{C1}\!\!+\!\!N_{2}E_{Cm}\!\!-\!\! \frac{E_{C1}}{|e|}(C_S V_S\!\!+\!\!C_{11}V_{G,1}\!\!+\!\!C_{12}V_{G,2}) \nonumber \\
&&+\frac{E_{Cm}}{|e|}(C_D V_D+C_{22}V_{G,2}+C_{21}V_{G,1})
\label{eqmu1}
\end{eqnarray}
\noindent where $C_{ij}$ is the capacitance between gate $j$ and
dot $i$, $C_{S}$ ($C_{D}$) is the capacitance from dot 1 (2) to
the source (drain), $E_{Ci}$ is the charging energy of the
individual dot $i$ and $E_{Cm}$ is the electrostatic coupling
energy~\footnote{Note that in ~\textcite{WielRMP2003} the
cross-capacitance terms ($C_{12}$ and $C_{21}$) are neglected.
However, they generally are significant in lateral dots.}. The
coupling energy $E_{Cm}$ is the change in the energy of one dot
when an electron is added to the other dot. One can obtain $\mu
_{2}(N_{1},N_{2})$ by simply interchanging 1 and 2 and also $C_D
V_D$ and $C_S V_S$ in Eq.~\ref{eqmu1}.

The solid lines in Fig.~\ref{Fig:DDelchem} depict the
electrochemical potentials around the triple points for low
source-drain bias. The diagrams schematically show the level
arrangement at different positions ($\mu _{1}(N_{1},N_{2})$ and
$\mu _{2}(N_{1},N_{2})$ are shown in short form $(N_{1},N_{2})$ in
the left and right dot respectively). In this case, we have
assumed the tunnel coupling to be small (i.e. negligible with
respect to the electrostatic coupling energy). This is called the
weak-coupling regime.

When the tunnel coupling $t_c$ becomes significant, the electrons are
not fully localized anymore in single dots but rather occupy
molecular orbitals that span both dots~\cite{WielRMP2003}. The molecular \textit{bonding} orbital, $\psi_B$, and the \textit{antibonding} orbital, $\psi_A$, are superpositions of the single-dot states in the left dot, $\phi_1$, and the right dot, $\phi_2$:
\begin{eqnarray}
	\psi_B &=& \alpha \phi_1 + \beta \phi_2,\\
	\psi_A &=& \beta \phi_1 - \alpha \phi_2.
\end{eqnarray}
When the single-dot states are aligned, the energy of bonding orbital is lower by $\left| t_c \right|$ than the energy of the single-dot orbitals, and the energy of the antibonding orbital is higher by the same amount.

The tunnel coupling is revealed in the charging diagram by a bending of
the honeycomb lines near the triple points, as depicted by the
dashed lines in Fig.~\ref{Fig:DDelchem}. The value of the tunnel
coupling can be determined experimentally from such a plot
by measuring the bending of the
lines~\cite{PioroPRB2005,HuttelPRB2005}, or by measuring the
charge distribution as a function of detuning between left and
right dot potentials~\cite{DiCarlo_sensing,PettaPRL2004}.

Note that, when drawing the diagram in Fig.~\ref{Fig:DDelchem}, we
assumed the electrons to be ``spinless''. Therefore, the first
electron can enter the molecular bonding orbital which takes $\left| t_c \right|$ less energy than in the case of weak tunnel coupling. The
second has to move into the antibonding orbital because of the Pauli exclusion principle. Thus, the extra energy needed to add the second electron is $E_C+2\left| t_c \right|$ ($\left| t_c \right|$ more than for the weak-coupling case) and the second triple point is pushed to higher gate voltages with respect to the weak-coupling case.

When spin is taken into account, the orbitals become doubly degenerate and both electrons can
occupy the bonding orbital. Thus, it only takes $E_C$ extra energy to add the second electron. This changes the charging diagram
drastically; namely, the dashed line in the (1,1) region (in the upper right corner in Fig.~\ref{Fig:DDelchem}) moves to the other side of the triple point! This scenario is sketched in the inset of Fig.~\ref{Fig:DDelchem}.
However, experiments on double dots with large tunnel coupling~\cite{PioroPRB2005,HuttelPRB2005} reproduce
the main diagram of Fig.~\ref{Fig:DDelchem}, and not the diagram of the inset that includes spin.
The reason for this is that the Coulomb interaction is typically
one or two orders of magnitude larger than the tunnel coupling.
Therefore, when the double dot is occupied by two electrons, the
electrons are again strongly localized. The orbital energy of the
two-electron system is then equal to the sum of the two single-dot
orbitals, which is the same as the sum of the bonding and the
antibonding orbital. When the second electron is added, the tunnel coupling energy that was gained by the first electron has to be ``paid back'', which causes the second triple point to appear at higher gate voltages than in the weak-coupling case. Therefore, Fig.~\ref{Fig:DDelchem} is
recovered when spin is included.

\subsubsection{High bias regime: bias triangles}
When the source-drain bias voltage is increased, two different
types of tunneling can occur. Up to now, we have only discussed
tunneling between aligned levels, where the initial and final
electronic state by definition have the same energy. This is
termed \textit{elastic} tunneling. However, tunneling can also
occur when there is an energy mismatch between the initial and
final state (levels are misaligned), in which case the process is
called \textit{inelastic}. For inelastic tunneling to take place,
energy exchange with the environment is required to compensate for
the energy mismatch, since the process as a whole has to conserve
energy. One important example of energy exchange is the
\textit{absorption} of one or more photons under microwave or
radio-frequency radiation, leading to photon-assisted
tunneling~\cite{WielRMP2003}. Energy \textit{emission} usually
takes place through phonons in the surrounding lattice. Note that
at cryogenic temperatures, the number of photons and phonons in
thermal equilibrium is usually negligle. Since inelastic tunneling
is a second-order process, the inelastic tunneling rate is in
general much lower than the elastic tunneling rate. However, when
there are no aligned levels elastic tunneling is suppressed and
inelastic tunneling dominates the electron transport.

The rate of inelastic tunneling between the dots is highly
sensitive to the density of states and the occupation probability
of photons and phonons. Therefore, a double dot system can be used
as a probe of the semiconductor environment~\cite{WielRMP2003} or
as a noise detector~\cite{AguadoPRL2000,OnacPRL2006}. The energy
window that is being probed is determined by the misalignment
between the levels in the two dots. Since this misalignment is
easily tuned by gate voltages, a wide range of the energy spectrum
can be investigated with very high resolution (typically of
order~1~$\mu$eV).

\begin{figure}[htb]
\includegraphics[width=3.4in, clip=true]{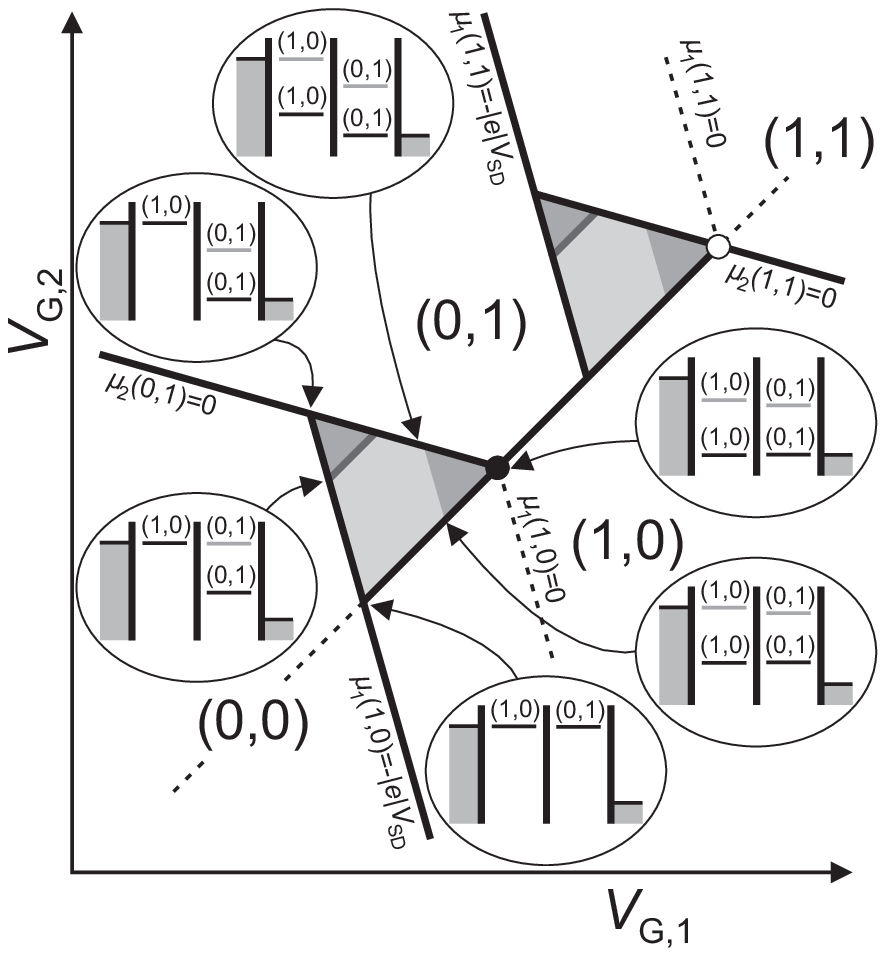}
\caption{The triple points for an applied source-drain bias
$V_{SD}$ and the drain kept at ground. A triangle is formed from
each triple point. Within this ``bias triangle'' charge transport
through the dot is energetically allowed. The grey lines and
regions in the triangles illustrate the gate voltages at which
transitions involving excited state levels play a role.
Electrochemical potentials corresponding to transitions involving
an excited state are shown in grey in the level diagrams.}
\label{Fig:DDbiastriangles}
\end{figure}

When the source-drain bias voltage is increased, the triple points
evolve into ``bias triangles'', as depicted in
Fig.~\ref{Fig:DDbiastriangles} for weak tunnel coupling. The
electron numbers refer to the triple points where the first
electrons are added to the double dot system; however, the
following discussion is valid for any number of electrons on
either dot.

To understand the electron transport within such a triangle, we
first look at the three legs. Along the base leg, $\mu_{1}(N_{1}+1,N_{2})$=$\mu _{2}(N_{1},N_{2}+1)$, and elastic
tunneling occurs. Moving along this same slope anywhere in the
plot will not change the relative alignment of the levels in the
two dots, but only change their (common) alignment with respect to
the source and drain.

Moving upwards along the left leg of the triangle, $\mu _{1}$ is
fixed ($\mu _{1}(1,0)$ is aligned with the source electrochemical
potential) and only $\mu _{2}$ is changed. At the bottom of the
triangle, the levels corresponding to transitions involving only
the ground states ($\mu _{1}(1,0)$ and $\mu _{2}(0,1)$, the black
levels in the diagrams) are aligned and elastic tunneling is
possible. Then, as $\mu _{2}$ is pulled down the levels become
misaligned and only inelastic tunneling can take place. Generally,
this will cause the current to drop. When $\mu _{2}$ is pulled
down so much that a level corresponding to a transition involving
an excited state (grey level of dot 2 in the diagrams) enters the
bias window, elastic tunneling becomes possible again, leading to
a rise in the current. When we move from this point into the
triangle, along a line parallel to the base of the triangle, these
levels remain aligned. Therefore, a line of elastic tunneling is
observed parallel to the base of the triangle (depicted as a dark
grey line). Going even farther up along the left leg of the
triangle, levels are again misaligned and only inelastic current
flows. Beyond the top of the triangle, $\mu _{2}(0,1)$ falls below
the drain electrochemical potential and the system is in Coulomb
blockade.

Moving down the upper leg from the top of the triangle, $\mu _{2}$
is fixed ($\mu _{2}(0,1)$ remains aligned with the drain
electrochemical potential) and $\mu _{1}$ is pulled down. When a
level corresponding to a transition involving an excited state in
the left dot is pulled into the bias window (grey level in left
dot in the diagrams), there are two paths available for electrons
tunneling from the source onto the first dot. A different current
can therefore be expected in the grey part in the upper right
corner of the triangle.

When the source-drain bias is inverted, the electrons move through
the dot in the opposite direction and the roles of dot 1 and dot 2 are
reversed.

In principle, the different lines and regions of elastic and
inelastic tunneling allow the full energy-level spectrum to be
determined of both dots. However, the visibility of lines and
regions depends strongly on factors such as the relative heights
of the three tunnel barriers, the efficiency of inelastic
tunneling processes (which again depends on the environment) and
relaxation within the dots. For example, if relaxation in the
first dot is much slower than the typical time for interdot
tunneling, elastic current involving excited states in both dot 1
and 2 can be observed. Another example: if the tunnel barrier
between the source and dot 1 is much higher than the other two, the
tunnel process from source to dot 1 dominates the behaviour of the
system and only excited states of dot 1 will be resolved in the
current spectrum. In practice, the system should be tuned such
that the parameter of interest has the strongest effect on the
current pattern, while the other factors can be neglected.

From a high source-drain bias measurement as discussed here, all four gate
capacitances can be deduced. With those, all the energy scales
such as charging energies, electrostatic coupling energy, tunnel
coupling and energy level spacing can be calibrated.

\begin{figure}[htb]
\includegraphics[width=3.4in, clip=true]{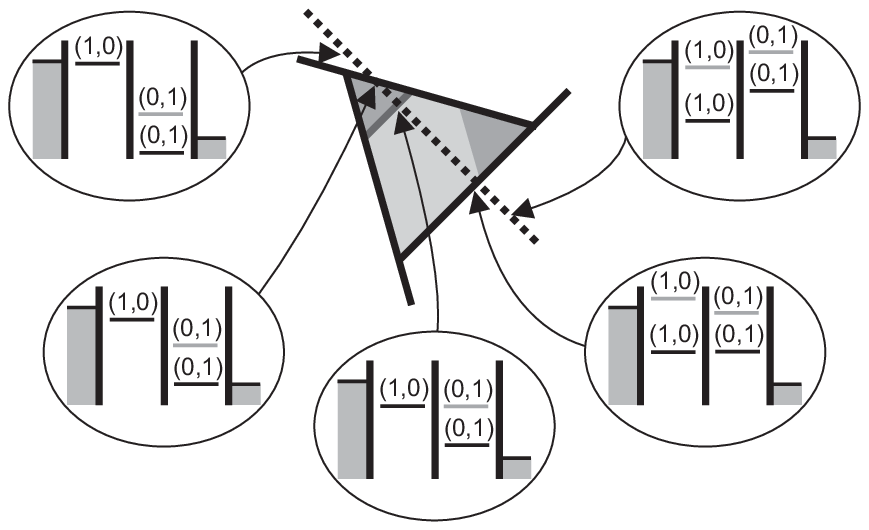}
\caption{Level diagrams for different detunings $\varepsilon$ between dot 1 and
dot 2 (dotted line in the bias triangle). Note that the average of
the levels in the two dots is kept constant, and only the
difference between the levels is changed.} \label{Fig:DDdetuning}
\end{figure}

In many of the experiments that are discussed in the following
sections, the levels in the two dots are detuned with respect to
each other, while keeping the average of the two at a constant
level. This is achieved by changing the gate voltages along a line
exactly \textit{perpendicular} to the base of the bias triangle.
The resulting axis is commonly referred to as the
\textit{detuning} axis, which we denote by $\varepsilon$. Figure~\ref{Fig:DDdetuning} displays the
level arrangements as a function of $\varepsilon$.

\subsection{Spin states in two-electron double dots}
\label{Sec:DDspinstates}
The physics of one and two-electron spin states in single dots was discussed in section
\ref{Section:SingleDotSpin}. In double quantum dots, electrons can
be transferred from one quantum dot to the other by changing the
electrostatic potentials using gate voltages. These interdot
charge transitions conserve electron spin and are governed by spin
selection rules, leading to a phenomenon called Pauli spin
blockade. In order to understand this spin blockade, we first
examine the spin states in the double dot system and the possible
transitions between these spin states, while neglecting processes
that lead to mixing of these spin states. Such mixing terms will
be introduced later, in section \ref{Section:STmixing}.

We focus on the two-electron regime, which has been the focus of
many recent double dot experiments. We work in the region of the
charge stability diagram where the occupancy of the double dot can
be (0,1), (1,1), or (0,2).

For (0,1) and (0,2) spin states, the spin physics is identical to
the single dot case since the left quantum dot is not occupied. We
briefly repeat the description of the single-dot states, as
discussed in Section~\ref{Section:SingleDotSpin}. In the (0,1)
charge state, the right dot contains a single electron. At zero
magnetic field, the two spin states are degenerate. A finite
magnetic field results in a Zeeman splitting between the spin-up
and spin-down electrons, with
$E_{\downarrow}$=$E_{\uparrow}$+$E_Z$, where $E_Z$ is the Zeeman
energy. In the (0,2) charge state, there are four possible spin
states: the singlet, denoted by $S(0,2)$, and the three triplets
$T_+(0,2)$, $T_0(0,2)$ and $T_-(0,2)$. The spin parts of the
wavefunctions of these states are:
\begin{eqnarray}
    S(0,2)&=&(\ket{\ua_2\da_2}\!-\!\ket{\da_2\ua_2})/\sqrt{2} \\
    T_+(0,2)&=&\ket{\ua_2\ua_2} \\
    T_0(0,2)&=&(\ket{\ua_2\da_2}\!+\!\ket{\da_2\ua_2})/\sqrt{2} \\
    T_-(0,2)&=&\ket{\da_2\da_2},
\end{eqnarray}
where the subscript denotes the dot in which the electron resides.
At zero magnetic field, the triplets $T(0,2)$ are separated by
$E_{ST}$ from the singlet ground state $S(0,2)$. An in-plane magnetic field
Zeeman splits the triplet spin states. As in the single dot case,
a perpendicular magnetic field tunes $E_{ST}$ and also Zeeman
splits the triplet states.

\begin{figure}[htb]
\includegraphics[width=3.2in, clip=true]{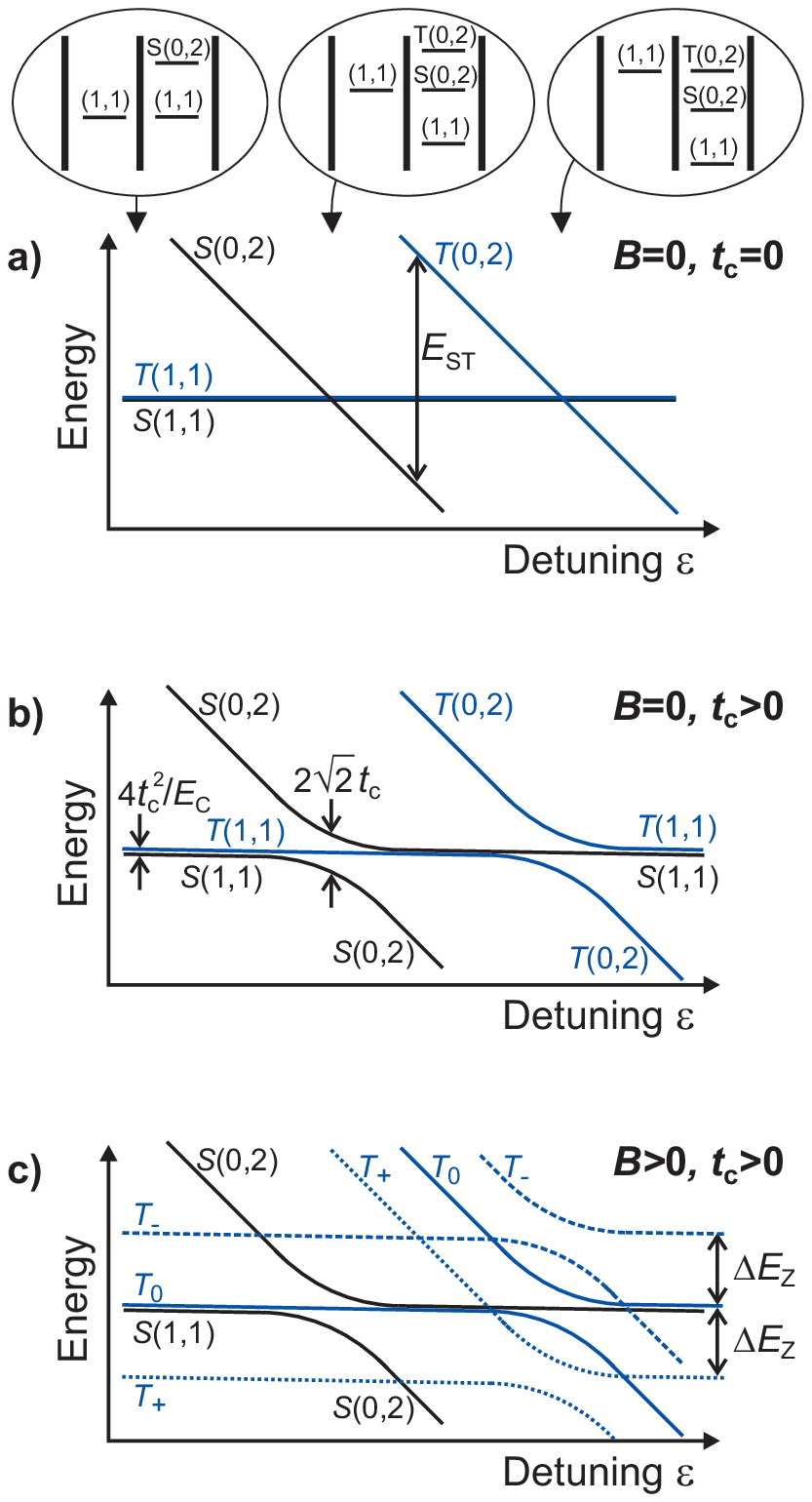}
\caption{(Color in online edition) Energies of the two-electron spin singlet and triplet
levels in a double dot as a function of detuning $\varepsilon$ between the
levels in the two dots for the case of \textbf{(a)} $B$=0 and
negligible tunnel coupling $t_c$, \textbf{(b)} $B$=0 but
significantly high value for $t_c$, and \textbf{(c)} finite $B$
and significantly high value for $t_c$. The electrochemical potentials in the two dots are indicated for three values of detuning in the diagrams on top, for case (a). Here, (1,1) denotes the electrochemical potential of both degenerate states $S$(1,1) and $T$(1,1). Note that other, higher-energy single-dot states will lead to avoided crossings at even larger values of detuning.}
\label{fig:DDSpinEnergies}
\end{figure}

In the (1,1) charge state, the two-electron states are also spin
singlets and triplets, but with the electrons in different dots:
\begin{eqnarray}
    S(1,1)&=&(\ket{\ua_1\da_2}\!-\!\ket{\da_1\ua_2})/\sqrt{2} \\
    T_+(1,1)&=&\ket{\ua_1\ua_2} \\
    T_0(1,1)&=&(\ket{\ua_1\da_2}\!+\!\ket{\da_1\ua_2})/\sqrt{2} \\
    T_-(1,1)&=&\ket{\da_1\da_2},
\end{eqnarray}
The energy difference between the lowest-energy singlet
and triplet states, $J$, depends on the tunnel coupling $t_c$ and
the single-dot charging energy $E_C$. When the single-dot levels
in the two dots are aligned,  $J=4t_c^2/E_C$ in the Hubbard
approximation~\cite{BurkardPRB1999,BurkardThesis}.
Figure~\ref{fig:DDSpinEnergies}a depicts the energies of the
two-electron spin states as a function of detuning between the two
dots, for the case of negligibly small tunnel coupling. Since the
three triplet states are degenerate, we denote them here by
$T(1,1)$ and $T(0,2)$. The diagrams indicate the electrochemical potentials in left and right dot for three values of $\varepsilon$.

Due to the tunnel coupling the (1,1) and (0,2) charge states
hybridize. In the case of spinless electrons, this would simply
result in an avoided crossing between the (1,1) and (0,2) charge
states that is characterized by a tunnel splitting, $2\sqrt{2}
t_c$. However, since the interdot transitions preserve spin, the
(1,1) singlet (triplet) states only couple to (0,2)
singlet (triplet) states. As a result, the ground state singlets
hybridize at a different value of detuning than the triplets, as depicted
in Figure~\ref{fig:DDSpinEnergies}b. At $B$=0, $E_{ST}$ is typically in the range 0.4-1~meV in electrostatically defined dots in GaAs.
This pushes the avoided crossings of the triplets far away from
the avoided crossing of the singlets, which has two interesting consequences.
First, the singlet-triplet energy difference $J$ strongly depends on detuning, allowing simple electrical control over $J$.
Second, the charge distribution of the singlet and triplet states are very different over a wide
range of detunings. For example, at the value of detuning where the
singlets have an avoided crossing, the electrons are in the charge state
(\ket{(1,1)}+\ket{(0,2)})/$\sqrt{2}$ in case they form a spin
singlet, but almost fully in the charge state \ket{(1,1)} if they
form a spin triplet. This spin-dependent charge distribution
allows readout of the spin state through charge sensing~\cite{TaylerNaturePhysics2005,EngelPRL2004}.

In a finite magnetic field, the triplet states are split by the
Zeeman energy. Figure~\ref{fig:DDSpinEnergies}c shows the energy
levels for a Zeeman splitting that exceeds the tunnel coupling.
The application of a large magnetic field can be used to decouple
the $T_+$ and $T_-$ triplet states from the $T_0$ triplet state,
thus confining the relevant state space to $S$ and $T_0$.

The singlet-triplet energy difference $J$ is often referred to as the \textit{exchange energy}. In the strict sense of the word, exchange energy refers to the difference in Coulomb energy between states whose orbital wavefunctions differ only in their symmetry (symmetric for a spin singlet and antisymmetric for a spin triplet)~\cite{ashcroft}. In the case of two electrons in a double dot, $J$ can also include a large contribution due to hybridization between the (1,1) and the (0,2) and (2,0) charge states, especially near one of the avoided crossings. In this sense, one could argue that $J$ is not a true exchange energy. However, the double dot spin system can still be described by the Heisenberg spin Hamiltonian $H= J \vec{S}_1 \vec{S}_2$, where $\vec{S}_{1,2}$ are the electron spin operators. Therefore, $J$ acts as an \textit{effective} exchange coupling. To avoid confusion, we minimize reference to $J$ as `exchange energy' in this review.

\subsection{Pauli spin blockade}
\label{Sec:PauliBlockade}
The conservation of spin in electron tunneling leads to current
rectification in dc transport in the two electron regime. This
effect, termed spin blockade or Pauli blockade, was first observed
in experiments on vertically coupled quantum
dots~\cite{OnoSpinBlock}. Later experiments in few-electron
lateral dots combined charge sensing and transport to study the
effect~\cite{JohnsonSpinBlock}. Measurements of transport in the
Pauli blockade regime provided some of the first indications that
the hyperfine interaction plays an important role in the electron
spin dynamics. Pauli blockade has also been utilized to implement
spin-to-charge conversion for read out of the spin state of
electrons in double quantum dots.

\begin{figure}[hbt]
\includegraphics[width=3.4in]{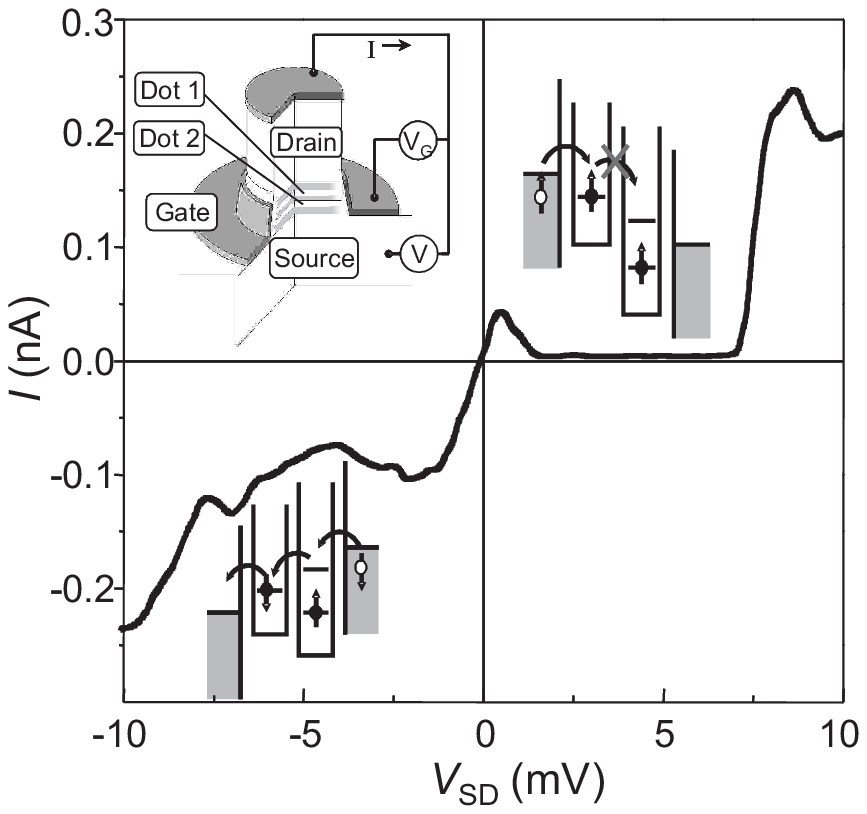}
\caption{Current ($I$) measured as a function of
source-drain voltage ($V$) in a vertical double dot system.
Non-zero current is measured over the entire range of negative
voltage. For positive bias, current is blocked in the range
2$<$V$<$7 mV. At bias voltages exceeding 7 mV, the (0,2) triplet
state becomes accessible and Pauli blockade is lifted. Insets:
Device schematic and energy level configuration at positive and
negative bias voltages. Data reproduced from ~\textcite{OnoSpinBlock}.}
\label{Fig:Ono1}
\end{figure}

The origin of Pauli blockade is schematically illustrated in the
insets of Fig.~\ref{Fig:Ono1}. At negative bias electrons are
transferred through the device in the sequence
(0,1)$\rightarrow$(0,2)$\rightarrow$(1,1)$\rightarrow$(0,1). In
this cycle the right dot always contains a single electron. Assume
this electron is spin-up. Then, in the transition
(0,1)$\rightarrow$(0,2) the right dot can only accept a spin-down
electron from the leads due to Pauli exclusion, and a $S(0,2)$
state is formed. Similarly, only a spin-up electron can be added
if the first electron is spin-down. From $S(0,2)$, one electron
can tunnel to the left dot and then out to the left lead.

In contrast, when the bias voltage is positive charge transport
proceeds in the sequence
(0,1)$\rightarrow$(1,1)$\rightarrow$(0,2)$\rightarrow$(0,1) and
the left dot can be filled from the Fermi sea with either a
spin-up or a spin-down electron, \textit{regardless of the spin of
the electron in the right dot}. If the two electrons form a
singlet state $S(1,1)$, the electron in the left dot can transfer
to the right dot forming $S(0,2)$. However, if the electrons form
one of the triplet states $T(1,1)$, the electron in the left dot
will not be able to tunnel to the right dot because $T(0,2)$ is
too high in energy. The system will remain stuck in a (1,1) charge
state until the electron spin relaxes. Since the $T_1$ time can
approach milliseconds, the current in this direction is negligible
and the dot is said to be in \textit{spin blockade}. Because it is
the Pauli exclusion principle that forbids the electrons to make a
transition from a $T(1,1)$ state to $S(0,2)$, this blockade is
also referred to as Pauli blockade.

The spin blockade effect leads to current rectification in dc
transport. Figure \ref{Fig:Ono1} shows an I-V curve taken from a
vertical double dot. Non-zero current is observed for negative
voltages. For positive bias voltage, spin blockade is observed in
the range 2-7~mV. Once the bias voltage exceeds the
singlet-triplet splitting $E_{ST}$ of the (0,2) charge state, also
the $T(0,2)$ state is energetically accessible from $T(1,1)$ and
the blockade is lifted. A theoretical model reproduces the observed current pattern~\cite{FranssonPRB2006}.

Note that the spin blockade can be lifted by photon-assisted tunneling~\cite{SanchezPSS2006}; the photon then supplies the energy needed to make the transition from $T(1,1)$ to $T(0,2)$. Interestingly, it is predicted that for a suitable choice of the applied photon frequency, the resulting pumped current can have a large spin polarization~\cite{CotaPRL2005,SanchezPRB2006}.

\begin{figure}[htb]
\begin{center}
\includegraphics[width=8.6cm]{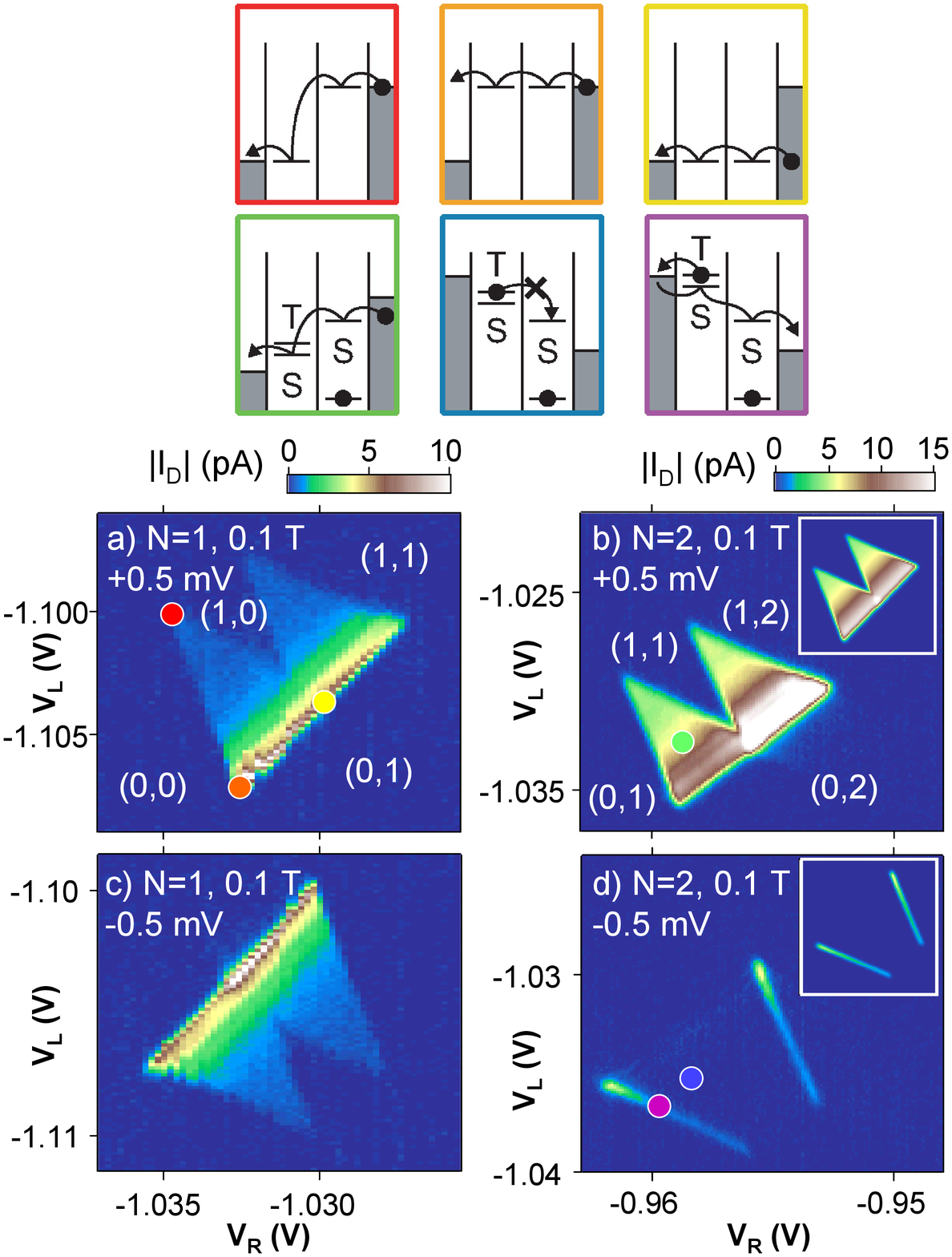}
\end{center}
\caption{(Color in online edition) Double dot current measured as a
function of $V_L$ and $V_R$ in the one and two-electron regime. In
the one-electron regime (a, c) the finite bias triangles at
negative bias mimic the positive bias data, except for an overall
change in the sign in the current. However, in the two-electron
regime, charge transport shows a striking asymmetry when the sign
of the bias voltage is changed. At negative bias in the
two-electron regime charge transport is blocked, except near the
edges of the finite bias triangles, where exchange of electrons
with the leads lifts the spin blockade. Insets show simple rate
equation predictions of charge transport. Data
reproduced from ~\textcite{JohnsonSpinBlock}.}
\label{Fig:Johnson:Transport}
\end{figure}

Pauli blockade has also been observed in lateral double dot
systems. In these systems the tunnel rates and offset energies are
easily tuned. Moreover, devices equipped with a charge sensor can
be used to measure the average occupancy of the double dot during
charge transport (see Fig.\ \ref{fig:structure} for a device
image). Figure \ref{Fig:Johnson:Transport} shows experimental data
from measuring current as a function of $V_L$ and $V_R$ in the one-
and two-electron regimes for both signs of bias
\cite{JohnsonSpinBlock}. Apart from a change in the sign of
current when the voltage is changed, the one-electron data are
mirror images of each other for positive and negative bias. This
is in contrast with data acquired in the two-electron regime,
where current flows freely for positive bias but is strongly
suppressed at negative bias due to Pauli exclusion (the voltage
bias convention in this paper is opposite to Ref.\
\cite{OnoSpinBlock}, so blockade is observed at negative bias). At
negative bias, current is only observed along the edges of the
bias triangles, where an electron can be exchanged with the leads
lifting the spin blockade (see diagrams in Fig.\ \ref{Fig:Johnson:Transport}).

\begin{figure}[hbt]
\begin{center}
\includegraphics[width=8.6cm]{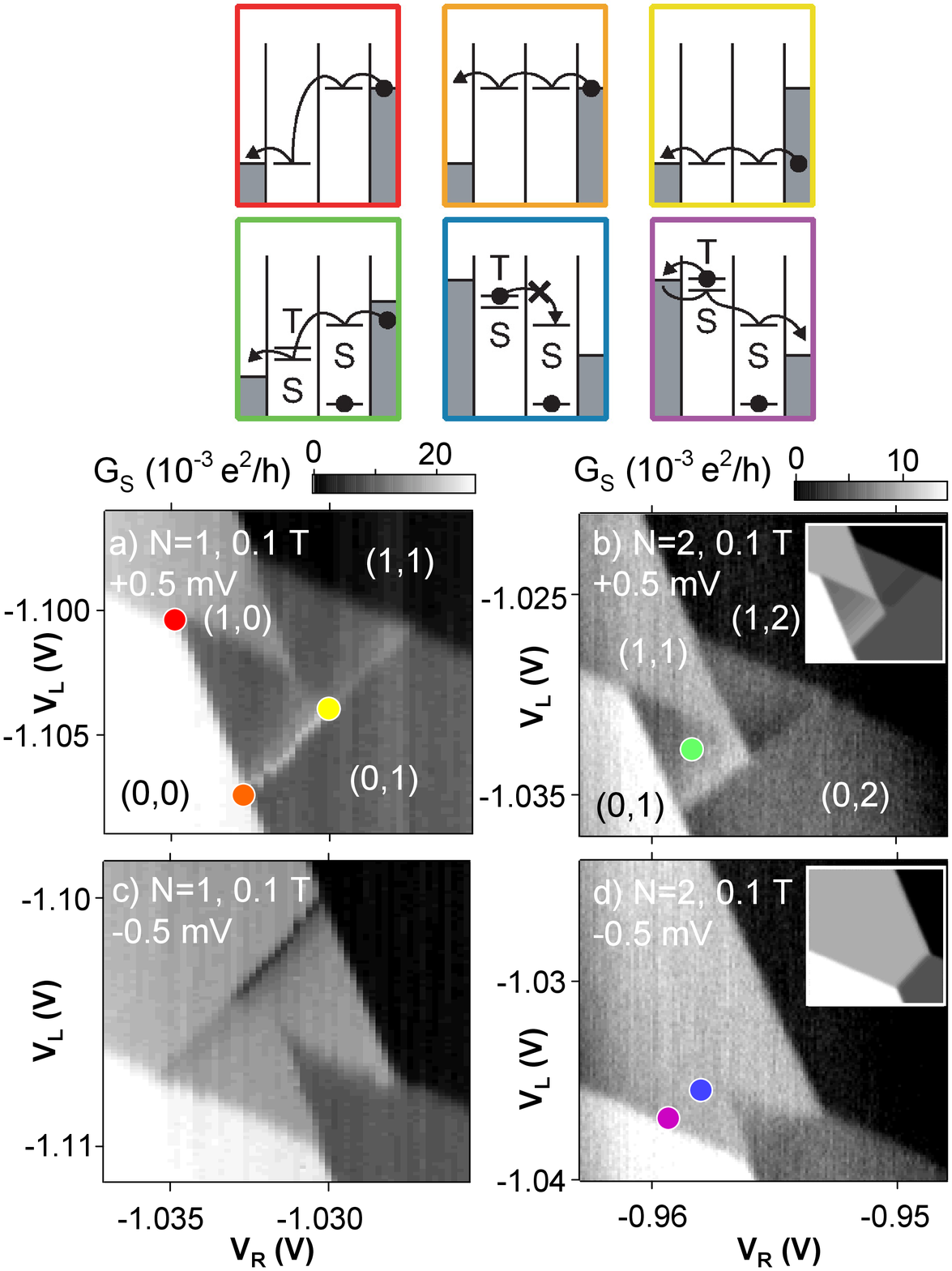}
\end{center}
\caption{(Color in online edition) Charge sensor conductance, $g_s$
measured as a function of $V_L$ and $V_R$ in the one and
two-electron regimes. The charge sensor conductance in the
lower-left finite bias triangle in the one electron regime is a
weighted average of the (0,0), (0,1), and (1,0) charge sensing
signals. In the two-electron Pauli blockade regime, the charge
sensor conductance in the finite bias triangles is pinned to the
(1,1) charge sensing value. This indicates that charge transport
is blocked by the (1,1)$\rightarrow$(0,2) charge transition.
Insets show simple rate equation predictions of charge sensor
conductance. Data
reproduced from ~\textcite{JohnsonSpinBlock}.}
\label{Fig:Johnson:Sensing}
\end{figure}

Charge sensing measurements of the time-averaged occupancy of the
double quantum dot during transport directly demonstrate that the
current rectification is due to a blocked inter-dot charge
transition. Figure \ref{Fig:Johnson:Sensing} shows the charge
sensor conductance measured as a function of $V_L$ and $V_R$ in
the one- and two-electron regimes for both positive and negative
bias. For the two-electron case at positive bias charge transport
in the lower-left bias triangle proceeds in the sequence
(0,1)$\rightarrow$(0,2)$\rightarrow$(1,1)$\rightarrow$(0,1). The
charge sensing signal in the finite bias triangles is a weighted
average of the (0,1), (0,2) and (1,1) charge sensing levels. At
negative bias in the two-electron regime charge transport in the
lower-left bias triangle follows the sequence
(0,1)$\rightarrow$(1,1)$\rightarrow$(0,2)$\rightarrow$(0,1). The
data in Fig.\ \ref{Fig:Johnson:Sensing}(d) show that the charge
sensing conductance in the finite bias triangles is practically
identical to the background (1,1) charge sensing signal. These
data indicate that the charge transition from (1,1) to (0,2) is
the limiting step in the current: precisely what is expected for a
double dot in spin blockade.

\subsection{Hyperfine interaction in a double dot: Singlet-Triplet mixing}
\label{Section:STmixing} Early experiments in semiconducting
heterostructures in the quantum Hall regime demonstrated that spin
polarized currents could be used to polarize the nuclei in the
substrate~\cite{wald94,dixon97}. These measurements gave a clear indication that
electronic effects can have a strong influence on the nuclear spin
system. So far we have ignored the consequences of the hyperfine
interaction in few-electron quantum dots, but early indications of
this important interaction were visible in the first Pauli
blockade experiments by Ono \textit{et al.} \cite{OnoSpinBlock}.
In this section we review several recent experiments that have
shown that hyperfine effect can have profound consequences on the
electron spin dynamics in GaAs quantum dots.

In GaAs quantum dots each electron spin is coupled to a bath of
nuclear spins through the contact hyperfine interaction (see
Section \ref{Section:Hyperfine}). The importance of the hyperfine
field becomes apparent when considering two spatially separated
electron spins in a double dot structure. Each electron has a
distinct orbital wavefunction and averages over a different set of
nuclei. As a result, each electron experiences a slightly
different nuclear field. The difference in the nuclear fields,
$\Delta$$B_{N,z}$, couples the singlet and triplet spin states. For
example, the z-component of the nuclear field couples $S(1,1)$ and
$T_0(1,1)$, with the Hamiltonian (in the basis ($S(1,1), T_0(1,1)$):
\begin{equation}
H= \left(\begin{array}{c c}
0 & g \mu_B \Delta B_{N,z}\\
g \mu_B \Delta B_{N,z} & 0 \end{array}\right).
\end{equation}
Since $S$ and $T_0$ are not eigenstates of this Hamiltonian, the off-diagonal terms will drive rotations between $S$ and $T_0$.
Similarly, the x-component and the y-component of the nuclear
field mix $T_+(1,1)$ and $T_-(1,1)$ with $S(1,1)$.

\begin{figure}[htb]
\includegraphics[width=8.5cm]{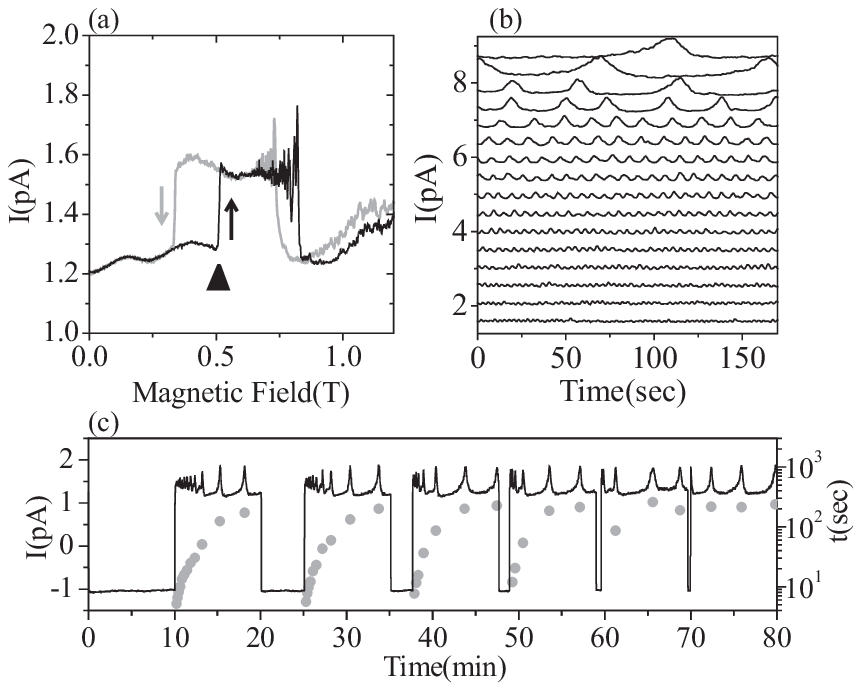}
\caption{(a) Pauli blockade leakage current as a function of
magnetic field for increasing and decreasing magnetic field
sweeps. (b) Leakage current as a function of time for fields in
the range of 0.7 T (bottom trace) to 0.85 T (top trace). (c)
Transient behavior of the leakage current measured by moving in
and out of the Pauli blockade regime using $V_{SD}$. Data
reproduced from ~\textcite{OnoPRL2004}.} \label{Fig:OnoOsc}
\end{figure}

Figure.~\ref{Fig:OnoOsc}(a) shows measurements of the ``leakage
current'' in the Pauli spin blockade regime in vertical double dots
as a function of magnetic field for two different sweep
directions~\cite{OnoPRL2004}. Upon increasing the magnetic field,
the leakage current was nearly constant until $B$=0.5~T, where a
sudden increase in the leakage current was measured. The leakage
current decreased suddenly for fields exceeding 0.9~T.
Measurements of the leakage current for the opposite magnetic
field sweep direction showed hysteretic behavior. The amount of
hysteresis decreased for slower magnetic field sweep rates. In the
high leakage current regime ($B\approx$ 0.7 T), the leakage
current showed surprising oscillations in time. The frequency of
these oscillations was a sensitive function of the external field
(see Fig.\ \ref{Fig:OnoOsc} (b)). By moving in and out of the
Pauli spin blockade regime using gate voltages, Ono \textit{et
al.} determined that the oscillatory time dependence of the
leakage current developed on a 5 minute timescale. Moreover, the
leakage current was modified by the application of cw radiation at
the $^{71}$Ga or $^{69}$Ga NMR lines. All these aspects indicate
that the nuclear spins play a major role in the observed
behaviour.

\begin{figure}[tb]
\includegraphics[width=3.4in, clip=true]{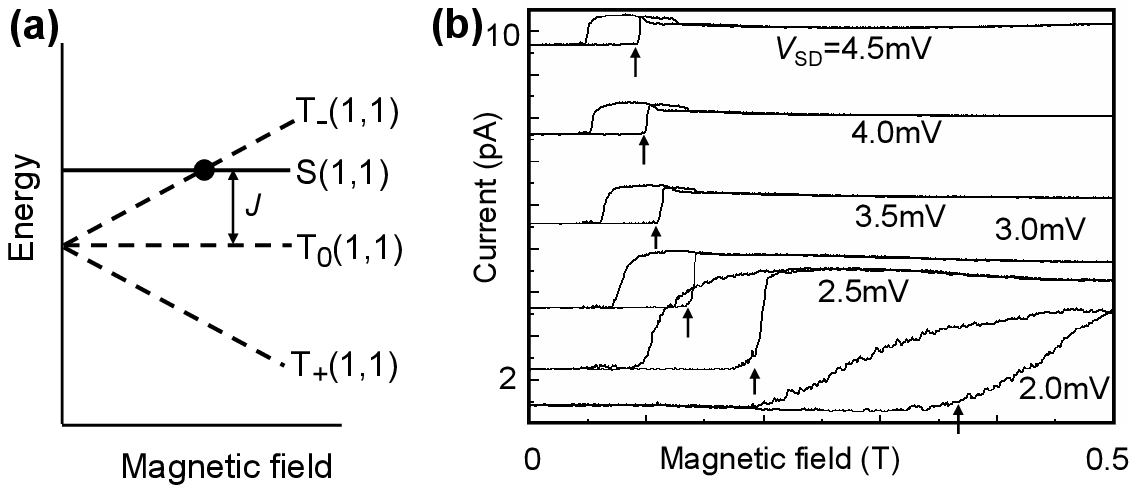}
\caption{\textbf{(a)} Effect of the Zeeman energy on $S(1,1)$ and $T(1,1)$ states,
which are separated by an energy $J$. The $S(1,1)$ and
$T_-(1,1)$ become degenerate when the Zeeman energy is
equal to $J$. \textbf{(b)} Magnetic field
dependence of the leakage current measured at various values of
source-drain voltages, $V_{SD}$, in the Pauli blockade region. Each
curve is offset by 1~pA to the top. Data reproduced from
~\textcite{Tarucha06}.} \label{Fig:DDOno2}
\end{figure}

The leakage current in the Pauli spin blockade region occurs due
to spin relaxation from $T_-(1,1)$ to $S(1,1)$ and the hysteretic
behavior observed in Fig.~\ref{Fig:OnoOsc}a can be explained in
terms of triplet-to-singlet relaxation via hyperfine-induced flip-flops with the spins of the lattice nuclei in the dot.
In the measured device, the detuning between the two dots corresponds to a point just to the right of the avoided crossing between $S(1,1)$ and $S(0,2)$in Fig.~\ref{fig:DDSpinEnergies}b. Here, $S(1,1)$ is slightly higher in energy than $T(1,1)$. This energy separation is about 10~$\mu$eV in the measured device. At zero magnetic field, this energy mismatch makes the ``flip-flop'' mechanism between electron and nuclear spins inefficient. However, the energy difference is compensated by the Zeeman energy at a magnetic field
of about 0.5~T, which is comparable to the magnetic field
where a current step is observed (indicated by a triangle in
Fig.~\ref{Fig:OnoOsc}a). On approaching this particular magnetic
field, $T(_-1,1)$ and $S(1,1)$ become degenerate (see
Fig.~\ref{Fig:DDOno2}a). Then, the hyperfine-induced $T_-(1,1)$-to-$S(1,1)$ relaxation becomes efficient, because energy as well as spin is
conserved in ``flip-flops'' between the electronic and nuclear spin systems. Many such flip-flops lead to a finite
nuclear spin polarization, which acts back on the electron as an effective magnetic field~(see Section~\ref{Sec:spin-orbit}. Because
the nuclear spin has a long lifetime (on the order of minutes), a
nuclear spin polarization accumulates to sustain the $T_-$-$S$
degeneracy condition on sweeping down the external
field~\cite{OnoPRL2004}. An \textit{increasing} nuclear field thus compensates the \textit{decreasing} external magnetic field; in other words, the effective
magnetic field resulting from nuclear spin polarization
\textit{adds} to the external field. From the considerations in
Appendix~\ref{App:nuclearfield}, we see that this implies that the electron spin is
changed by $\Delta S_z$=+1; this is consistent with
hyperfine-induced transitions from $T_-$ to $S$. The hyperfine
interactions are thus the origin of the hysteretic loop. We note
that the similar effect was well studied in ESR experiments on
two-dimensional electron gases~\cite{dobers88,Teraoka04}.

More detailed experiments on the hysteretic behavior are performed
for a vertical double dot, as shown in
Fig.~\ref{Fig:DDOno2}b~\cite{Tarucha06}. The observed hysteretic
behavior significantly depends on the source-drain voltage, that
is, the hysteretic loop becomes small and shifts to the lower
field for the higher source-drain voltage $V_{SD}$. This is well
understood in terms of the decrease of singlet-triplet energy splitting, which is
estimated from the threshold field (arrows) as a measure:
increasing $V_{SD}$ increases the detuning between two
dots. As can be seen from Fig.~\ref{fig:DDSpinEnergies}b, this decreases the energy difference between $T(1,1)$ and $S(1,1)$ and therefore a smaller magnetic field is needed to compensate for it.

\begin{figure}[htb]
\begin{center}
\includegraphics[width=6cm]{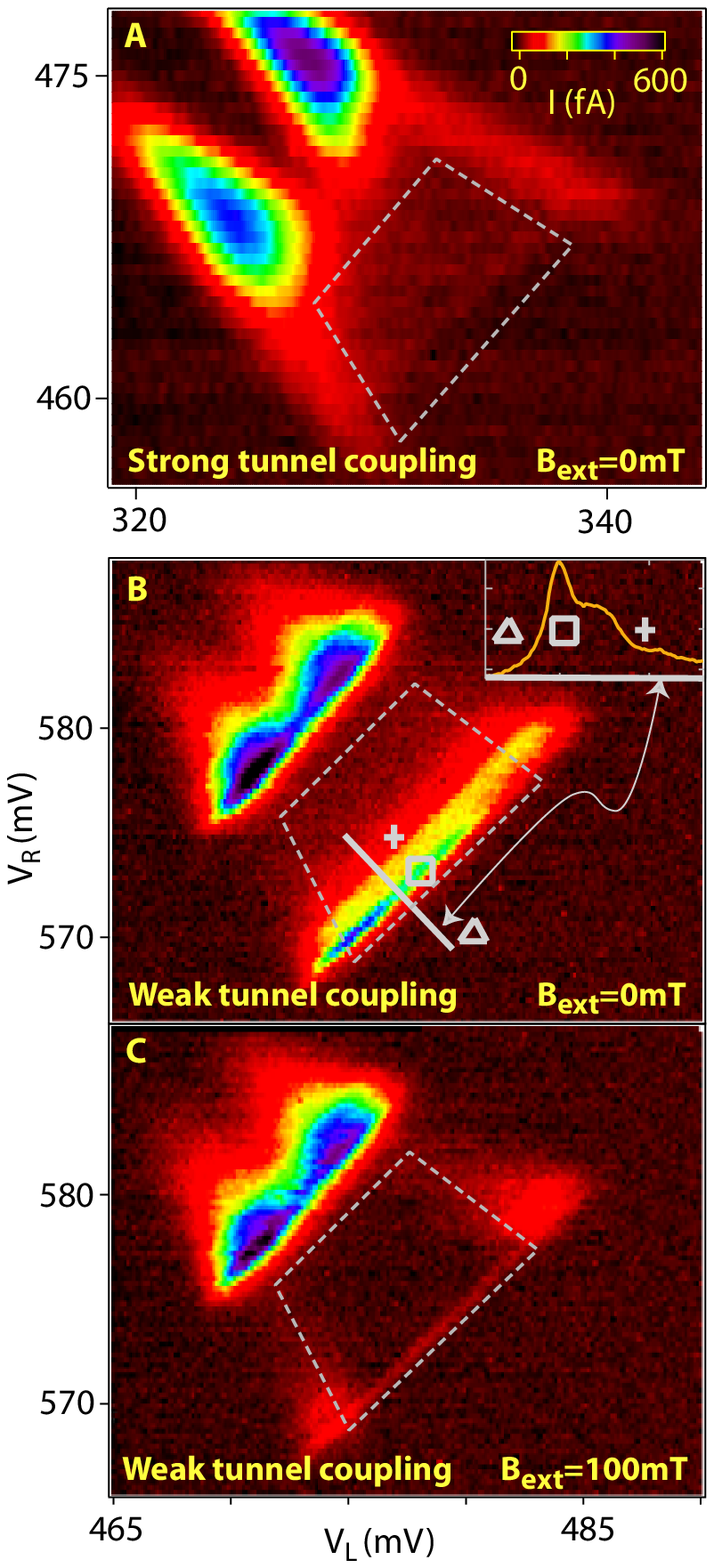}
\end{center}
\caption{(Color in online edition) Transport in the Pauli blockade regime
as a function of $V_L$ and $V_R$. (a) In the limit of strong
tunnel coupling current suppression due to Pauli blockade is
observed. For weak tunnel coupling, the Pauli blockade leakage
current displays a striking magnetic field dependence. At $B$=0 mT
(b), Pauli blockade is lifted near the (1,1)-(0,2) charge
transition (near zero detuning). In contrast, for $B$=100 mT (c),
current is suppressed due to Pauli blockade. Data
reproduced from ~\textcite{koppens05}.}
\label{Fig:KoppensLeakage}
\end{figure}

Further insight into the role of the hyperfine interaction on the
electron spin dynamics was gained in experiments on lateral
quantum dots~\cite{koppens05}. These experiments measured the
Pauli spin blockade leakage current as a function of the external
magnetic field and of the exchange splitting separating the (1,1)
singlet and triplet spin states. Figure~\ref{Fig:KoppensLeakage}
explores the tunnel coupling and magnetic field dependence of the
Pauli blockade in plots of the double dot current as a function of
$V_L$ and $V_R$. For strong interdot tunnel couplings current
rectification due to Pauli blockade is observed
(Fig.~\ref{Fig:KoppensLeakage}a). When the tunnel coupling is
reduced, the Pauli blockade is lifted and a substantial current
starts to flow, as shown in Fig.~\ref{Fig:KoppensLeakage}b.
Increasing the magnetic field to 100~ mT quenches this leakage
current (see Fig.~\ref{Fig:KoppensLeakage}c). In all cases, a
large current is observed when the voltage bias exceeds the (0,2)
singlet-triplet energy difference $E_{ST}$.

These data can be explained by considering the dependence of the
two-electron spin states on magnetic field and exchange splitting,
as illustrated in Fig.~\ref{fig:DDSpinEnergies} (see
also~\cite{jouravlev05,coish05}). For small tunnel coupling
(Fig.~\ref{fig:DDSpinEnergies}a), the singlet $S(1,1)$ and the three triplets $T(1,1)$ are nearly
degenerate over the entire range of detuning. Increasing the
tunnel coupling results in a finite exchange splitting between
$S(1,1)$ and all $T(1,1)$ states (Fig.~\ref{fig:DDSpinEnergies}b). The
inhomogeneous hyperfine fields mix $S(1,1)$ and $T(1,1)$ when the
energy splitting between these states is less than or comparable
to the nuclear field scale, $E_{nuc}$$\sim$100 neV. This condition
is achieved over the entire range of detunings for small tunnel
coupling but only at large detuning for strong tunnel coupling. An
external field splits off the $m_S=\pm$1 triplet states, $T_+$ and
$T_-$ by the Zeeman energy (Fig.~\ref{fig:DDSpinEnergies}c). When
$B<B_{N}$ these states also rapidly mix with $S(1,1)$ due to the
inhomogeneous hyperfine fields. However, when $B>B_{N}$ the
$T_+$ and $T_-$ states do not mix with $S(1,1)$ anymore, and
the spin blockade is recovered.

\begin{figure}[htb]
\includegraphics[width=8.6cm]{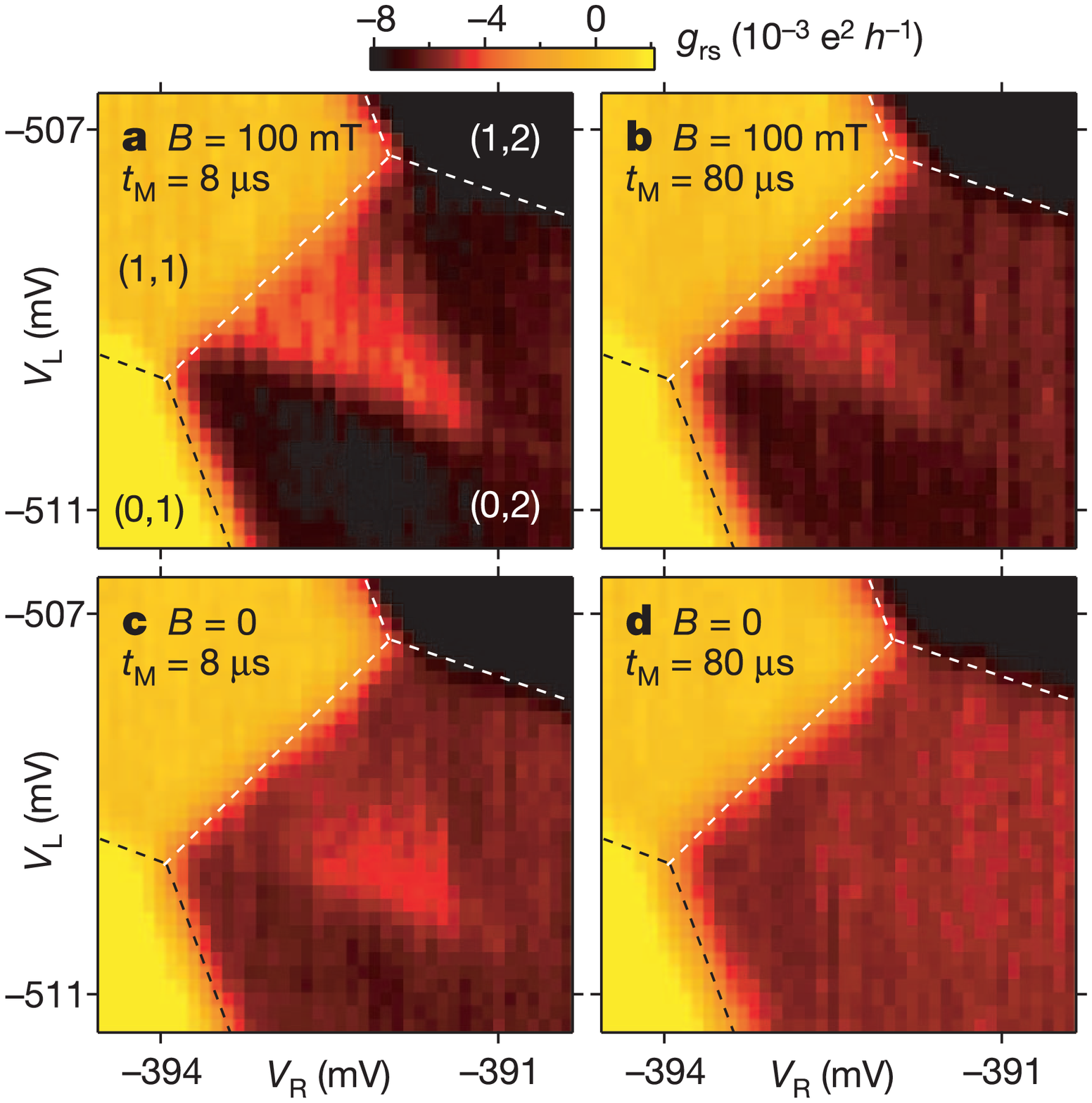}
\caption{(Color in online edition) Charge sensor conductance, $g_s$,
measured as a function of $V_L$ and $V_R$ using the $T_1$ pulse
sequence. The triangular shaped region in the (0,2) region of the
charge stability diagram, termed the ``pulse triangle", is due to
spin blocked interdot charge transitions. A relaxation time is
determined by measuring the decay of this signal as a function of
time (see (a),(b)). $T_1$ shows a strong dependence on magnetic
field. This is apparent in the $B$=0 mT data of (c), where near
zero detuning the spin states have completely relaxed on 8~$\mu$s
timescales. For long times, $\tau_M$=80 $\mu$s and $B$=0 mT the
spin states have completely relaxed and the pulse triangle is
absent. Data reproduced from ~\textcite{johnson05}.}
\label{Fig:Johnson:Triangle}
\end{figure}

Time-resolved techniques have been used to measure the hyperfine-induced relaxation
of a spin triplet state in a two-electron double quantum
dot~\cite{Petta_ST,johnson05}. These experiments used pulsed gate
techniques to prepare a spin triplet state and then measure the
decay of that spin state using spin-to-charge conversion. The
pulse experiment is performed near the (1,1)-(0,2) region of the
charge stability diagram. Gates are set in (0,1) to empty the left
dot. A pulse then shifts the gate voltages to the (1,1) region of
the charge stability diagram. A spin-up or spin-down electron
enters the left dot forming a spin singlet or spin triplet state.
A spin triplet state is formed 75$\%$ of the time. To measure the relaxation time $T_1$ of the spin triplet state, a
third pulse is applied to the device which tilts the double well
potential so that $S(0,2)$ is the ground state. In order for the
left electron to tunnel to the right dot, the (1,1) triplet state
must spin relax to $S(1,1)$ and then tunnel to $S(0,2)$. By
measuring the occupancy of the double dot as a function of the
time spent in the biased configuration the spin relaxation time can be
determined.

Representative data are shown in Fig.\ \ref{Fig:Johnson:Triangle}
as a function of magnetic field and time in the biased
configuration, $\tau_M$. In Fig.\ \ref{Fig:Johnson:Triangle} (a)
$g_s$ is plotted as a function of $V_L$ and $V_R$ with $B$=100 mT
and $\tau_M$= 8 $\mu$s. A triangular shaped signal (``pulse
triangle") appears in the (0,2) region of the charge stability
diagram, which is indicative of spin blocked transitions. For
$B$=100~mT and $\tau_m$= 80~$\mu$s the signal in the pulse
triangle reduces to a value approaching the background $S(0,2)$
charge sensing level, indicating that $\tau_M$$\sim$$\tau_{ST}$
and the spin blocked triplet states have relaxed to (1,1)S and
tunneled to $S(0,2)$. In addition to the observed time dependence
a strong magnetic field effect is observed. Reducing $B$ from 100
mT to 0 mT quenches the triplet state signal near the interdot
charge transition for the $\tau_M$=8~$\mu$s data, which implies
that spin relaxation is much faster near zero field at small
detunings. Finally, with $B$=0~mT and $\tau_M$=80~$\mu$s the
signal in the pulse triangle is completely absent indicating
complete spin relaxation.

\begin{figure}[tb]
\includegraphics[width=8.6cm]{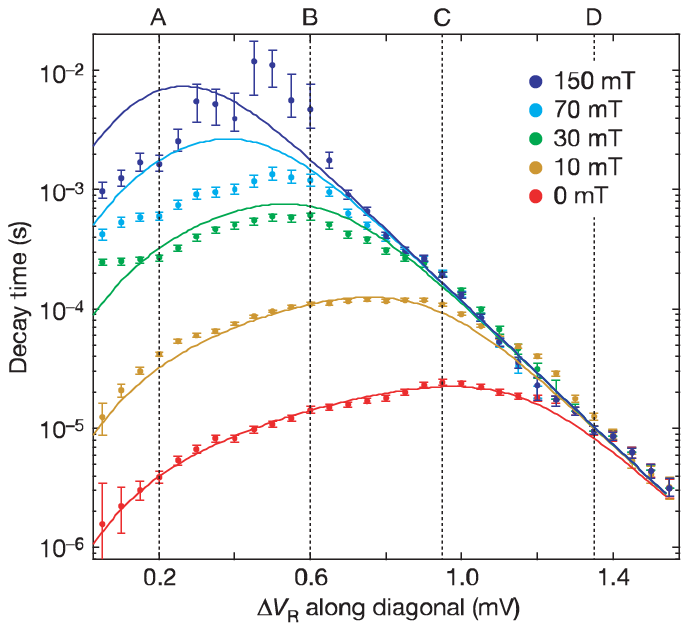}
\caption{(Color in online edition) Spin relaxation time $T_1$ of the double-dot spin triplet plotted as a
function of detuning for a range of external magnetic fields. At
low detunings, a strong magnetic field dependence is observed due
to hyperfine driven spin relaxation. At large detunings, spin
relaxation occurs due to coupling to the leads, and is independent
of magnetic field. Data are fit using a simple model of hyperfine
driven relaxation and thermally activated coupling to the leads. Data
reproduced from ~\textcite{johnson05}.}
\label{Fig:Johnson:Decay}
\end{figure}

The full dependence of the spin triplet relaxation time $T_1$ as a function
of magnetic field and detuning is plotted in Fig.\
\ref{Fig:Johnson:Decay}. At small detunings near the interdot
charge transition, $T_1$ displays a strong dependence on magnetic field. Simply increasing the field from 0 to 100~mT extends
$T_1$ from microsecond to millisecond timescales. At
larger values of the detuning, $T_1$ is nearly independent of magnetic field. This indicates that hyperfine-mediated spin relaxation is no longer dominant, but that relaxation is instead due to a coupling to the leads (which is independent of magnetic field). Experimental data are fit using a simple
model of spin relaxation from $T(1,1)$ to $S(1,1)$ followed by
inelastic decay from $S(1,1)$ to $S(0,2)$. The model assumes
hyperfine driven spin relaxation as well as a spin relaxation
contribution from coupling to the leads at large detunings. Best
fits to the model give $B_{N}$=2.8~mT, which is consistent with the
estimated number of nuclei in the dot.


\section{Coherent spin manipulation}
\label{Section:Coherent}

\subsection{Single-spin manipulation: ESR}
\label{Sec:esr} 
\begin{figure}[htb]
\includegraphics[width=7cm]{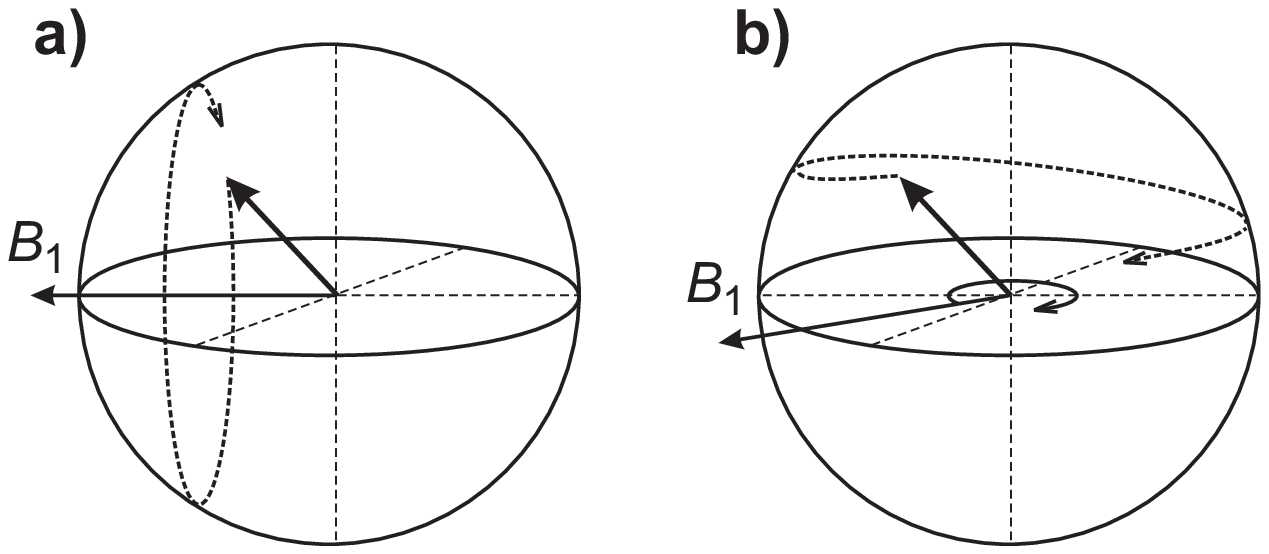}
\caption{Motion of the electron spin during a spin resonance
experiment. (a) The motion as seen in a reference frame that
rotates about the $\hat{z}$ axis at the same frequency $f_{ac}$ as
the spin itself and the resonant rotating magnetic field, $B_1$.
Naturally, the rotating field $\vec{B_1}$ lies along a fixed axis
in this rotating reference frame. An observer in the rotating
frame will see the spin simply precess about $\vec{B}_1$, with a
rate $\omega_1$. (b) An observer in the lab reference frame sees
the spin spiral down over the surface of the Bloch sphere.}
\label{fig:nutation}
\end{figure}

A variety of techniques can be used to
coherently drive transitions between the Zeeman split levels of a
single electron. The most well-known approach is electron spin
resonance (ESR), whereby a rotating magnetic field, $B_{1}$, is
applied perpendicularly to the static field $B$ along $\hat{z}$,
and on-resonance with the spin-flip transition energy ($f_{ac}=g
\mu_B B / h$), as illustrated in
Fig.~\ref{fig:nutation}\cite{poole}. Alternatively, spin rotations
could be realized by electrical or optical excitation.
Electric fields can couple spin states through the spin-orbit
interaction~\cite{DobrowolskaPRL1982,kato03a,debald05,schulte05,golovach06} (see also
Section~\ref{Sec:spin-orbit}), by making use of an inhomogeneous
static magnetic field~\cite{tokura06}, or by $g$-tensor
modulation~\cite{kato03b}. Optical excitation can induce spin
flips via Raman transitions~\cite{imamoglu99} or the optical Stark effect~\cite{gupta01}. To date,
\emph{driven} coherent rotations of a single spin in a solid have
only been realized using ESR, and only in a few specific systems ~\cite{xiao04,rugar04,jelezko04,hansonprb2006}, including in a quantum
dot~\cite{koppens06}. In addition, the \emph{free} precession of
an electron spin in a quantum dot has been observed with optical
techniques~\cite{dutt05,greilich06a}.

\begin{figure}[htb]
\includegraphics[width=7cm]{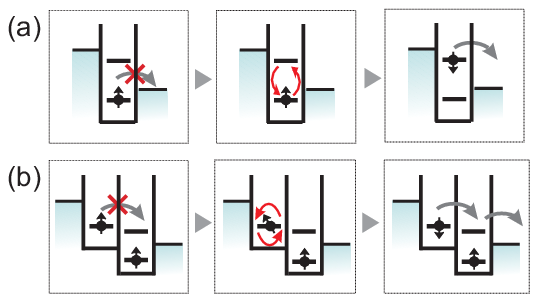}
\caption{(Color in online edition) Schematic diagrams of a single quantum dot and a double
quantum dot, illustrating electrical detection of ESR. In both
cases, transport through the system is blocked, but the blockade
is lifted when the ESR condition is satisfied and the spin of the
electron is flipped. (a) The two electrochemical potential levels
shown are the spin-up and spin-down levels of the lowest
single-electron orbital. The system is in Coulomb blockade. (b)
The levels shown are as in the discussion of spin blockade in
double dots in Section~\ref{Sec:PauliBlockade}. The Zeeman
sublevels are not shown.} \label{fig:ESR_detection}
\end{figure}

The quantum dot ESR experiment was realized by Koppens \textit{et al.}~\cite{koppens06}, and is inspired by the idea of Engel and Loss to tune a single quantum dot to Coulomb blockade with the electrochemical potential alignment as
shown in Fig.~\ref{fig:ESR_detection}a, such that the Coulomb
blockade is lifted when the electron spin is repeatedly flipped~\cite{engel01,engel02}.
In practice, this requires excitation in the microwave regime, as
the Zeeman splitting must be well above the thermal energy.
Furthermore, the alternating electric fields that are unavoidably
also generated along with the alternating magnetic field, can kick
the electron out of the dot via photon-assisted tunneling (PAT)
processes~\cite{platero04}. In early attempts to detect ESR, PAT
processes and heating of the electron reservoirs lifted the
blockade long before enough power was applied to lift the blockade
by ESR~\cite{HansonThesis2005}. Efforts to suppress the electric
field component while maximizing the magnetic component, via
optimized cavities~\cite{simovic06} or microfabricated
striplines~\cite{koppens06}, have so far not been sufficient to
overcome this problem.

Instead, ESR detection in quantum dots has been realized using two
quantum dots in series, tuned to the spin blockade regime described in
section~\ref{Sec:PauliBlockade}. The two dots are weakly coupled,
and subject to a static magnetic field $B$, such that the $T_{0}$
state is mixed with the singlet but the $T_{\pm}$ states are not.
Current is then blocked as soon as the double dot is occupied by
two electrons with parallel spins (one electron in each dot), but
the blockade is lifted when the spin in the left or the right dot
is flipped (Fig.~\ref{fig:ESR_detection} b).

In this double dot ESR detection scheme, the relevant transition
occurs between the two dots. This transition is not affected by
temperature broadening of the leads. As a result, ESR detection
can be done with Zeeman splittings much below the thermal energy,
and thus with experimentally much more accessible frequencies.
Furthermore, by applying a large voltage bias across the double
dot structure, photon-assisted tunneling processes can be greatly
suppressed.

\begin{figure}[!htb]
\includegraphics[width=7.8cm]{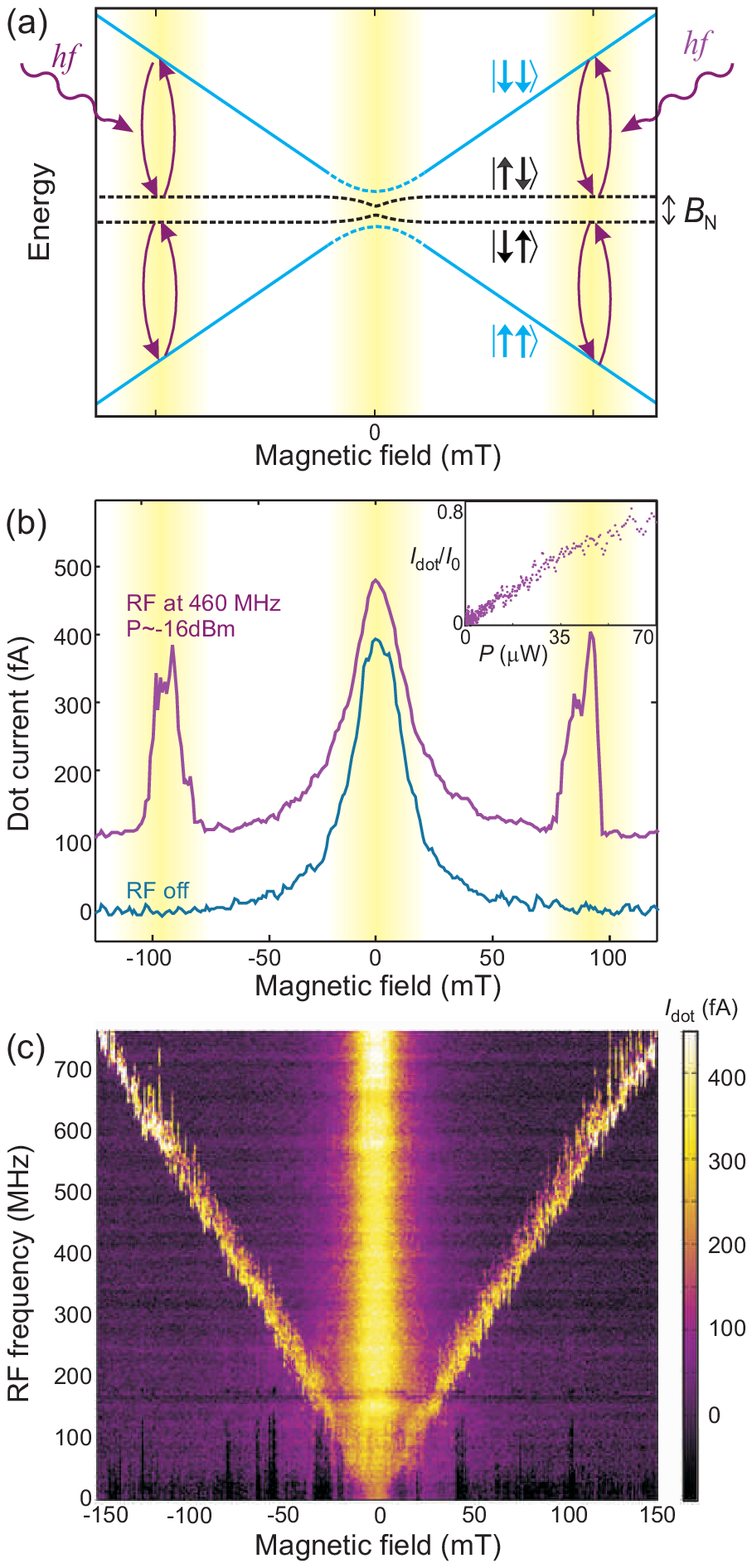}
\caption{(Color in online edition) (a) Energy levels of the two-electron spin states in the
double dot. ESR can drive transitions between the states with
parallel spins to the states with anti-parallel spins, thereby
lifting spin blockade. (b) Measured current through two quantum
dots in the spin blockade regime in the absence (blue) or presence
(pink) of an AC magnetic field. The pink curve is offset by 100 fA for clarity. 
At zero-field, all three triplets are admixed with the singlet, so here the current is never
blocked. With the AC field turned on, two satellite peaks develop
at the electron spin resonance condition. Inset: the amplitude of
the ESR peaks increases linearly with RF power ($\propto
B_{ac}^2$) before saturation occurs, as predicted~\cite{engel01}.
(c) Measured current (in color-scale) through the two dots as a
function of static magnetic field and excitation frequency. Data reproduced from ~\textcite{koppens06}.}
\label{fig:ESR_spectroscopy}
\end{figure}

The ESR response is seen clearly in transport measurements through
the double dot. When the static magnetic field is swept, clear
peaks in the current develop at the resonant field when an AC
magnetic field is turned on, as seen in
Fig.~\ref{fig:ESR_spectroscopy} (the alternating magnetic field
$B_{ac}$ can be decomposed into a component with amplitude
$B_1=B_{ac}/2$ rotating in the same direction as the spin
precession and responsible for ESR, and a component rotating the
opposite way, which hardly affects the spin because it is very far
off-resonance). The linear dependence of the satellite peak location on the RF frequency,
which is the characteristic signature of ESR, is clearly seen.
The characteristic signature of ESR is the linear dependence
of the satellite peak location on the RF frequency which is
clearly seen in the data when the RF frequency is varied from 10
to 750 MHz. A linear fit through the top of the peaks gives a
$g$-factor with modulus $0.35\pm0.01$, which is similar to the
values obtained from high-bias transport measurements in single
dots (see Section~\ref{Section:SingleDotSpin}).

\begin{figure}[htb]
\includegraphics[width=3.4in]{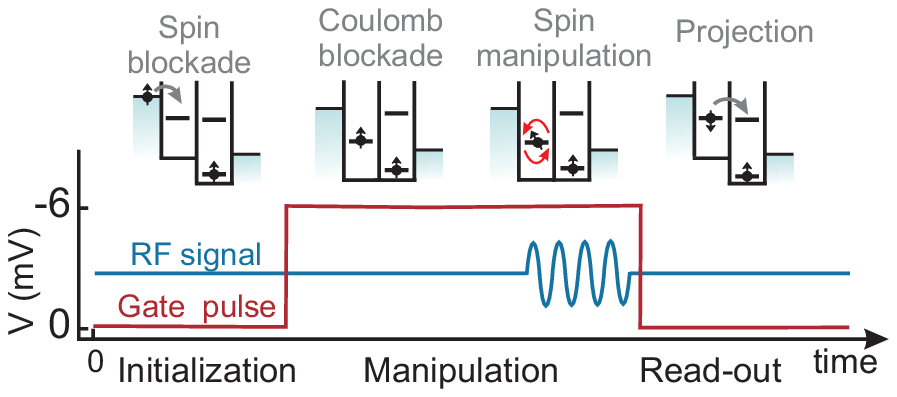}
\caption{(Color in online edition) The control cycle for coherent manipulation of the
electron spin via electron spin resonance.}
\label{fig:ESR_Rabi_concept}
\end{figure}

In order to observe coherent single-spin rotations, the system is
pulsed into Coulomb blockade while $B_{ac}$ is applied. This
eliminates decoherence induced by tunnel events from the left to
the right dot during the spin rotations. The experiment then
consists of three stages (Fig.~\ref{fig:ESR_Rabi_concept}):
initialization through spin blockade in a statistical mixture of
$\uparrow\uparrow$ and $\downarrow\downarrow$, manipulation by a
RF burst in Coulomb blockade, and detection by pulsing back for
projection (onto $S(0,2)$) and tunneling to the lead. If one of
the electrons is rotated over $(2n+1)\pi$ (with integer $n$), the
two-electron state has evolved to $\uparrow\downarrow$ (or
$\downarrow\uparrow$), giving a maximum contribution to the
current (as before, when the two spins are anti-parallel, one
electron charge moves through the dots). However, no electron flow
is expected after rotations of $2n \pi$, where two parallel spins
are in the two dots after the RF burst.

\begin{figure}[htb]
\includegraphics[width=8cm]{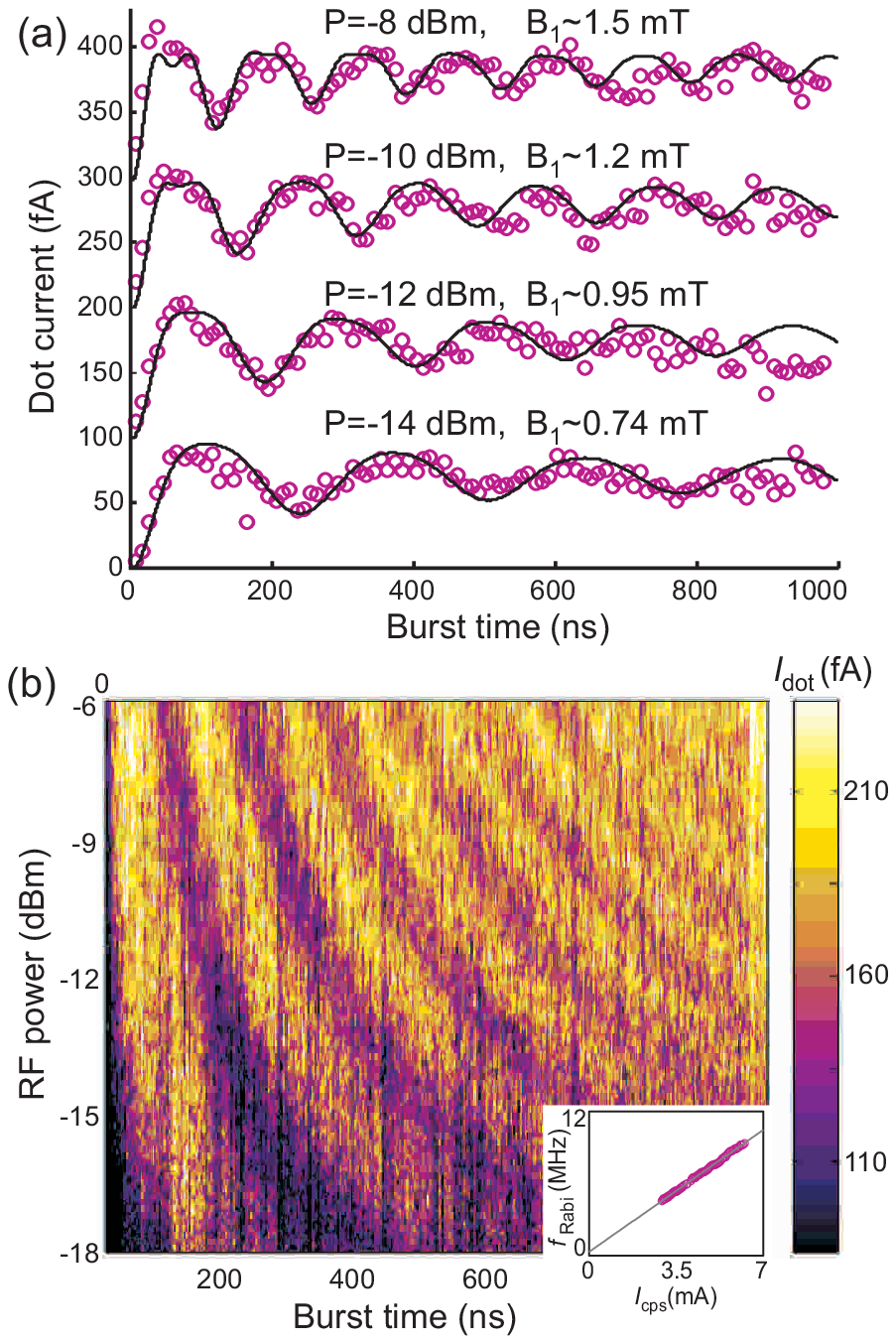}
\caption{(Color in online edition) Coherent single-spin rotations. (a) The dot current --
reflecting the spin state at the end of the RF burst -- oscillates
as a function of RF burst length (curves offset by 100 fA for
clarity). The period of the oscillation increases and is more
strongly damped for decreasing RF power ($P$ represents the
estimated power applied to the on-chip stripline). Each
measurement point is averaged over 15 seconds. The solid lines are
obtained from numerical computation of the time evolution of the
electron spins, using a simple Hamiltonian that includes $B$,
$B_{1}$ and a Gaussian distribution of nuclear fields in each of
the two dots. (b) The oscillating dot current (in colorscale) is displayed over a wide
range of RF powers (the sweep axis) and burst durations. The
dependence of the extracted Rabi frequency $f_{Rabi}$ on RF power
is shown in the inset. Data reproduced from~\textcite{koppens06}.} 
\label{fig:ESR_Rabi}
\end{figure}

The measured dot current oscillates periodically with the RF burst
length (Fig.~\ref{fig:ESR_Rabi}), demonstrating driven, coherent
electron spin rotations, or Rabi oscillations. A key signature of
the Rabi process is a linear dependence of the Rabi frequency on
the RF burst amplitude, $B_1$ ($f_{Rabi}= g \mu_B B_1/h$). This is
verified by extracting the Rabi frequency from a fit of the
current oscillations of Fig.~\ref{fig:ESR_Rabi}b with a sinusoid,
which gives the expected linear behavior
(Fig.~\ref{fig:ESR_Rabi}b, inset). The maximum $B_1$ that could be
reached in the experiment was $\sim 2$ mT, corresponding to
$\pi/2$ rotations of only 25 ns (i.e. a Rabi period of $\sim 100$
ns). The main limitation that prevented the use of larger $B_1$'s
was still photon-assisted tunneling, even in this double dot
detection scheme. From the spread in the nuclear field and the RF field strengths that could be applied, a fidelity of $75\%$ was estimated for intended $180^\circ$ rotations~\cite{koppens06}.

The oscillations in Fig.~\ref{fig:ESR_Rabi}b remain visible
throughout the entire measurement range, up to $1 \mu$s. This is
striking, because the Rabi period of $> 100$ ns is much
longer than the time-averaged coherence time $T_2^*$ of roughly 25
ns, caused by the nuclear field fluctuations (see
section~\ref{Section:Hyperfine}). The slow damping of the
oscillations is only possible because the nuclear field fluctuates
very slowly compared to the timescale of spin rotations and
because other mechanisms, such as the spin-orbit interaction,
disturb the electron spin coherence only on even longer
timescales.

Finally, we note that in this first ESR experiment, the excitation
was on-resonance with either the spin in the left dot or the spin
in the right dot, or with both, depending on the value of the
random nuclear fields in each of the two dots. In all cases,
blockade is lifted and ESR is detected. In future experiments,
controllable addressing of the spins in the two dots separately
can be achieved through a gradient in either the static or the
oscillating magnetic field. Such gradient fields can be created
relatively easily using a ferromagnet or an asymmetric stripline.
Alternatively, the resonance frequency of the spins can be
selectively shifted using local g-factor
engineering~\cite{salis01,jiang01}.

\subsection{Manipulation of coupled electron spins}
\label{Section:Coherent:TwoSpin} It has been shown that single spin
rotations combined with two-qubit operations can be used to create
basic quantum gates. For example, Loss and DiVincenzo have shown
that a XOR gate is implemented by combining single-spin rotations
with $\sqrt{SWAP}$ operations~\cite{LossDiVincenzo}. In the previous
section experiments demonstrating single-spin manipulation were
reviewed. To implement more complicated gate sequences, two-qubit
interactions are required. In this section we review experiments by Petta~\textit{et al.} that have used fast control of the singlet-triplet energy splitting in a double dot system to demonstrate a
$\sqrt{SWAP}$ operation and implement a singlet-triplet spin echo
pulse sequence, leading to microsecond dephasing times
\cite{petta05}.

A few-electron double quantum dot is used to isolate two electron
spins (the device is similar to that shown in Fig.~\ref{fig:structure}). The device is operated in the vicinity of the
(1,1) - (0,2) charge transition (see Fig.~\ref{fig:DDSpinEnergies}). The absolute number of electrons
in the double dot is determined through charge sensing with the QPC.

\begin{figure}[htb]
\includegraphics[width=8.6cm]{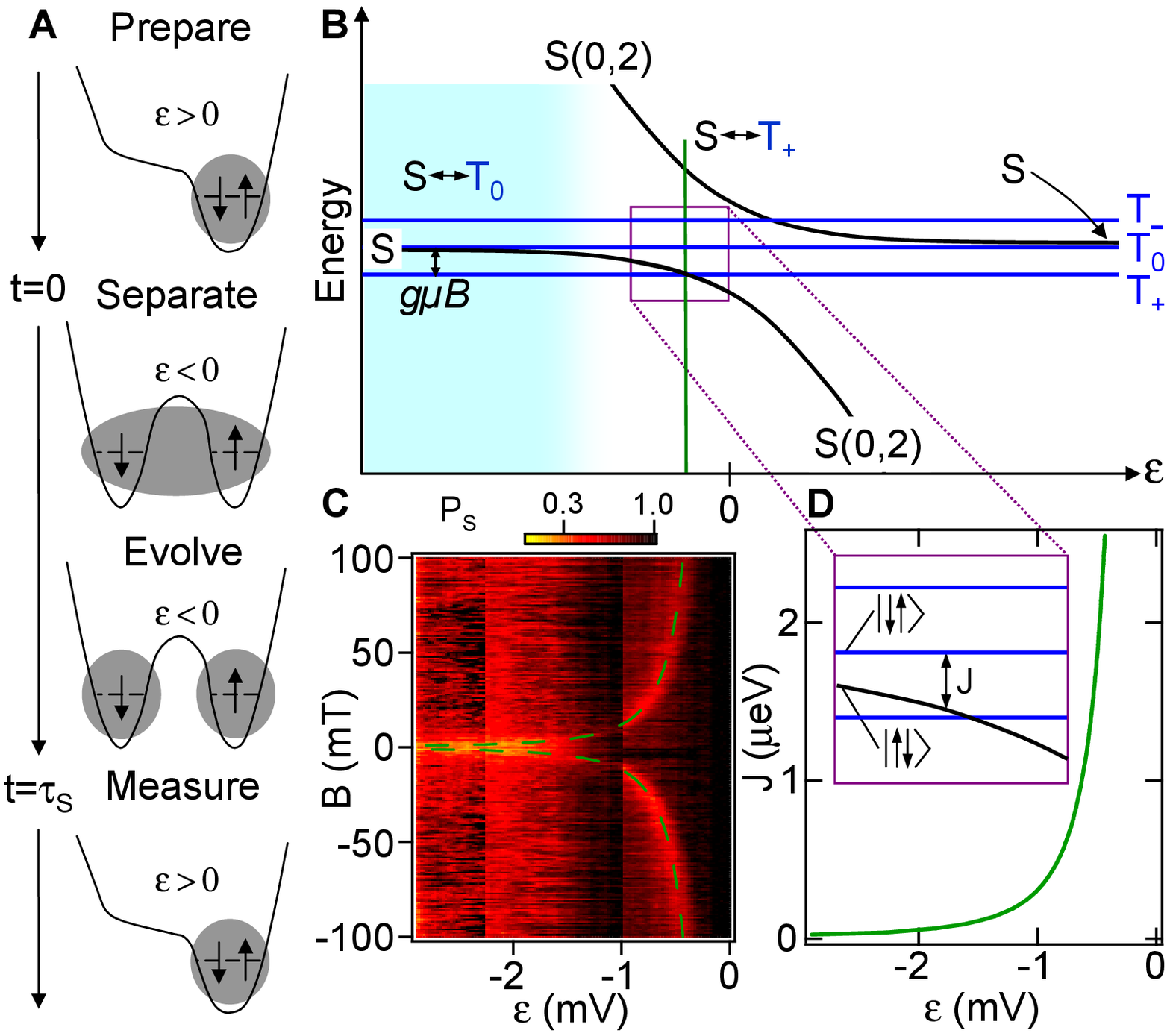}
\caption{(Color in online edition) (a) Schematic representation of pulse sequence used to
measure the singlet state decay. (b) Energy of the two-electron
spin states as a function of detuning for the singlet-triplet
qubit. Zero detuning is defined here as the value for which the energies of $S$(1,1) and $S$(0,2) are equal. At positive detunings, the ground
state is $S$(0,2). For negative detunings, and at finite fields,
$S$ and $T_{0}$ are nearly degenerate. At zero
magnetic field, the triplet states are degenerate. (c) Singlet state
probability measured as a function of detuning and magnetic field
for $\tau_S$=200 ns $\gg$$T_2^*$. (d) Hybridization of the (1,1)
and (0,2) charge states results in a gate-voltage-tunable energy
splitting, $J(\varepsilon)$. Data
reproduced from ~\textcite{petta05}.}\vspace{-0.5 cm} \label{Fig:PettaST}
\end{figure}

The energy of the two-electron spin states as a function of
detuning is illustrated in Fig.~\ref{Fig:PettaST}(b) (see also Fig.~\ref{fig:DDSpinEnergies} for a zoom-out). At
positive detuning the ground state is $S$(0,2). The triplets,
$T_{+,0,-}$(0,2) are off-scale in this plot ($E_{ST}$ = 0.4~meV). For sufficiently negative detunings, $S$(1,1) and
$T_0$(1,1) are nearly degenerate. An external magnetic field splits off
$T_+$(1,1) and $T_-$(1,1) by the Zeeman energy. Near $\varepsilon$=0,
the singlet states $S$(1,1) and $S$(0,2) are hybridized due to the
interdot tunnel coupling $t_c$. This hybridization results in an
energy splitting $J$($\varepsilon$) between $T_0$(1,1) and $S$(1,1) that
is a sensitive function of the detuning.

This energy level diagram can be mapped out experimentally by
measuring the decay of a initially prepared singlet state as a
function of magnetic field and detuning. The pulse sequence is
schematically shown in Fig.~\ref{Fig:PettaST}(a) (see also Fig.~\ref{fig:DDSpinEnergies}). The singlet
state, $S$(0,2), is prepared at positive detuning. A pulse is
applied to the device which lowers the detuning, so that the two electrons forming the spin singlet state are separated (one electron in each dot, $S(1,1)$). The spins are then held in the separated
configuration for a time $\tau_s$$>>$$T_2^*$. At locations in the
energy level diagram where $S$ is nearly degenerate with one of the
triplet states fast spin mixing will occur, thereby reducing the singlet occupation $P_S$. Fig.~\ref{Fig:PettaST}(c) shows $P_S$ as a function of $B_{ext}$ and $\varepsilon$. A strong magnetic field dependent signal is observed,
corresponding to the $S$(1,1)-$T_+$(1,1) degeneracy. For detunings
more negative than $-1.5$ mV, $S$(1,1) and $T_0$(1,1) are nearly degenerate
resulting in a reduced singlet state probability. $J(\varepsilon)$ is
extracted from the $S$(1,1)-$T_+$(1,1) degeneracy and is plotted in
Fig.~\ref{Fig:PettaST}(d). As can be seen from this figure, a shift in detuning of just a few mVs reduces $J$
from a few $\mu$eV to well below 100 neV.

Hyperfine fields were shown in Section \ref{Section:STmixing} to
lead to current leakage in the Pauli blockade regime and to enhanced
low-field spin relaxation rates. One relevant question for quantum information processing is how long two spatially separated electron spins retain coherence in this solid state environment. To directly
measure this time a two-electron spin singlet state is prepared, then the electron spins are spatially separated, and finally
correlations between the electron spins are measured at a later time. This
experiment is performed using fast electrical control of $J$. In the spatially separated (1,1)
configuration the electron spins experience distinct hyperfine
fields. In a semiclassical picture, the electron spins precess
about the local hyperfine fields. Spatial variations in $B_{N}$, $\Delta B_{N}$
result in different spin precession rates for the spatially
separated electron spins. This drives a rotation between $S$(1,1)
and the triplet states. To measure the rotation rate in the hyperfine
fields the separation time $\tau_s$ is varied.

The rotation rate in the presence of the hyperfine fields is
determined by performing spin-to-charge conversion after a
separation time $\tau_S$. Detuning is increased and the double
well potential is tilted so that $S$(0,2) is the ground state. A
separated singlet state $S$(1,1) will adiabatically follow to
$S$(0,2), while the triplets $T_{+,0,-}$(1,1) will remain in a spin
blocked (1,1) charge state for a long time, $T_1$. A charge sensing
signal of (0,2) indicates that the separated spins remain in the
singlet state, while a charge signal of (1,1) indicates that the
separated spins rotated into a triplet state.

\begin{figure}[htb]
\includegraphics[width=7cm]{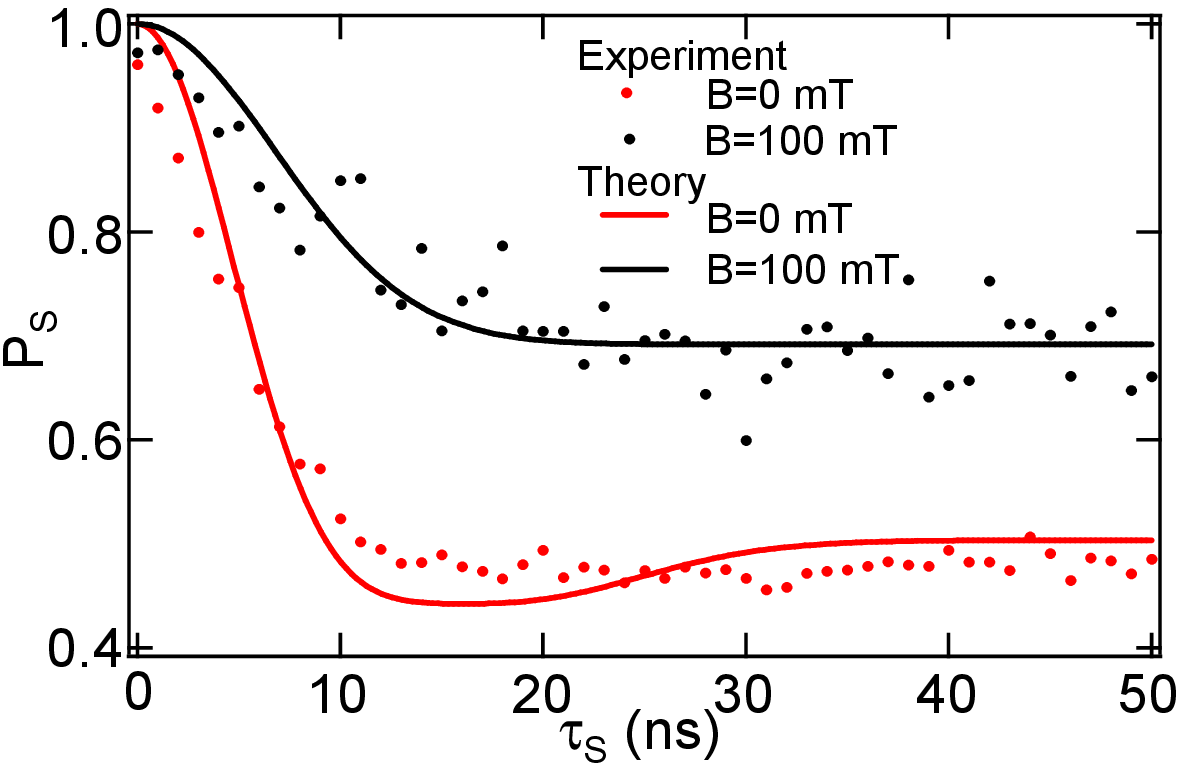}
\caption{(Color in online edition) Singlet state probability, $P_{S}$, measured as a
function of separation time, $\tau_{S}$. Data points are acquired
at $B$=0 and $B$=100 mT. Solid lines are best fits to the data
using a semiclassical model of the hyperfine interaction. Data
reproduced from ~\textcite{petta05}.}
\label{Fig:SingletDecay}
\end{figure}

\begin{figure*}[htb]
\includegraphics[width=14cm]{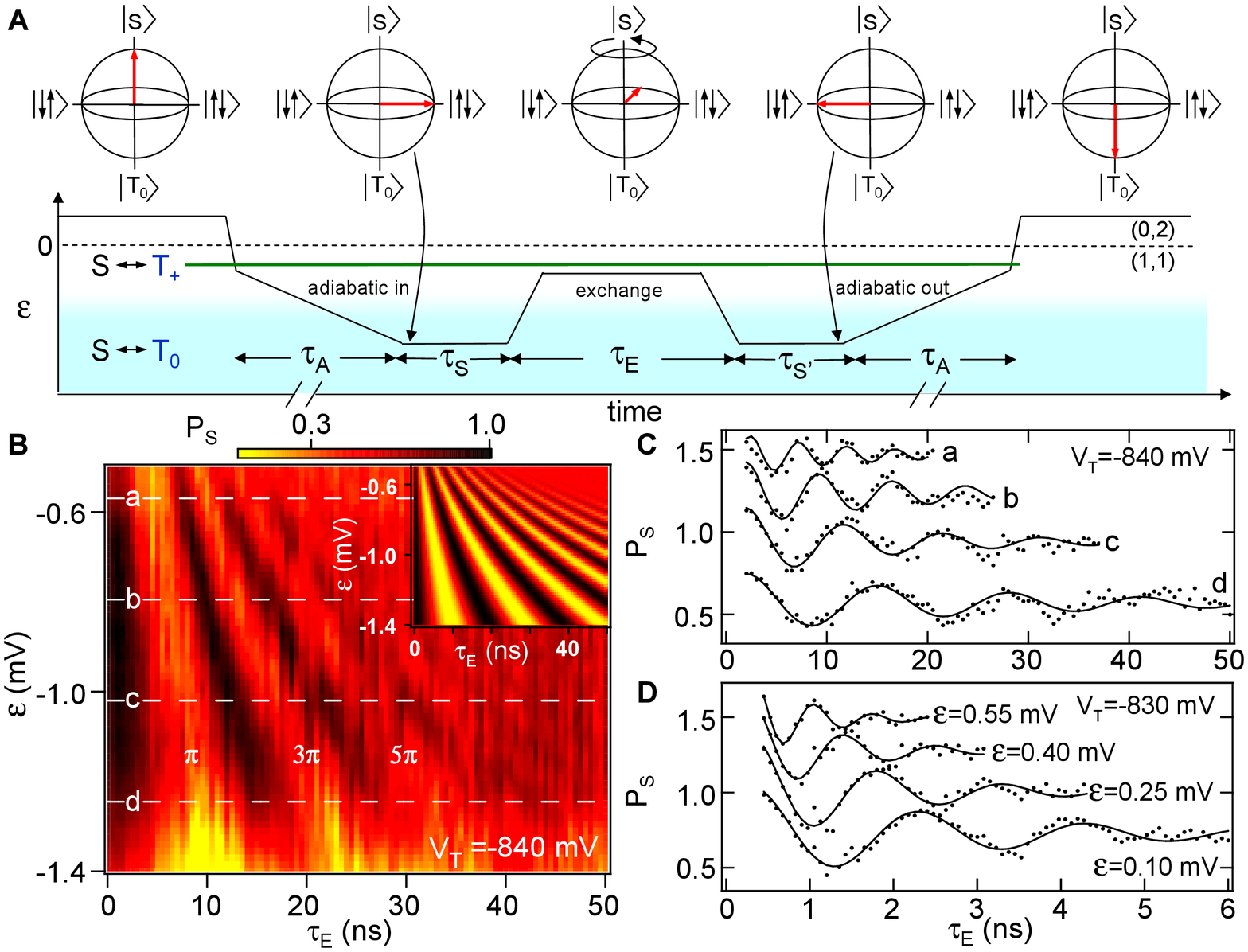}
\caption{(Color in online edition) Coherent two-electron spin state rotations. (a) Pulse
sequence. (b) Singlet state probability, $P_{S}$, measured as a
function of the pulse time, $\tau_{E}$ and detuning,
$\varepsilon$. (c) Horizontal cuts through the data in (b) show clear
oscillations in $P_{S}$. (d) By increasing the tunnel coupling, a
fast $\sqrt{SWAP}$ operation time of $\sim$180 ps is achieved. Data
reproduced from ~\textcite{petta05}.}
\label{Fig:SWAP}
\end{figure*}

Figure~\ref{Fig:SingletDecay} shows the singlet state probability
as a function of separation time $\tau_S$, $P(\tau_S)$ for $B$=0
and $B$=100 mT. For $\tau_S$$\ll$$T_2^*$ we find $P_S$$\sim$1.
$P_S$ exhibits a Gaussian decay on a 10 ns timescale and has long
time saturation values of 0.5 (0.7) for $B$=0 ($B$=100 mT). The
data are fit using a simple semiclassical model of the hyperfine
fields assuming an average over many nuclear spin configurations.
Best fits to the data give $B_{N}$=2.3 mT and $T_2^*$=10 ns. The
theoretical curves account for a measurement contrast of
$\sim$60$\%$. Long time $P_S$ values reflect the spin state
degeneracy at zero and finite fields. The measurement shows that
the separated spins lose coherence in $\sim$ 10 ns.

Two electron spins can be manipulated by fast control of the singlet-triplet energy splitting $J$. The Hamiltonian of the two-electron system in the basis ($|S \rangle$, $|T_0 \rangle$) can be approximated for zero and negative detuning by
\begin{equation}
H= \left(\begin{array}{c c}
-J(\varepsilon) & g \mu_B \Delta B_{N,z}\\
g \mu_B \Delta B_{N,z} & 0 \end{array}\right).
	\label{Eq:STHamiltonian}
\end{equation}
Note that $|S \rangle$ and $|T_0 \rangle$ are defined as the lowest-energy spin singlet state and spin $T_0$-triplet state, respectively. Whereas the $|T_0 \rangle$ state is almost a pure (1,1) orbital state in the region of interest, the state $|S \rangle$ has an orbital character that changes with detuning due to the hybridization of $S$(1,1) and $S$(0,2) (see Fig.~\ref{Fig:PettaST}(b)). Since the difference between the nuclear fields in the dots, $\Delta B_{N}$, acts on $S$(1,1) but not on $S$(0,2), the Hamiltonian~(\ref{Eq:STHamiltonian}) is not exact. However, it is a very good approximation in the regime where $t_c \gg g \mu_B \Delta B_{N}$.\footnote{For $J \gg g \mu_B \Delta B_{N}$, $|S \rangle$ and $|T_0 \rangle$ are good eigenstates. The mixing term $g \mu_B \Delta B_{N}$ only has an effect when $J \lesssim g \mu_B \Delta B_{N}$. Thus, if $|S \rangle$ is almost equal to $S$(1,1) for $J \lesssim g \mu_B \Delta B_{N}$, the use of Hamiltonian~(\ref{Eq:STHamiltonian}) is valid. If $t_c \gg g \mu_B \Delta B_{N}$, the condition $J \lesssim g \mu_B \Delta B_{N}$ is only met far away from the avoided crossing. Here, $J$ is of order $t^2_c/\left| \varepsilon \right|$, and the condition can thus be rewritten as $t^2_c/\left| \varepsilon \right| \lesssim g \mu_B \Delta B_{N}$. Since $t_c \gg g \mu_B \Delta B_{N}$, it follows that $t_c/\left| \varepsilon \right| \ll 1$. Therefore, the weight of the $S$(0,2)-component in $|S \rangle$, given by $\approx t_c/(2\sqrt{\varepsilon^2 + t^2_c})$, is indeed negligible in this regime.}

To visualize the effects of $J$ and $\Delta B_{N}$ we draw the two-electron spin states
using a Bloch sphere representation in Fig.\ \ref{Fig:SWAP}. The
effect of $J$ in this representation is to
rotate the Bloch vector about the z-axis of the Bloch sphere. An
initially prepared $|\!\!\ua\da \rangle$ spin state
will rotate into a $|\!\!\da\ua \rangle$ spin state in
a time $\tau_E$=$\pi$$\hbar$/$J(\varepsilon)$. This is a SWAP
operation. Leaving $J$ ``on'' for half of this time performs a $\sqrt{SWAP}$ operation. 

$\sqrt{SWAP}$ combined with single-spin rotations can be used to create arbitrary quantum gates. In fact, this two-spin operation allows universal quantum computing by itself, when the logical qubit is encoded in three spins~\cite{DiVincenzoNature2000}. If an inhomogenous effective magnetic field is present, encoding a qubit in just two spins is sufficient for creating any quantum gate using just the exchange interaction~\cite{Levy}. In this system, the qubit basis states are the singlet and the $T_0$ triplet state. Note that a Loss-DiVincenzo $\sqrt{SWAP}$ operation corresponds to a single-qubit rotation in the singlet-triplet basis.

\begin{figure*}[ht]
\includegraphics[width=14cm]{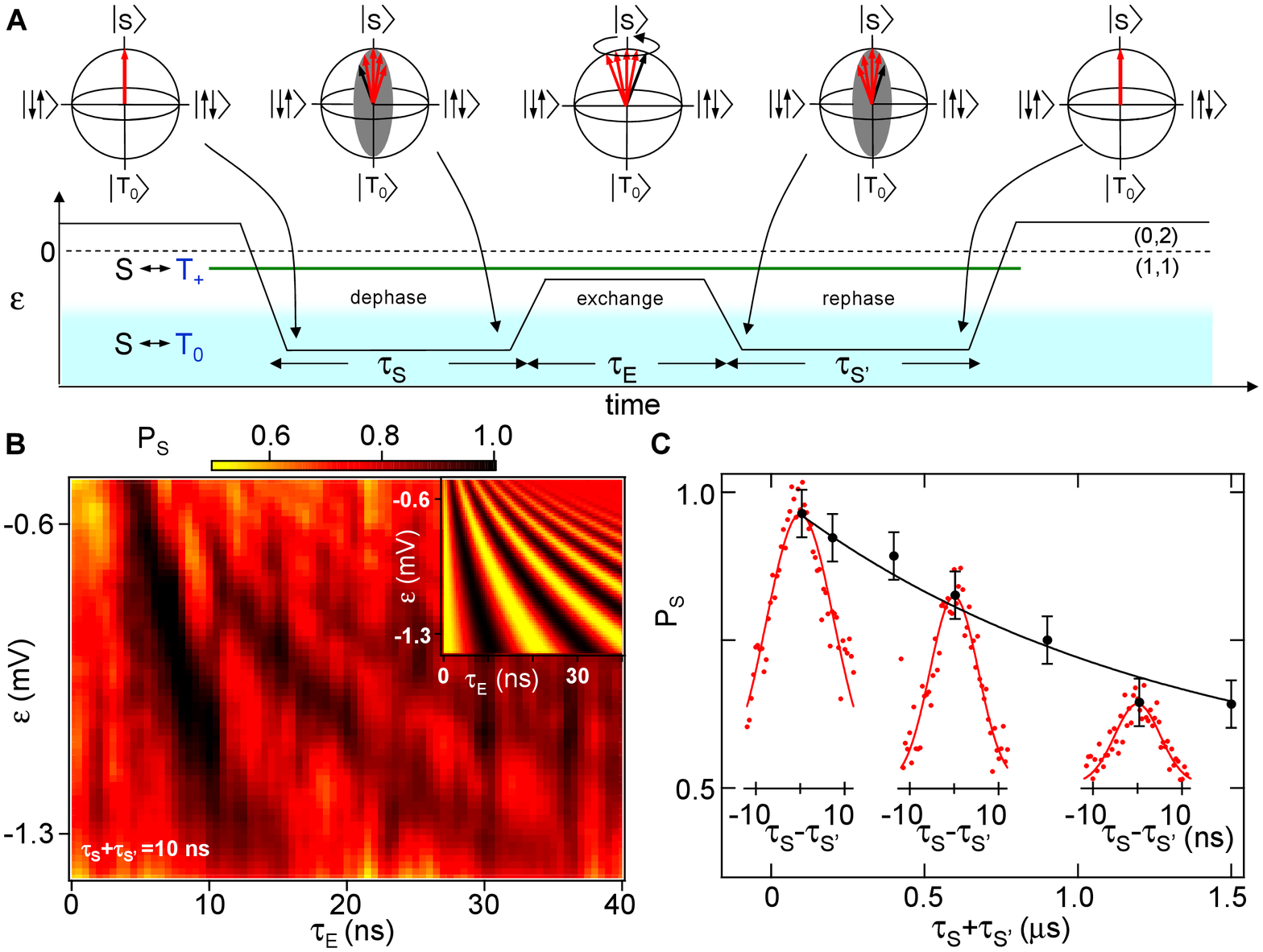}
\caption{(Color in online edition) Error correction. (a) Singlet-triplet spin echo pulse
sequence. (b) Singlet state probability, $P_{S}$, measured as a
function of pulse time, $\tau_{E}$ and detuning,
$\varepsilon$. Clear singlet state recoveries are observed for $\pi$,
3$\pi$, 5$\pi$ exchange pulses. (c) The singlet state recovery
persists out to $T_2$=1.2 $\mu$s. Data
reproduced from ~\textcite{petta05}.} \label{Fig:Echo}
\end{figure*}

A SWAP operation has been implemented using fast control of $J$. The pulse sequence is illustrated
in Fig.~\ref{Fig:SWAP}a. The system is prepared at positive
detuning in $S$(0,2). The singlet is then spatially separated by
making the detuning more negative; this is at first done fast with respect to the hyperfine mixing
rate $T_2^*$ to avoid mixing with the $T_+$ state. Once beyond the $S$-$T_+$ degeneracy, the detuning is lowered further, but now slowly with respect to $T_2^*$. This prepares the system in the ground state of the
hyperfine fields, here defined $|\!\!\ua\da \rangle$. This
state is an eigenstate of the nuclear fields and is insensitive to
hyperfine fluctuations.

To perform coherent two-electron spin rotations a pulse is
applied to the system which increases the energy splitting $J$ between $S$ and $T_0$. This drives a z-axis rotation in the Bloch sphere representation by an angle $\theta$. The rotation is then turned off
by again lowering the detuning. A spin state projection measurement is
performed by reversing the initialization process, thereby mapping $|\!\!\ua\da \rangle$$\rightarrow$
$|S(1,1) \rangle$ and $|\!\!\da\ua \rangle$ $\rightarrow$
$|T_0(1,1) \rangle$. Spin-to-charge conversion is then used to
determine the spin state.

Figure \ref{Fig:SWAP}(b) shows the measured singlet state
probability as a function of the rotation pulse time $\tau_E$ and
$\varepsilon$ during the rotation pulse. $P_S$ shows clear
oscillations as a function of both $\varepsilon$ and $\tau_E$. The
period of the oscillations agrees well with a theoretical
calculation obtained using a calibration of $J(\varepsilon)$ from the
$S(1,1)$-$T_+(1,1)$ resonance condition. Horizontal cuts through
the data are shown in Fig.~\ref{Fig:SWAP}(c). By increasing
$t_c$ and hence $J$, a fast $\sqrt{SWAP}$ operation time of 180 ps
is obtained (see Fig.~\ref{Fig:SWAP}(d)).

Fast control of the $J$ can be harnessed to
implement a singlet-triplet spin echo pulse sequence. As shown in
Fig.~\ref{Fig:SingletDecay}, hyperfine fields lead to fast
dephasing of the spin singlet state. In the Bloch sphere
representation, $B_{N}$ drives a random x-axis rotation. Since
$B_{N}$ is a fluctuating quantity, this rotation rate will vary from one experimental run to the
next. However, since the nuclear spin dynamics are much slower
than the electron spin dynamics, the hyperfine dephasing can be
reversed using a spin-echo pulse sequence.

The spin-echo pulse sequence is illustrated in Fig.~\ref{Fig:Echo}(a). The singlet state, $S$(0,2) is prepared at
positive detuning. The detuning is decreased quickly with respect
to $B_{N}$ but slowly compared to $t_c$, creating a (1,1)
singlet state. Each spin evolves in the presence of the hyperfine
fields during the separation time $\tau_S$, which in the Bloch
sphere representation corresponds to an x-axis rotation. An
exchange pulse of angle $\pi$ is applied to the system, which
rotates the Bloch vector about the z-axis of the Bloch sphere.
Exchange is turned off and the spins evolve for a time
$\tau_{S'}$. During this time, the hyperfine fields rotate the
Bloch vector back towards $S$(1,1), refocusing the spin singlet
state.

Figure \ref{Fig:Echo} (b) shows $P_S$ as a function of $\varepsilon$
and $\tau_E$ in the spin echo pulse sequence. $P_S$ shows clear
oscillations as a function of $\tau_E$. For $\pi$, 3 $\pi$, and 5
$\pi$ pulses clear singlet state recoveries are observed. To
determine the coherence time we set $\tau_S$=$\tau_{S'}$ and vary
the total separation time $t_{tot}$=$\tau_S$+$\tau_{S'}$. Figure
\ref{Fig:Echo}(c) shows $P_S$ as a function of $\tau_S$-$\tau_{S'}$ for
increasing $t_{tot}$. A singlet state recovery is observed for
$t_{tot}$ exceeding 1 microsecond. A fit to the singlet state
decay using an exponential form leads to a best fit $T_2$=1.2
microseconds. Remarkably, this spin echo pulse sequence extends
the coherence time by a factor of 100. Experiments are currently
being performed to determine the physical origin of the 1.2
microsecond decay. Possible sources of the decay are nuclear spin
evolution (see Section~\ref{Section:Hyperfine}), charge dephasing~\cite{Hu06}, and decay to $T_+$(1,1).

\section{Perspectives}

This review has described the spin physics of few-electron quantum dots. Much of this work can be evaluated within the context of spin-based classical or quantum information processing. In this context, the state-of-the-art can be best summarized by making a comparison with the first five ``DiVincenzo criteria'' \cite{DiVincenzo_criteria}, applied to the Loss-DiVincenzo proposal for encoding a logical qubit in a single electron spin~\cite{LossDiVincenzo}.

1. Have a scalable
physical system with well-defined qubits. Electron spins are
certainly well-defined qubits. The Zeeman energy difference
between the qubit states can be made much larger than the thermal
energy. The states can be measured using transport spectroscopy
(see section III). Concerning scalability it is difficult to make
predictions. In principle, circuits of solid state devices are
scalable, but evidently many practical problems will have to be surmounted.

2. Be able to initialize to a simple fiducial state such as \ket{0000...}.
By waiting until relaxation takes place at low temperature and in high magnetic field, the many-qubit ground state will be occupied with probability close to 1. Another option is to use the energy difference between the states or the different coupling to the reservoir to induce spin-selective tunneling from the reservoir onto the dot.

3. Have long coherence times. The $T_{2}$-coherence time has not been
determined extensively, but already, a lower bound of $\sim 1 \mu$s has been established at 100~mT. How quantum coherence scales with the size of the system is an interesting open question. The coherence times of qubits can be prolonged by error correction, one of the holy grails in this field. This can be done effectively only when many manipulations are allowed before decoherence takes place. The rule of thumb is that the coherence time should be at least 10$^4$ times longer than the time for a typical one- or two qubit operation.
 
4. Have a universal set of quantum gates. The Loss and DiVincenzo proposal provides two gates which together allow for universal quantum computing. Single-qubit rotations have been implemented by ESR, with a fidelity of $\sim 75\%$, and a duration of 25 ns for a $\pi/2$ rotation. The two-qubit gate is based on the two-spin \mbox{\sc SWAP} operation, which has been demonstrated as well, combined with single-spin rotations. The $\mbox{\sc SWAP}$ has already been operated at sub-ns levels (180 ps for a $\sqrt{\mbox{\sc SWAP}}$ gate), although the fidelity is yet unclear. These are only the first experimental results and further improvements are expected.

5. Permit high quantum efficiency, qubit-specific measurements. The procedure of spin to charge conversion and measuring the charge is a highly efficient measurement of the qubit state. It allows for a single-shot readout measurement with demonstrated fidelities already exceeding $90\%$. An optimization of experimental parameters can certainly increase this to $>99\%$. 
We note also that the QPC charge meter is a fairly
simple device that can be integrated easily in quantum dot
circuits.

We see that qubits defined by single electron spins in quantum dots largely satisfy the DiVincenzo criteria. As an alternative, it is also possible to encode the logical qubit in a combination of spins. For instance, when the logical qubit is encoded in three spins instead of 
a single spin, the exchange interaction by itself is sufficient for universal quantum computation~\cite{DiVincenzoNature2000}. Adding a difference in Zeeman energy between the two dots reduces the number of spins per logical qubit to two~\cite{Levy}. Coherent operations on this so-called \textit{singlet-triplet} qubit have already been experimentally demonstrated (see Section~\ref{Section:Coherent:TwoSpin}). These two- and three-electron qubit encodings eliminate the need for the technologically challenging single-spin rotations. Many more variations for encoding qubits in several electron spins have been proposed, each having its own advantages and drawbacks~\cite{wulidar02a, wulidar02b,ByrdLidar,Meier,Kyriakidis2005,TaylerNaturePhysics2005,hanson06}. In the end, the best implementation for a given system will depend on many factors that are hard to oversee at this stage.

In the near future, the natural continuation of the recent work will be to combine the various components (readout, ESR and exchange gate) in a single experiment. This may allow for new experiments exploring quantum coherence in the solid state, for instance involving non-local entanglement and testing Bell's inequalities. As another example, the precise role of quantum measurements may be investigated in this system as well.

On a longer timescale, the main challenges are scalability and coherence. Scalability is mostly a practical issue. The coherence challenge provides a number of very interesting open questions. The coherence time is currently limited by the randomness in the nuclear spin system. If this randomness is 
suppressed the coherence time will become longer. Polarization of the nuclei turns out not to be very efficient, except for polarizations $> 99.9\%$. As an alternative, the nuclear spins could be put and kept in a particular, known quantum state~\cite{giedke06,klauser06,stepanenko06}. 

It is yet unknown if nuclear spins can indeed be controlled up to
a high level of accuracy. A completely different approach would be
using a different material. The isotopes of the III-V
semiconductors all have a non-zero nuclear spin. In contrast, the
group IV semiconductors do have isotopes with zero nuclear spin.
If spin qubits are realized in a material that is isotopically
purified to for instance $^{28}$Si or $^{12}$C only, the hyperfine interaction is completely absent.

We believe that the techniques and physics described in this review will 
prove valuable regardless of the type of quantum dot that is used to confine the 
electrons. The unprecedented level of control over single electron spins will 
enable exploration of new regimes and pave the way for tests of simple quantum 
protocols in the solid state.

\section*{Acknowledgments}
We acknowledge the collaboration with many colleagues, in particular those from our institutes in Tokyo, Delft and at Harvard. We thank David Awschalom, Jeroen Elzerman, Joshua Folk, Toshimasa Fujisawa, Toshiaki Hayashi, Yoshiro Hirayama, Alex Johnson, Frank Koppens, Daniel Loss, Mikhail Lukin, Charlie Marcus, Tristan Meunier, Katja Nowack, Keiji Ono, Rogerio de Sousa, Mike Stopa, Jacob Taylor, Ivo Vink, Laurens Willems van Beveren, Wilfred van der Wiel, Stu Wolf and Amir Yacoby.

The authors acknowledge financial support from the DARPA-QuIST program.
RH, LPK, and LMKV acknowledge support from the Dutch Organization for Fundamental Research on Matter (FOM) and the Netherlands Organization for Scientific Research (NWO). RH acknowledges support from CNSI, AFOSR, and CNID. JRP acknowledges support from the ARO/ARDA/DTO STIC program.
ST acknowledges financial support from the Grant-in-Aid for Scientific Research A (No. 40302799), the Special Coordination Funds for Promoting 
Science and Technology, MEXT, CREST-JST.

\appendix
\section{Sign of the ground state spin and the nuclear fields in GaAs}
\label{App:nuclearfield} In this Appendix we derive the sign of
the ground state of electron and nuclear spins in GaAs and the
sign and magnitude of the effective magnetic field felt by
electrons due to thermal and dynamical nuclear polarization.
\subsection{Sign of the spin ground states}
We define the spin to be `up' if it is oriented in the direction
of the externally applied magnetic field $B_z$ along the $z$-axis.
In other words, an electron with spin \textbf{\textit{S}} is
spin-up if the quantum number for the $z$-component of the spin, $S_z$, is positive. The
magnetic moments associated with the electron spin
\textbf{\textit{S}} and the nuclear spin \textbf{\textit{I}} are
\begin{eqnarray}
     \mbox{\boldmath$\mu$}_S =& -g_S \frac{|e|}{2m_e}\mbox{\boldmath$S$};\ \  &\mu_{S,z} = -g_S \mu_B S_z\\
     \mbox{\boldmath$\mu$}_I =& g_I \frac{|e| }{2m_p}\mbox{\boldmath$I$};\ \  &\mu_{I,z} = g_I \mu_N I_z
\end{eqnarray}
where $\mu_B$ and $\mu_N$ are the Bohr magneton (57.9~$\mu$eV/T)
and the nuclear magneton (3.15~neV/T), respectively (note that in our notation, the spin angular momentum along $z$ is given by $\hbar S_z$). The difference in the 
sign of the magnetic moments is due to the difference in the
polarity of the electron and proton charge. The Zeeman energy is
given by $E_{Z}=-\mbox{\boldmath$\mu\!\cdot\! B$}$. Since both
free electrons and protons have a positive $g$-factor, the spins
in the ground states of a free electron (spin-down) and a proton
(spin-up) are anti-parallel to eachother.

The nuclear $g$-factors of the isotopes in GaAs are all positive:
$g_I(^{69}Ga)$ =+1.344, $g_I(^{71}Ga)\!$  =+1.708 and
$g_I(^{75}As)\!$ =+0.960. The electron $g$-factor in GaAs is
negative ($g_S$=-0.44). Hence, both the nuclei and the electrons
in the ground state in GaAs have their spin aligned parallel to
the external field, i.e. they are spin-up.

\subsection{Sign and magnitude of the thermal nuclear field}
The two Ga-isotopes, $^{69}$Ga (60.11\% abundance) and $^{71}$Ga
(38.89\% abundance), and $^{75}$As all have nuclear spin 3/2. We
can calculate the thermal average of the spin $<\!\!I\!\!>$ of
each isotope using the Maxwell-Boltzmann distribution. For
example, at 10 T and 20 mK, $<\!\!I\!\!>_{^{69}Ga}$=+0.30,
$<\!\!I\!\!>_{^{71}Ga}$=+0.38 and $<\!\!I\!\!>_{^{75}As}$=+0.22.
Then, following Paget \textit{et al.}~\cite{paget77} we
approximate the effective field, generated by the polarization of
isotope $\alpha$ through the hyperfine contact interaction, by
\begin{equation}
    B_{N,\alpha}=b_N(\alpha)\left\langle I\right\rangle_\alpha
    \label{Bbrelation}
\end{equation}
with $b_N(^{69}Ga)$=-0.91 T, $b_N(^{71}Ga)$=-0.78 T and $b_N(^{75}As)$=-1.84 T (formula 2.17-19 of ~\textcite{paget77}). Since $\left\langle I\right\rangle_\alpha$ is always positive in thermal equilibrium, we derive from equation \ref{Bbrelation} that the thermal nuclear field acts \textit{against} the applied field. 

\subsection{Sign of the dynamic nuclear field}
A nuclear polarization can build up dynamically via flip-flop
processes, where an electron and a nucleus flip their spin
simultaneously. Because of the large energy mismatch between
nuclear and electron Zeeman energy, a flip-flop process where an
electron spin is excited is very unlikely, since the required
energy is not available in the system ($\Delta
E_{Z,nucl}\!\ll\!k_BT\!\ll\!\Delta E_{Z,el}$). Therefore, we only
consider the flip-flop processes where the electron flips its spin
from down to up ($\Delta S_z$=+1), thereby \textit{releasing} the
Zeeman energy. This brings the nucleus to a different spin state
with $\Delta I_z$=-1. Many of these processes can dynamically
build up a considerable polarization, whose sign is opposite to
that of the thermal nuclear field. This has already been observed
in the ESR experiments on 2DEGs (see e.g. ~\textcite{dobers88}),
where the excited electron spin relaxes via a flip-flop process.
The external field at which the ESR field is resonant shifts to
lower values after many of these processes, indicating that indeed
this nuclear field \textit{adds} to the external field.

\bibliography{hanson_rmp07}
\end{document}